\documentclass[longauth]{aa} 

\usepackage{graphicx}
\usepackage{txfonts}
\usepackage{subfig}
\usepackage{hyperref}
\hypersetup{colorlinks=true,urlcolor=blue, linkcolor=blue, citecolor=blue}
\usepackage{gensymb}

\begin{document}

    \title{The SPHERE view of the Chamaeleon I star-forming region\thanks{Based on data collected at the European Southern Observatory,
Chile (ESO Programs 096.C-0333(A), 198.C-0209(F), 098.C-0486(A), 098.C-0760(B), 099.C-0147(B), 099.C-0891(B), 0100.C-0329(A), 0101.C-0303(A), 0102.C-0243(A), 2102.C-5050(A), 1100.C-0481(D), 1100.C-0481(Q), 1104.C-0415(A)), 1104.C-0415(E), 106.21HJ.001.}}

   \subtitle{The full census of planet-forming disks with GTO and DESTINYS programs}

   \author{C. Ginski\inst{1,2,3} 
   \and A. Garufi\inst{4}
   \and M. Benisty\inst{5,6}
   \and R. Tazaki\inst{6,7,3}
   \and C. Dominik\inst{3}
   \and \'{A}. Ribas\inst{8}
   \and N. Engler\inst{9}
   \and T. Birnstiel\inst{10,11}
   \and G. Chauvin\inst{5,6}
   \and G. Columba\inst{12}
   \and S. Facchini\inst{13}
   \and A. Goncharov\inst{1}
   \and J. Hagelberg\inst{14}
   \and T. Henning\inst{15}
   \and M. Hogerheijde\inst{2}
   \and R.~G. van Holstein\inst{16}
   \and J. Huang\inst{17}
   \and T. Muto\inst{18}
   \and P. Pinilla\inst{19}
   \and K. Kanagawa\inst{20}
   \and S. Kim\inst{21}
   \and N. Kurtovic\inst{22}
   \and M. Langlois\inst{23}
   \and C. Manara\inst{24}
   \and J. Milli\inst{6}
   \and M. Momose\inst{25}
   \and R. Orihara\inst{25}
   \and N. Pawellek\inst{26}
   \and C. Pinte\inst{27,6}
   \and C. Rab\inst{10,22}
   \and T.~O.~B. Schmidt\inst{28}
   \and F. Snik\inst{2}
   \and Z. Wahhaj\inst{16}
   \and J. Williams\inst{29}
   \and A. Zurlo\inst{30,31}
   }

   \institute{School of Natural Sciences, Center for Astronomy, University of Galway, Galway, H91 CF50, Ireland\\
              \email{christian.ginski@universityofgalway.ie}
             \and Leiden Observatory, Leiden University, P.O. Box 9513, 2300 RA Leiden, The Netherlands
             \and Anton Pannekoek Institute for Astronomy, University of Amsterdam, Science Park 904, 1098 XH Amsterdam, The Netherlands
             \and INAF, Osservatorio Astrofisico di Arcetri, Largo Enrico Fermi 5, I-50125, Firenze, Italy
             \and Universit\'{e} C\^{o}te d'Azur, Observatoire de la C\^{o}te d'Azur, CNRS, Laboratoire Lagrange, Bd de l'Observatoire, CS 34229, F-06304 Nice cedex 4, France
             \and Universit\'{e} Grenoble Alpes, CNRS, Institut de Plan\'{e}tologie et d’Astrophysique (IPAG), F-38000
Grenoble, France
             \and Astronomical Institute, Tohoku University, Sendai 980-8578, Japan
             \and Institute of Astronomy, University of Cambridge, Madingley Road, Cambridge CB3 0HA, UK
\and ETH  Zurich,  Institute  for  Particle  Physics  and  Astrophysics,  Wolfgang-Pauli-Strasse 27, 8093 Zurich, Switzerland
\and
        University Observatory, Faculty of Physics, Ludwig-Maximilians-Universität München, Scheinerstr. 1, 81679 Munich, Germany
        \and
        Exzellenzcluster ORIGINS, Boltzmannstr. 2, D-85748 Garching, Germany
        \and
        Department of Physics and Astronomy "Galileo Galilei" - University of Padova, Vicolo dell’Osservatorio 3, 35122 Padova, Italy
        \and
        Dipartimento di Fisica, Universit\`a degli Studi di Milano, Via Celoria, 16, Milano, I-20133, Italy
        \and
        Observatoire de Gen\`eve, Universit\'e de Gen\`eve, 51 Ch. des Maillettes, 1290, Sauverny, Switzerland
        \and Max-Planck-Institut f\"{u}r Astronomie, K\"{o}nigstuhl 17, 69117, Heidelberg, Germany.
             \and European Southern Observatory, Alonso de C\'{o}rdova 3107, Vitacura, Casilla 19001, Santiago, Chile  
             \and Department of Astronomy, University of Michigan, 323 West Hall, 1085 S. University Avenue, Ann Arbor, MI 48109, USA
\and Division of Liberal Arts, Kogakuin University, 1-24-2 Nishi-Shinjuku, Shinjuku-ku, Tokyo 163-8677, Japan
\and Mullard Space Science Laboratory, University College London, Holmbury St Mary, Dorking, Surrey RH5 6NT, UK 
\and Department of Earth and Planetary Sciences, Tokyo Institute of Technology, 2-12-1 Ookayama, Meguro-ku, Tokyo 152-8551, Japan
\and Department of Astronomy, Tsinghua University, Beijing 100084, China
\and Max-Planck-Institut für extraterrestrische Physik, Giessenbachstrasse 1, 85748 Garching, Germany 
\and Centre de Recherche Astrophysique de Lyon, CNRS, UCBL, ENS Lyon, UMR 5574, F-69230, Saint-Genis-Laval, France 
\and European Southern Observatory, Karl-Schwarzschild-Strasse 2, 85748, Garching bei München, Germany
\and College of Science, Ibaraki University, 2-1-1 Bunkyo, Mito, Ibaraki 310-8512, Japan
\and Department of Astrophysics, University of Vienna, T\"urkenschanzstrasse 17, 1180 Vienna, Austria
\and School of Physics and Astronomy, Monash University, Clayton Vic 3800, Australia
             \and Hamburger Sternwarte, Gojenbergsweg 112, 21029, Hamburg, Germany
             \and Institute for Astronomy, University of Hawai’i at Manoa, Honolulu, HI 96822, USA
             \and
              Instituto de Estudios Astrof\'isicos, Facultad de Ingenier\'ia y Ciencias, Universidad Diego Portales, Av. Ej\'ercito Libertador 441, Santiago, Chile
              \and
              Millennium Nucleus on Young Exoplanets and their Moons (YEMS), Chile 
             }
   \date{Received September 15, 1996; accepted March 16, 1997}

 
  \abstract
   {The past few years have seen a revolution in the study of circumstellar disks. New instrumentation in the near-infrared and (sub)millimeter regimes have allowed us to routinely spatially resolve disks around young stars of nearby star-forming regions. As a result, we have found that substructures with scales of $\sim$10\,au in disks are common. We have also revealed a zoo of different morphologies, sizes, and luminosities that is as complex as the diversity of architectures found in evolved exoplanet systems.}
   {We   study disk evolutionary trends as they appear in scattered light observations. Scattered light traces the    micron-sized particles at the disk surface that are well coupled to the gas. This means that scattered light observations can be used to trace the distribution of the disk gas and its interaction with embedded perturbers.}
   {We used VLT/SPHERE to observe 20 systems in the Cha I cloud in polarized scattered light in the near-infrared. We combined the scattered light observations with existing literature data on stellar properties and with archival ALMA continuum data to study trends with system age and dust mass. We also connected resolved near-infrared observations with the spectral energy distributions of the systems.}
   {In 13 of the 20 systems included in this study we detected resolved scattered light signals from circumstellar dust. For the CR\,Cha, CT\,Cha, CV\,Cha, SY\,Cha, SZ\,Cha, and VZ\,Cha systems we present the first detailed descriptions of the disks in scattered light. The observations found typically smooth or faint disks, often with little substructure, with the notable exceptions of SZ\,Cha, which shows an extended multiple-ringed disk, and WW\,Cha, which shows interaction with the cloud environment. New high S/N K-band observations of the HD\,97048 system in our survey reveal a significant brightness asymmetry that may point to disk misalignment and subsequent shadowing of outer disk regions, possibly related to the suggested planet candidate in the disk.
   We resolve for the first time the stellar binary in the CS\,Cha system. Multiple wavelength observations of the disk around CS\,Cha have revealed that the system contains small, compact dust grains that may be strongly settled, consistent with numerical studies of circumbinary disks. \\
   We find  in our sample that there is a strong anti-correlation between the presence of a (close) stellar companion and the detection of circumstellar material with five of our seven nondetections located in binary systems. We also find a correlation between disk mass, as inferred from millimeter observations, and the detection of scattered light signal. 
   Finally, we find a tentative correlation between relative disk-to-star brightness in scattered light and the presence of a dust cavity in the inner (unresolved) disk, as traced by the system spectral energy distribution. At the same time, faint disks in our sample are generally younger than 2\,Myr. 
   }
   {}


   \maketitle
%

\section{Introduction}

Circumstellar disks around young stars are the sites of planet formation.  The sizes, masses, and compositions of the planets as well as the architecture of the planetary systems being formed are a result of the properties of the planet-forming disks and the processes acting in these disks. However, it is  still unclear  how the statistics of planetary system architecture and planet populations relate to the observed properties of planet-forming disks.
The past few years have seen a surge in spatially resolved  observations of such disks, using ALMA at submillimeter wavelengths and high-contrast imaging techniques at optical and near-IR wavelengths.  Initially, these observations   focused on the largest and brightest disks, mostly around Herbig AeBe stars, for example the HD\,135344\,B system (\citealt{Andrews2011,Muto2012,Garufi2014,Marel2016,Stolker2016}), the HD\,100546 system (\citealt{Grady2001,Pineda2014,Walsh2014,Garufi2016}), or the HD\,97048 system (\citealt{Walsh2016,Ginski2016,Plas2017}). \\
In the ALMA wavelength range a number of surveys of individual regions with limited spatial resolution and sensitivity have been carried out, for example Chameleon\,I (\citealt{Pascucci2016}), Lupus (\citealt{Ansdell2016}), $\sigma$\,Orionis (\citealt{Ansdell2017}), Taurus (\citealt{Long2019}), and Ophiuchus (\citealt{Cieza2019}).
In near-infrared scattered light, to date such dedicated surveys of nearby star-forming regions are missing, with the majority of the field focusing on the study of individual objects. A first dedicated survey of T Tauri stars from multiple regions was carried out in the Disks around T Tauri Stars with SPHERE (DARTTS-S, \citealt{Avenhaus2018,Garufi2020}) program. Furthermore, \cite{Garufi2018} conducted a large literature study, including 58 Herbig and T Tauri stars. However, all of these studies were inhomogeneous, in the sense that they included systems from multiple star-forming regions, and were biased toward bright and extended disks. Even so, their initial results suggest already some intriguing trends. Among other things, they find an anti-correlation between the amount of near-infrared excess and the disk brightness in scattered light and the prevalence of spiral structures in older disks around more massive stars. However, to fully access the complementary information provided by scattered light and millimeter-wavelength observations we need   scattered light surveys of individual star-forming regions.\\
In this study we present polarimetric scattered light observations of 20 members of the Chameleon~I cloud (Cha\,I hereafter), conducted with      Spectro-Polarimetric High-contrast Exoplanet REsearch  (VLT/SPHERE; \citealt{Beuzit2019}).
The Chameleon complex is one of only a few nearby star-forming regions that is ideally suited for the study of young stellar objects.  Located at a distance of $\sim$180$\pm$10\,pc \citep{Voirin2018}, it is close enough that direct imaging observations are able to resolve circumstellar disks on spatial scales of a few au.\\
In   section 2 we describe the composition of the sample and put it in the context of the young star population of Cha\,I. In section~\ref{data-reduction} we describe the observations and the data reduction. We   first discuss the polarization of the stellar light and its implications in section~\ref{sec: stellar-pol}, and then describe the measurements taken in the polarimetric images, as well as global and individual disk properties in sections~\ref{sec: ellipse-fitting}, \ref{sec: contrast}, and \ref{sec: ind systems}. We highlight the first resolved detection of the inner binary in the CS\,Cha system in section~\ref{sec: CSCha-orbit}, and discuss our results in section~\ref{sec: discussion}. 

\section{Sample and stellar properties} \label{sec: stellar_properties}

Our sample consists of 20 members of Cha\,I, which we summarize in Table~\ref{tab: sample}.
The main selection criterion was the optical brightness of the targets (Gmag < 13\,mag), which allowed them to be used as natural guide stars for the SPHERE adaptive optics system. Additionally, we only observed sources for which the spectral energy distribution indicated excess emission in the infrared, suggestive of the presence of a dust disk around the star. The sample contains 90\% of the Cha\,I sources that can be observed by SPHERE. We are therefore nearly complete within the population of solar-like and intermediate-mass stars, with the aforementioned technical threshold translating into a stellar mass of approximately 0.5 M$_\odot$.\\   
A large fraction of our targets (11 out of 20) are part of a stellar system. We illustrate the configuration of the systems in Table~\ref{tab: sample}. The CS\,Cha and WW\,Cha systems feature a circumbinary disk. In both cases, the central stars are in tight orbits and the binary nature has been determined spectroscopically. The CT\,Cha, HP\,Cha, Sz\,41, and SZ\,Cha systems have a circumprimary disk configuration with wide, low-mass companions (which is itself a binary in HP Cha). DI\,Cha is a quadruple system with a close companion (named D) and a binary system (BC) at large separations. Finally, the CHX\,22, WX\,Cha, and WY\,Cha systems are known close visual binaries with mass ratios closer to unity. 

All stellar properties were computed based on the recent $Gaia$ DR3 parallax \citep{Gaia2023}. In four cases indicated in Table~\ref{tab: sample}, the $Gaia$ parallactic measurement is either uncertain or absent. Three of these cases are close binaries. The stellar masses and ages were computed through a set of PMS tracks \citep{Siess2000, Bressan2012, Baraffe2015, Choi2016} from the effective temperature and the stellar luminosity calculated from a Phoenix model of the stellar photosphere \citep{Hauschildt1999} scaled to the de-reddened optical photometry from the literature. The mass accretion rates were calculated from the accretion luminosities updated to $Gaia$ DR3 by \cite{2016A&A...585A.136M} using the usual relation of $\dot{M}_{acc} = 1.25 \frac{L_{acc}R_{*}}{GM_{*}}$ (for example, \citealt{1998ApJ...495..385H}). Finally, a crude estimate of the disk dust mass was derived from the ALMA fluxes at 887\,$\mu m$ by \cite{Pascucci2016} using the $Gaia$ distance and the typical assumption (optically thin emission, dust temperature of 20 K, and dust opacity by \citealt{Beckwith1991}).

From Table~\ref{tab: sample}, the bias in stellar mass due to the stellar optical brightness is clear, with 0.5 M$_\odot$ (for VZ Cha and WY Cha) being the lower end of the distribution and 2.4 M$_\odot$ (HD97048) the upper end. Conversely, the disk dust mass range is large as it spans from $\sim$940\,M$_\oplus$ (HD97048) to less than 1 M$_\oplus$ (CHX 22). In particular, our sample contains the four objects with the highest dust masses in Cha\,I (WW\,Cha, CR\,Cha, SZ\,Cha, CS\,Cha) based on \citet{Pascucci2016}.

 
\begin{table*}
 \centering
 \caption{Properties of the sample. Columns are: target name, distance from $Gaia$ DR3, stellar mass, a crude estimate of the disk dust mass, mass accretion rate, age, and possible multiplicity of the target. The detailed derivation of these parameters is described in section~\ref{sec: stellar_properties}. The known multiplicity status of each system is indicated  in the last column. Brackets denote circumstellar or circumbinary disks. Stars   with a plus sign (+) indicate known wide companions, while stars without this sign indicate close companions.
 }
  \begin{tabular}{@{}lcccccccc@{}}
  \hline 
 Target         & d\,[pc]               & M$_*$\,[M$_\odot$]    & M$_\mathrm{dust}$\,[M$_\oplus$] & log $\dot{M}_\mathrm{acc}$ [M$_\odot$/yr] & age\,[Myr]                & Binary  \\
 \hline
 \smallskip
 CHX 18 N & 191.6 $\pm$ 0.03 & 0.9 & 13 & $-7.52$ & 0.8--1.0 & (*)\\
 \smallskip 
 CHX 22 & 191.2 $^{(1)}$ & 1.9$\pm$0.1 / 0.6$\pm$0.1 & <1 & $<-4.8$ & 2.8--3.3 & (**)\\
 \smallskip
 CR\,Cha        & 185.2 $\pm$ 0.4       & 1.4$\pm$0.1       & 198  & $-8.41$ &       0.8--1.0        & (*)\\ 
 \smallskip
 CS\,Cha        & 190 $^{(2)}$ & 1.3$\pm$0.1 / 0.6$\pm$0.1      & 84    & $-7.99$ & 3.0--5.3 & (**)*\\ 
 \smallskip
 CT\,Cha        & 190.0 $\pm$ 0.4       & 0.9$\pm$0.2       & 50        & $-6.53$ &       1.2--1.9        & (*)* \\ 
 \smallskip
 CV\,Cha        & 191.8 $\pm$ 0.5       & 2.1$\pm$0.2       & 29 & $-7.26$ & 1.3--1.5      & (*) \\ 
 \smallskip
 DI Cha & 189.0 $\pm$ 0.6 & 2.3$\pm$0.2 & 9 & $-7.39$ & 0.6--0.9 & (*)* + **\\
 \smallskip
 HD97048 & 184.4 $\pm$ 0.7 & 2.4$\pm$0.2 & 940  & $<-7.72$ & 3.6--4.4 & (*)\\
 \smallskip
 HP Cha & 187.3$\pm$1.3 & 1.7$\pm$0.1 & 80 & $<-5.63$  & 5.2--5.7 & (*)(**)\\
 \smallskip
 PDS 51 & 190 $^{(2)}$ & 0.9$\pm$0.1 / 0.7$\pm$0.1 & 2 & $-7.80$& 2.5--3.9 & (**) \\
 \smallskip
 RX J1106.3-7721 & 177.9 $^{(1)}$ & 2.9$\pm$0.3 & $-$ & $-$ & 1.0--1.4 & (*) \\
 \smallskip
 SY\,Cha        & 180.7 $\pm$ 0.4       & 0.7$\pm$0.1       & 50& $<-4.66$ & 1.5--2.0 & (*) \\ 
 \smallskip
 Sz 41 & 191.8 $\pm$ 0.4 & 0.6$\pm$0.1 / 0.5$\pm$0.1 & <1 & $-7.38$ & 0.4--0.7 & (*)* \\
 \smallskip
 Sz 45 & 188.9 $\pm$ 0.6 & 0.6$\pm$0.1 & 10 &  $-8.02$ & 1.6--2.2 & (*)\\
 \smallskip
 SZ\,Cha        & 190.2 $\pm$ 0.9       & 1.5$\pm$0.1       & 150       & $-7.56$ & 1.5--1.8 & (*)* \\  
 \smallskip
 TW Cha & 183.1 $\pm$ 0.4 & 0.7$\pm$0.1 & 23 & $-8.56$ & 1.5--2.2 & (*)\\
 \smallskip
 VZ\,Cha        & 191.1 $\pm$ 0.6       & 0.5$\pm$0.1       & 60 & $-7.14$ & 1.1--1.5      & (*) \\ 
 \smallskip
 WW\,Cha        & 188.8 $\pm$ 1.0       & 1.9$\pm$0.1       & 630 & $-6.28$      & 0.2--0.5      & (**) \\  
 \smallskip
 WX\,Cha        & 190.6 $^{(1)}$        & 0.5$\pm$0.1       & 10    & $-6.69$ &       0.9--1.2        & (*)* \\ 
 \smallskip
 WY\,Cha        & 174.5 $^{(1)}$        & 0.7$\pm$0.1       & 3    & $-8.36$ &       1.5--2.1 & (*)* + * \\  
 
 \hline
                        
\hline\end{tabular}
\tablefoot{$^{(1)}$: the $Gaia$ parallactic measurement has a large renormalized unit weighting error (RUWE $\gg$ 2) pointing to a very uncertain solution. $^{(2)}$: no $Gaia$ parallax is calculated, and a value of 190 pc is adopted.}
\label{tab: sample}
\end{table*}

\section{Observations and data reduction}
\label{data-reduction}

Observations were carried out with the IRDIS (Infra-Red Dual-beam Imager and Spectrograph, \citealt{2008SPIE.7014E..3LD}) and ZIMPOL \citep[Zurich IMaging POLarimeter;][]{Schmid2018} subsystems of VLT/SPHERE. IRDIS was operated in the dual-beam polarimetric imaging (DPI, \citealt{2014SPIE.9147E..1RL}) mode. 
This mode is used to detect linear polarization signals. For this purpose, a half-wave plate is used to rotate the astrophysical polarization signal in the frame of reference of two wire grid polarizers.
The two perpendicular polarization directions are taken simultaneously, such that an ideal subtraction of the (unpolarized) stellar signal is possible.
This observation mode is described in detail in \cite{2020A&A...633A..63D} and \cite{2020A&A...633A..64V}.\\
A detailed overview of the observation setup and atmospheric conditions is given in Appendix \ref{appendix: observations}. 
Single star targets, or systems with very low-mass companions were generally observed with a coronagraph in place, obscuring the central star.
In all but three cases we used the N\_ALC\_YJH\_S coronagraph for this purpose, which has a nominal inner working angle\footnote{The inner working angle of all coronagraphs is defined as the angular separation at which 50\% throughput is achieved.} of 92.5\,mas (\citealt{2009A&A...495..363M, 2011ExA....30...39C}). 
Since CV\,Cha was observed in J-band we used the N\_ALC\_YJ\_S coronagraph (optimized for shorter wavelengths), which has a slightly smaller inner working angle of 72.5\,mas.
For the K-band observations of HD\,97048 and SY\,Cha we used the N\_ALC\_KS coronagraph with an inner working angle of 150\,mas.
For the known binary systems with bright secondary stars PDS\,51, WX\,Cha, and WY\,Cha we did not employ a coronagraph but rather used short detector integration times (DIT) to avoid saturation.\\
All data were reduced using the IRDAP (IRDIS Data reduction for Accurate Polarimetry\footnote{https://irdap.readthedocs.io}) pipeline. The data reduction package is described in detail in \cite{2020A&A...633A..64V} and will thus only be briefly summarized here.
After initial baseline data reduction (sky-subtraction, flat-fielding, bad pixel masking), the images were centered using either a center calibration frame with calibrated satellite spots (coronagraphic data) or a Gaussian fit (noncoronagraphic data).
Each polarimetric cycle was then reduced by creating single and double difference images from the individual Q$^+$,Q$^-$,U$^+$, and U$^-$ images (signifying different rotation positions of the internal half-wave plate).
Instrumental polarization was then removed using a full Mueller Matrix model of the instrument + telescope (see \citealt{2020A&A...633A..63D} for details). After this step residual (astrophysical) stellar polarization could be measured and removed, using the method described by \cite{2011A&A...531A.102C}.
This was done in order to remove as much stellar signal as possible thereby enabling a clear view of the surrounding circumstellar material.\\
The so-created final Stokes Q and U images were then used to generate radial Stokes images Q$_\phi$ and U$_\phi$, following \cite{2006A&A...452..657S}, but with a flipped sign convention more appropriate for disk scattered light observations as discussed in \cite{2020A&A...633A..63D}.
Q$_\phi$ contains as positive values all azimuthally aligned polarized signal and as negative values all radially aligned polarization signal. 
U$_\phi$ contains all polarization signal 45$^\circ$ offset from the radial or azimuthal direction. 
Since a circumstellar disk will predominantly show single scattering (unless viewed under very high inclinations), the Q$_\phi$ image contains most of the actual signal, while the U$_\phi$ image can in principle be regarded as a convenient noise map.
All IRDIS observations were flux calibrated by short noncoronagraphic\footnote{By that we mean that the central star has been moved off-axis from the coronagraphic mask so that the central part of its PSF is not blocked anymore.} flux observations. Neutral density filters of varying throughput were inserted to prevent saturation of the bright central star. Flux observations were typically performed before and after the main science sequence and with shorter DIT.\\
In addition to the IRDIS near-infrared observations CS\,Cha polarimetric observations with SPHERE/ZIMPOL in the optical were performed on 2018 December 23 in imaging mode P1 and on 2019 January 20 in field-stabilized mode P2. Images were taken in the slow-polarimetry mode using different filters in two arms of ZIMPOL: the I\_PRIM filter in the arm 1 and R\_PRIM filter in the arm 2. The star was placed behind a semi-transparent coronagraphic mask (V\_CLC\_MT\_WF) with a radius of 77.5\,mas. In total, 20 polarimetric $QU$ cycles (120 exposures with DIT = 10 s) were recorded during the first night and 4 $QU$ cycles (32 exposures with DIT = 58 s) during the second night. Each cycle consisted of four consecutive measurements with different HWP offset angles of $0\degree$, $22.5\degree$, $45\degree$, and $67.5\degree$ switching the Stokes parameters $+Q$, $-Q$, $+U$, and $-U$, respectively. At the beginning and the end of science observations flux measurements were performed with the star offset from the coronagraphic mask using the neutral density filter ND\_4.0 and DIT = 20 s (first night) and DIT = 80 s (second night). \\
The ZIMPOL data were reduced with the data reduction pipeline developed at ETH Zurich. The preprocessing and calibration of raw frames included subtraction of the bias and dark frames, flat-fielding, and correction for the modulation and demodulation efficiency. The instrumental polarization was corrected by normalizing the fluxes in the frames of two opposite polarization states measured in the annulus with inner radius of 100 pixels and outer radius of 200 pixels as described in \citep{Engler2017}. All frames were centered by fitting 2D Gaussian function to the intensity gradients of the stellar profile. The final format of the reduced $Q$ and $U$ images is $1024\times 1024$ pixels with the pixel size approximately $3.6\times 3.6$ mas on sky.
The ZIMPOL data of CS\,Cha are discussed in section~\ref{CSCha-discussion} and section~\ref{sec: CSCha-orbit}.\\
In addition to the near-infrared and optical data, we present complementary ALMA observations of the SY\,Cha systems. SY\,Cha was observed with ALMA in Band 6 during Cycle 6 as a part of project 2018.1.00689.S.  The observations were done with the antenna configuration of C43-6 (15.25~min on source), C43-7 (15.25~min on source), and C43-9 (72.75~min on source) with the longest baseline of 16.2~km.  The details of the observations and data reduction are presented in \cite{Orihara2023}.  In this paper, we present the continuum image at the central frequency of 225~GHz with the beam size of 0\farcs145$\times$0\farcs109 and with the RMS noise level of 0.0175~mJy/beam.

\begin{table}
 \centering
 \caption{Degree of linear polarization (DoLP) and angle of linear polarization (AoLP) of the stellar light measured from the SPHERE/IRDIS observations. The CS\,Cha and HD\,97048 data points were obtained in the J-band. For HP\,Cha  only   the primary star in the system was considered.}
  \begin{tabular}{@{}lcc@{}}
  \hline 
 System         &  DoLP [\%] & AoLP [$^\circ$] \\
 \hline
 CHX 18 N & 0.37$\pm$0.04 & 132.4$\pm$4.7 \\
 CHX 22 & 1.33$\pm$0.17 & 133.2$\pm$4.4 \\
 CR\,Cha        & 1.84$\pm$0.02 & 124.8$\pm$0.6                 \\
 CS\,Cha        & 0.34$\pm$0.02 & 141.1$\pm$5.4                 \\
 CT\,Cha        & 0.99$\pm$0.10 & 124.6$\pm$3.8                 \\
 CV\,Cha        & 0.94$\pm$0.39 & 109.0$\pm$9.1                 \\
 DI Cha & 1.42$\pm$0.05 & 143.7$\pm$1.1\\
 HD97048 & 1.40$\pm$0.25 & 132.4$\pm$5.3 \\
 HP Cha & 0.63$\pm$0.10 & 168.3$\pm$3.6 \\
 PDS 51 & 3.16$\pm$0.16 & 129.4$\pm$1.3 \\
 RX J1106.3-7721 & 1.44$\pm$0.06 & 146.1$\pm$1.2 \\
 SY\,Cha        & 0.91$\pm$0.08 & 83.8$\pm$2.4                  \\
 Sz 41 & 1.00$\pm$0.72 & 107.6$\pm$18.7 \\
 Sz 45 & 0.90$\pm$0.14 & 130.3$\pm$2.9 \\
 SZ\,Cha        & 1.37$\pm$0.03 & 118.8$\pm$0.2                 \\
 TW Cha & 0.43$\pm$ 0.12 & 148.3$\pm$9.8 \\
 VZ\,Cha        & 0.41$\pm$0.01 & 141.4$\pm$5.3                 \\
 WW\,Cha        & 1.11$\pm$0.04 & 164.2$\pm$0.9                 \\
 WX\,Cha        & 1.97$\pm$0.28 & 134.7$\pm$5.6                 \\
 WY\,Cha        & 0.27$\pm$0.02 & 131.8$\pm$3.7                 \\
 
\hline\end{tabular}
\label{tab: star_pol}
\end{table}

\begin{figure}
\centering
\includegraphics[width=9cm]{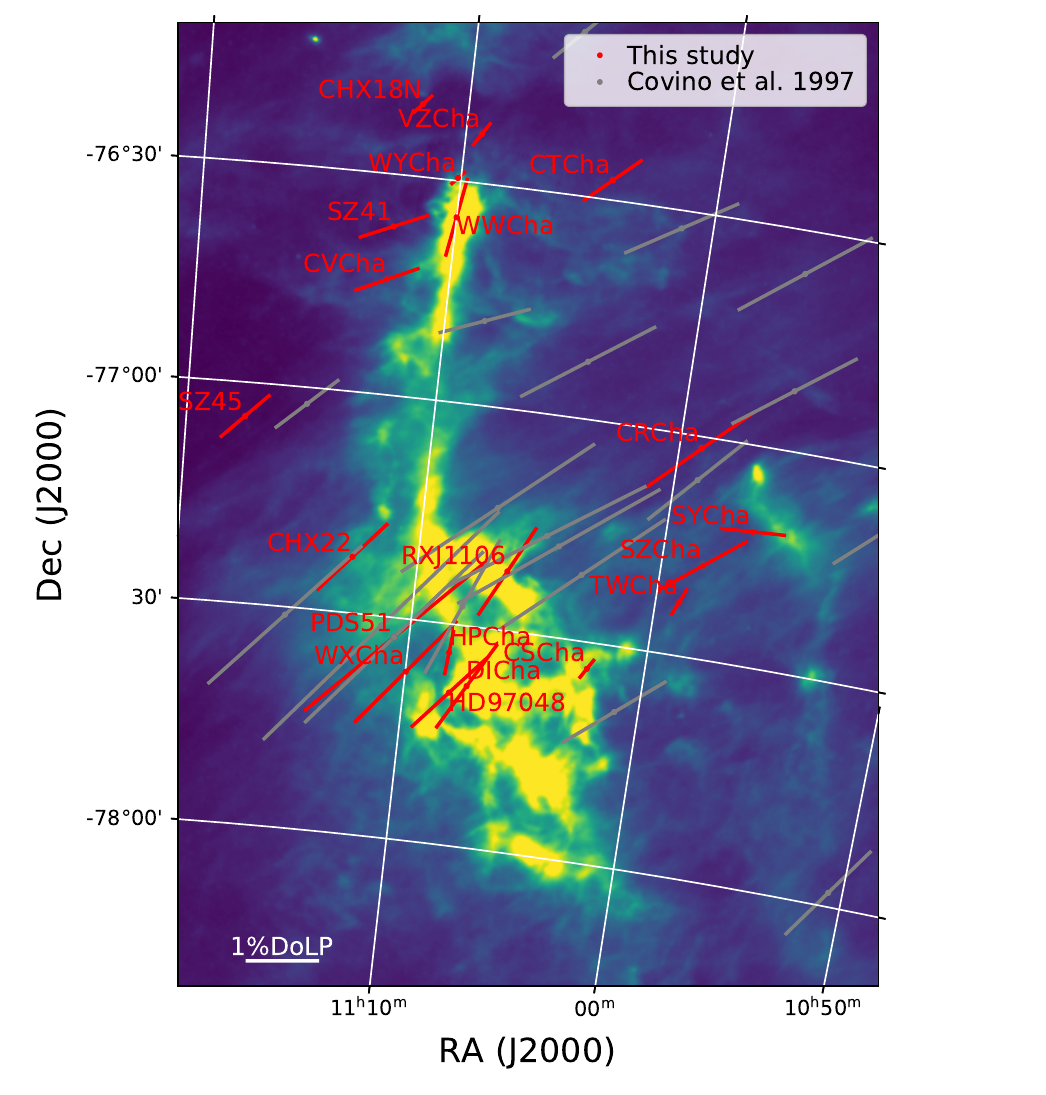}
\caption[]{Degree of linear polarization and angle of linear polarization of our target sources (red) and measurements by \cite{Covino1997} using background stars behind the Cha\,I cloud (gray).
Optical measurements by \cite{Covino1997} were extrapolated to the near-infrared regime by Serkowski's law. Polarization vectors are overlayed on a Herschel/SPIRE image at 160\,$\mu$m showing the dust in the Cha\,I cloud.}
\label{fig: app: pol_vetor}
\end{figure}

\section{Results on the global sample}
In this section, we describe our sample as a whole. An analysis of the individual sources is given in section\,\ref{sec: ind systems}. All the targets in our sample show some stellar (unresolved) polarization (see section\,\ref{sec: stellar-pol}). However, resolved polarized light is detected in only 13 of the 20 targets. This resolved signal is described in section\,\ref{sec: ellipse-fitting} and related to the disk and stellar properties in section\,\ref{sec: contrast}. 

\subsection{Stellar polarization}
\label{sec: stellar-pol}
Following \cite{2021A&A...647A..21V}, we used the IRDAP pipeline to measure the degree of linear polarization of the (unresolved) stellar light, which may also include a contribution from the inner disk regions ($<$8\,au). We concentrated in all cases on the primary star in the system. The results are given in Table~\ref{tab: star_pol}. In Figure~\ref{fig: app: pol_vetor}, we overplotted the position of our target systems as well as the amount and angle of linear polarization on a Herschel/SPIRE map of the Cha\,I cloud complex (showing thermal dust emission at 160\,$\mu m$).
For comparison we plotted background stars (behind the Cha\,I cloud) with optical polarization measurements from \cite{Covino1997}. We note that the angle of linear polarization is similar for all background stars. This is consistent with the stellar light being polarized by interstellar dust grains, with the majority of the dust column density likely located in the Cha\,I cloud complex itself. 

The majority (17 out of 20) of our target systems shows an angle of stellar polarization that is consistent with the background stars observed by \cite{Covino1997}. This is an indication that in these systems the source of the polarized stellar light is likely dominated by the same mechanism (i.e., scattering of interstellar dust). The amount of stellar polarization is generally slightly lower than for the background stars observed by \cite{Covino1997}. This can be due to the difference in wavelength since linear polarization due to interstellar dust is typically lower in the near-infrared (\citealt{1975ApJ...196..261S}) than in the optical. However, we also see a weak correlation with the distances of the individual systems listed in table~\ref{tab: sample}. In particular, the degree of linear polarization is the lowest for the CS\,Cha system and the WY\,Cha system. WY\,Cha is the most nearby system in our sample, while this is somewhat uncertain for CS\,Cha due to the only recently resolved binary nature of the system. However the nominal \textit{Gaia} parallax of CS\,Cha places it at $\sim$169\,pc (i.e., in front of the bulk population of the Cha\,I cloud). This fits well with the picture that the majority of the interstellar polarization in the \cite{Covino1997} sample is induced as the light passes through the dense Cha\,I dust cloud. CS\,Cha and WY\,Cha, may be located on the outer edges or slightly in front of the Cha\,I cloud complex given our line of sight. 

Three systems, HP\,Cha, SY\,Cha and WW\,Cha, show significantly different angles of linear polarization than the remaining sample. This may indicate that the dominant source of linear polarization is not interstellar dust scattering, but rather is induced locally; in other words, at angular separations smaller than the resolution element of SPHERE in the H-band ($\sim$40\,mas; i.e., inside of $\sim$8\,au).  
A possible explanation is an inner zone of the circumstellar disk. To cause a  polarization in the unresolved stellar light, this inner part of the disk needs to be inclined,  otherwise (in a face-on oriented disk) different polarization directions would cancel each other out. This scenario is discussed in great detail in \cite{2021A&A...647A..21V}. If the polarization signal is indeed caused by an unresolved inclined circumstellar disk, then the angle of polarization should be aligned with the minor axis of the disk, since the majority of the polarized light will be received from the disk ansae at scattering angles close to 90$^\circ$ (see \cite{2021A&A...647A..21V} for a radiative transfer model of such a configuration). We however need to caution that the degree of polarization that we measure for our target stars is likely a combination of local and interstellar effects. Thus detailed modelling of the local interstellar polarization would be necessary to derive constraints on the geometry of the inner disks in HP\,Cha, SY\,Cha and WW\,Cha. This is beyond the scope of this study. We also need to caution that for the other 17 systems, we can in principle not rule out that there is additional local polarization of the stellar light, since the minor axis of an inner disk may coincidentally line up with the general direction of the local interstellar polarization.  

\begin{figure*}
\centering
\includegraphics[width=0.999\textwidth]{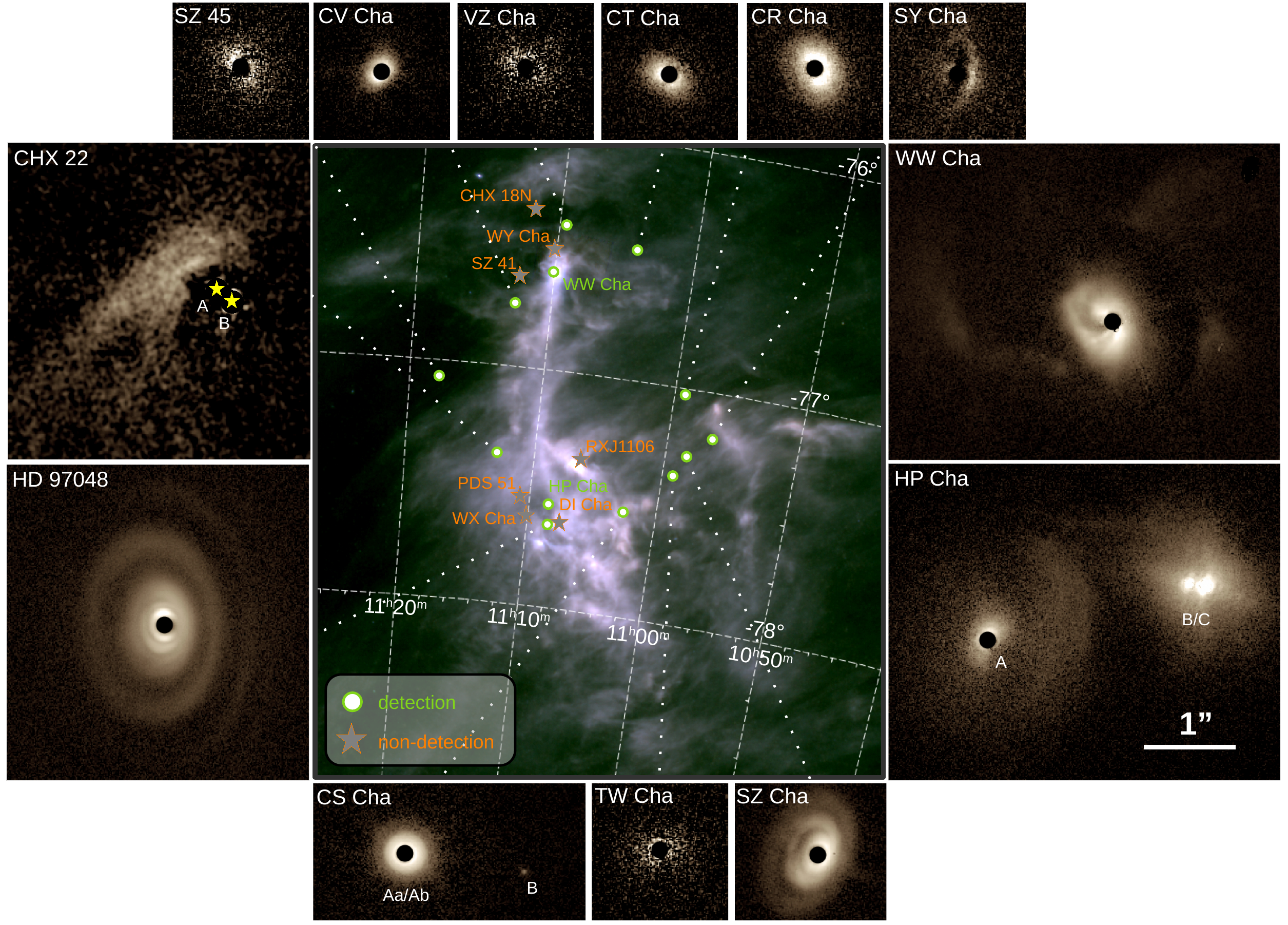}
\caption[]{Herschel/SPIRE RGB image of the Cha\,I star-forming region constructed from the 150$\mu m$, 250$\mu m$, and 500$\mu m$ channels  showing the interstellar dust (central panel).  SPHERE/IRDIS Q$_\phi$ images for all systems with extended circumstellar dust in our sample (surrounding panels). The shown SPHERE data were taken in the H-band, with the exceptions of the HD\,97048 and SY\,Cha system, for which we show K-band data, as well as the CS\,Cha and CV\,Cha systems for which we show J-band data. The position of each detected source within the Cha\,I cloud is indicated with white markers with green border; the positions of systems in our sample with nondetections of circumstellar dust are shown as   gray stars with orange borders. For the CS\,Cha, HP\,Cha, and CHX\,22 systems,   the positions of visible stellar multiple components are indicated.  }
\label{fig: all_detections}
\end{figure*}

\subsection{Disk geometry}
\label{sec: ellipse-fitting}
As is clear from Fig.\,\ref{fig: all_detections}, some resolved polarized signal is detected in 13 of our 20 targets. In turn, 12 of these show a polarized light pattern consistent with a disk detection \citep[with the only exception being CHX 22, see][]{Zhang2023}. The details of all individual systems are given in section\,\ref{sec: ind systems}. Here, we extract the aspect ratio of the detected disk in order to define an overall trend for the sample. 

The offset of ellipses, tracing iso-separation features on the disk, along the minor axis can be used to calculate the height of the $\tau=1$ scattering surface of the disk as was shown in \cite{deBoer2016} and \cite{Ginski2016}. While in the two aforementioned cases, multiple rings were present, this is not typically the case for the disks in our sample (with the exception of HD\,97048 and SZ\,Cha). However, we can still measure the offset of the ellipses that trace the outer edge of the disks as detected in scattered light, thereby determining the aspect ratio of the outermost region of the scattered light disk. We show in Appendix~\ref{app: geometric fitting} that this approach is valid, based on disk model images. 

\subsubsection{Fitting procedure}
\label{geometric-fitting-section}

For the ellipse fit we follow two separate approaches for the ringed disks and those without such substructure. In both cases we use a sliding aperture, to extract radial profiles in azimuthal bins of 1$^\circ$. The size of the individual apertures is chosen to reflect the resolution element in the images. For the ringed disks, we fit 1d-Gaussian profiles to extract the ring locations from the radial profiles. For the disks without ringed substructure, we trace the outer disk edge at which the disk signal drops below 3$\sigma$ as determined from the image background. After the initial measurement, we then bin the data in either case in azimuthal direction in increments of 10$^\circ$. We note that in case of low S/N, we excluded azimuthal disk regions for individual targets. For the uncertainties of the extracted data points, we use the full-width-at-half-maximum (FWHM) of the Gaussian for the ring-like structures. For the data points that trace the disk edge, we determine the slope of the radial profiles across the edge. The radial uncertainties $\Delta r$ are then derived as:

\begin{equation}
    \Delta r = \frac{3 * \delta F}{m_\mathrm{profile}}
\end{equation}

With $\delta F$ represents the standard deviation of the background flux in the image and $m_\mathrm{profile}$ the slope of the radial profile. A steep slope (i.e., a sharp disk edge) leads to small uncertainties, while a small slope (i.e., a ``fuzzy'' disk edge) leads to large uncertainties. For the azimuthal uncertainty, we use a value of 0.1$^\circ$, consistent with the calibration accuracy of our images. The extracted data points and associated uncertainties for both disk groups are shown in Figures~\ref{fig: ellipse-fit} and \ref{fig: ellipse-ring-fit} of Appendix \ref{app: geometric fitting}.

We use the extracted data points for a two-step fitting process. Initially, we use the least squares algorithm by \cite{oy1998NUMERICALLYSD} to find the best-fitting ellipse, without any prior constraints to the disk parameters. These best-fitting ellipses are overlayed in Figures~\ref{fig: ellipse-fit} and \ref{fig: ellipse-ring-fit}. We determine the uncertainty of this fit with a least squares Monte Carlo (LSMC) approach. We repeat the fit 10$^5$ times. Each time we draw the location of each data point from a normal distribution with the width of the data points uncertainty. We then use the standard deviation of each ellipse parameter across all runs as the uncertainty for that parameter. The results of these initial fits are shown in Table~\ref{tab: disk geometry}.
In Appendix~\ref{app: geometric fitting} we demonstrate that this approach reliably recovers the disk inclination and position angle if the disk inclination is larger than $\sim$10$^\circ$.

We then use a second step to refine the offset of the disk rings or outer disk edge from the stellar position. For this second step, we constrain the disk inclination and position angle, either to literature measurements from ALMA millimeter observations (typically gas observations) or based on the prior fitting step when no ALMA data is available. We additionally constrain the ring or disk-edge semi-major axis to the value found in the previous step. We then only allow ellipse offsets from the stellar position along the minor axis (i.e., consistent with offsets introduced solely by the projected disk height and not by disk asymmetry). Based on these parameters we generate a grid of elliptical annuli (for the rings) or apertures (for disks without rings), with offsets along the minor axis increasing in steps of 0.1 pixel (1.25\,mas). We then find the annulus or aperture for which the contained disk or ring flux is maximized. The uncertainties of the offset positions found in this way are found by finding the range of annulus or aperture positions within the flux uncertainty of the measurement. For the disk rings this is similar to the procedure employed to fit the ring locations for the HD\,97048 system in \cite{Ginski2016}. For the aspect ratio of the disk without rings we show the validity of this approach in Appendix~\ref{app: geometric fitting}. 
The offset and aspect ratio values found with this second fitting step are also given in Table~\ref{tab: disk geometry}. 
We note that for the faint disks around Sz\,45, TW\,Cha and VZ\,Cha the uncertainty of the disk inclination or position angle was large and no complementary ALMA data were available, thus we omitted the second fitting step in these cases.

\begin{table*}
 \centering
 \caption{Ellipse fitting results and system geometry from the literature. In addition to inclination and position angle,   the amount ($\Delta$u) and angle ($\alpha_\mathrm{u}$) of offset of the fitted ellipse center from the stellar position is given. The height of the disk at the radius of the fit is given only taking the offset component into account that is in the direction of the disk semi-minor axis. For the HD\,97048 and SZ\,Cha systems multiple rings were fit independent of each other. The \emph{fit} column   indicates whether the fit was performed based on the SPHERE data alone (\emph{SPHERE}), using the SPHERE data but with inclination constrained to within 2$\degree$ of the ALMA value (\emph{SHERE/con}), or if its based on ALMA (literature) data (\emph{ALMA}).}
  \begin{tabular}{@{}llccccccl@{}}
  \hline 
 System  & fit          & i ($^\circ$)  & PA ($^\circ$) & r (au)            & $\Delta$u (mas)    & $\alpha_\mathrm{u}$ ($^\circ$) & h (au)                   & ref. \\
 \hline
 CR\,Cha        & SPHERE       & 34.9 $\pm$ 5.9  &  29.8 $\pm$ 14.0  &  93.6 $\pm$ 4.1  &  23.2 $\pm$ 17.9  &  237.9  &  -2.7 $\pm$ 9.6  & this work \\
         & SPHERE/con   & 31.0 $\pm$ 1.4  &  36.2  &  93.1 $\pm$ 4.2  &  0.0 $\pm$ 15.9  &  -  &  0.0 $\pm$ 5.7  &   this work \\
         & ALMA                & 31.0$\pm$1.4   & 36.2$\pm$1.8  & -                     &        -                       &       -                       & - &     \cite{Kim2020}  \\
 CS\,Cha        & SPHERE       & 21.6 $\pm$ 6.4  &  257.7 $\pm$ 31.1  &  84.9 $\pm$ 10.8  &  6.3 $\pm$ 14.0  &  316.0  &  2.7 $\pm$ 8.4  & this work \\
         & SPHERE/con   &17.9 $\pm$ 0.1  &  82.6  &  85.4 $\pm$ 5.3  &  3.7 $\pm$ 14.7  &  -  &  2.3 $\pm$ 9.1  &  this work\\
         & ALMA        & 17.86$^{+0.05}_{-0.01}$        & 262.6             & -          &               -               &       -               & - &\cite{Kurtovic2022} \\
 CT\,Cha        & SPHERE       & 45.7 $\pm$ 5.0  &  59.0 $\pm$ 9.4  &  64.6 $\pm$ 3.9  &  33.0 $\pm$ 17.2  &  133.3  &  8.3 $\pm$ 6.5  & this work \\
         & SPHERE/con   &45.7 $\pm$ 5.0  &  59.0  &  64.6 $\pm$ 4.2  &  27.0 $\pm$ 19.6  &  -  &  7.2 $\pm$ 5.2  &  this work \\
 CV\,Cha        & SPHERE       & 43.0 $\pm$ 5.3  &  -46.7 $\pm$ 9.5  &  70.6 $\pm$ 4.0  &  42.6 $\pm$ 16.3  &  42.3  &  12.0 $\pm$ 6.2  & this work \\
         & SPHERE/con   &43.0 $\pm$ 5.3  &  -46.7  &  71.1 $\pm$ 4.6  &  52.7 $\pm$ 12.3  &  -  &  14.8 $\pm$ 3.7  & this work \\
 HD97048 & SPHERE       & 31.8 $\pm$ 7.1  &  -1.1 $\pm$ 15.7  &  45.4 $\pm$ 2.0  &  17.5 $\pm$ 10.4  &  119.9  &  5.3 $\pm$ 3.4  &  this work \\
         & SPHERE/con   &41.1 $\pm$ 0.9  &  4.5  &  45.4 $\pm$ 1.9  &  8.6 $\pm$ 1.2  &  -  &  2.4 $\pm$ 0.3  & this work \\
         & SPHERE       & 42.8 $\pm$ 3.5  &  2.3 $\pm$ 6.4  &  164.8 $\pm$ 6.4  &  129.2 $\pm$ 21.7  &  92.4  &  34.8 $\pm$ 11.0  & this work \\
         & SPHERE/con   &41.1 $\pm$ 0.9  &  4.5  &  164.6 $\pm$ 6.3  &  125.1 $\pm$ 12.3  &  -  &  35.1 $\pm$ 3.5  & this work \\
         & SPHERE       & 47.1 $\pm$ 5.0  &  -4.8 $\pm$ 4.0  &  285.1 $\pm$ 11.9  &  142.3 $\pm$ 88.5  &  298.2  &  27.3 $\pm$ 146.4  &  this work \\
         & SPHERE/con   & 41.1 $\pm$ 0.9  &  4.5  &  285.1 $\pm$ 11.7  &  196.2 $\pm$ 45.4  &  -  &  55.0 $\pm$ 12.8  & this work \\
         & ALMA         & 41.1$\pm$0.9 & 4.5$\pm$0.1 & & -& -& -& \cite{Plas2017} \\
 SY\,Cha        & SPHERE       & 60.4 $\pm$ 4.9  &  -8.7 $\pm$ 4.7  &  72.5 $\pm$ 6.6  &  43.7 $\pm$ 30.6  &  23.1  &  7.0 $\pm$ 15.5  & this work \\
         & SPHERE/con   &51.7 $\pm$ 1.2  &  344.7  &  72.6 $\pm$ 6.6  &  51.5 $\pm$ 41.7  &  -  &  11.9 $\pm$ 9.6  & this work \\
         & ALMA          & 51.7$\pm$1.2         & 344.7$\pm$1.7 & -                     & -                              & -                             &-&     \cite{Orihara2023}      \\
 Sz 45   & SPHERE       & 49.6 $\pm$ 5.8  &  215.0 $\pm$ 11.6  &  58.6 $\pm$ 3.3  &  26.3 $\pm$ 20.0  &  312.1  &  5.8 $\pm$ 8.3  & this work \\         
 SZ\,Cha        & SPHERE       & 46.8 $\pm$ 3.3  &  -22.9 $\pm$ 4.9  &  112.5 $\pm$ 4.9  &  91.4 $\pm$ 14.6  &  76.1  &  23.6 $\pm$ 4.8  &  this work \\
         & SPHERE/con   &42.1 $\pm$ 0.5  &  336.7  &  112.5 $\pm$ 4.7  &  147.0 $\pm$ 2.5  &  -  &  41.7 $\pm$ 0.8  & this work \\
         & SPHERE       & 38.8 $\pm$ 5.8  &  -11.5 $\pm$ 10.2  &  59.2 $\pm$ 3.1  &  64.3 $\pm$ 12.4  &  46.4  &  16.2 $\pm$ 5.2  &  this work \\
         & SPHERE/con   &42.1 $\pm$ 0.5  &  336.7  &  59.2 $\pm$ 3.1  &  61.3 $\pm$ 1.2  &  -  &  17.4 $\pm$ 0.4  & this work \\
         & ALMA          & 42.1$\pm$0.5         & 156.7$\pm$0.6 &       -               &        -                       &       -                       &       - & Hagelberg et al., in prep.            \\
 TW Cha  & SPHERE       & 37.3 $\pm$ 9.0  &  -73.4 $\pm$ 34.6  &  35.4 $\pm$ 3.1  &  29.4 $\pm$ 20.0  &  124.8  &  0.2 $\pm$ 16.3  & this work \\ 
 VZ\,Cha        & SPHERE       & 49.6 $\pm$ 5.1  &  55.7 $\pm$ 11.5  &  56.7 $\pm$ 3.4  &  46.3 $\pm$ 19.2  &  52.2  &  1.7 $\pm$ 9.8  & this work \\ 
 WW\,Cha        & SPHERE       & 44.1 $\pm$ 3.3  &  29.8 $\pm$ 4.8  &  138.9 $\pm$ 5.2  &  65.6 $\pm$ 16.4  &  119.5  &  17.8 $\pm$ 6.0  & this work \\
         & SPHERE/con   & 36.2 $\pm$ 1  &  34.0  &  138.1 $\pm$ 4.9  &  77.2 $\pm$ 20.8  &  -  &  24.7 $\pm$ 6.7  &  this work \\
         & ALMA          & 36.2             & 34.0              &       -           &     -                       &       -                       &       -& \cite{Kanagawa2021}             \\
 
\hline\end{tabular}
\label{tab: disk geometry}
\end{table*}

\subsubsection{Sample aspect ratio}

\begin{figure}
\centering
\includegraphics[width=9.0cm]{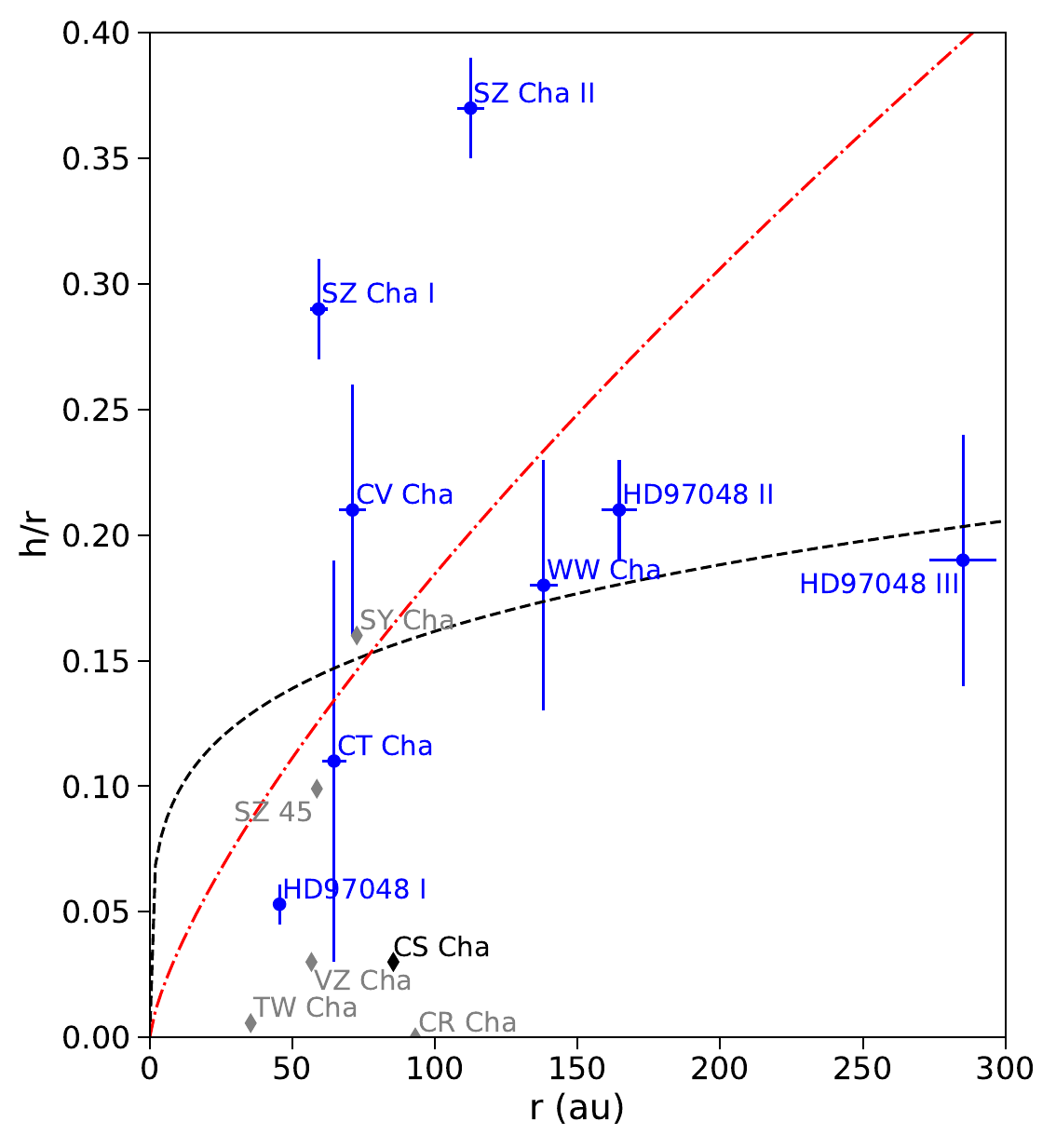}
\caption[]{Aspect ratio for all disks with at least one measured surface height. The black dashed line is the fit to several T Tauri stars done by \cite{Avenhaus2018}, while the red dash-dotted line is the fit to the SPHERE J-band data of HD\,97048 from \cite{Ginski2016}. Data points in gray and black have intrinsically very large uncertainties (not displayed, but shown in table~\ref{tab: disk geometry}). }
\label{fig: h_over_r}
\end{figure}

Using the fitting procedure described in section~\ref{geometric-fitting-section}, we calculated disk aspect ratios for 11 systems in our sample. However, we note that the aspect ratios for 6 of these systems (CR\,Cha, CS\,Cha, SY\,Cha, SZ\,45, TW\,Cha, VZ\,Cha) have large uncertainties (see Table~\ref{tab: disk geometry} for the disk height). We are plotting all measurements in Figure~\ref{fig: h_over_r} together with aspect ratio profiles from the literature for the extreme flaring Herbig star HD\,97048 (\citealt{Ginski2016}) and for a sample of T\,Tauri stars (\citealt{Avenhaus2018}).

Within our sample we have two disks with multiple ringed structures, HD\,97048 and SZ\,Cha. The disk around HD\,97048 is by far the most extended reaching out to $\sim$285\,au. In this work, we present new K-band measurements of this system. Compared to the J-band measurements from \cite{Ginski2016} we find at all separations lower aspect ratios. This is an expected effect due to the lower dust opacity at longer wavelengths. We note that the slope between 45\,au and 165\,au for the system is well consistent with the slope derived from the J-band observation. Similarly, we also find that for large separations the retrieved aspect ratio is no longer consistent with a single power law, likely due to a drop in dust surface density.
For the SZ\,Cha system we measure the aspect ratio in two individual rings. Our results indicate that this disk shows the most extreme aspect ratio within our sample with values of $\sim$0.29 and $\sim$0.37 at $\sim$59\,au and 113\,au, respectively. This is a much larger aspect ratio than found for either the HD\,97048 system or for the average sample of T\,Tauri stars from \cite{Avenhaus2018}.\footnote{While we get very consistent results for both stages of our fitting approach for the inner ring at 59\,au, we do see roughly a factor 2 increase in the aspect ratio for the outer ring and the constrained annulus grid method. This is likely caused by the somewhat ill defined separation of the outer ring signal from the inner ring signal along the minor axis, and in particular toward the forward scattering side of the disk.} The slope of the aspect ratio profile of SZ\,Cha is well consistent with the slope seen for HD\,97048.

We find meaningful individual measurements of the disk aspect ratio for the CT\,Cha, CV\,Cha and WW\,Cha systems. Of these WW\,Cha is well consistent with the average profile of T\,Tauri stars from \cite{Avenhaus2018}, while the disk around CT\,Cha appears vertically thinner. The disk around CV\,Cha shows a higher aspect ratio than expected from the literature profiles. There is considerable overlap within the uncertainties for all three systems, so it is not clear if they follow intrinsically similar aspect ratio profiles or if they are significantly different from one another. We note that as shown in Appendix~\ref{app: geometric fitting}, our edge-tracing approach to constrain the disk aspect ratio shows a systematic negative offset of $\sim$0.05. If we consider this offset then the CV\,Cha system becomes consistent with the measurement for the extremely flaring SZ\,Cha system, while WW\,Cha would be consistent with the profile for the HD\,97048 system. 

Generally we find that all disks with well-constrained aspect ratios show values larger than 0.15 at separations past 50\,au (including the bias correction of the measurement method). Conversely we find that for the sample of disks with large uncertainties in their recovered aspect ratios (gray and black data points in figure~\ref{fig: h_over_r}), the general trend is that they have low aspect ratios. Within this subsample, the CS\,Cha system is a special case. Since it is seen under very low inclination ($\sim$20$^\circ$) the result of our edge-tracing fitting approach is intrinsically very uncertain. So it may well be that the system exhibits a larger aspect ratio than what we recover. However, the remaining 5 systems with large uncertainties (CR\,Cha, SY\,Cha, SZ\,45, TW\,Cha, VZ\,Cha), are all seen under intermediate inclinations. In all of these cases, the large uncertainties originate in the comparatively low S/N of these disk detections. As well discussed in section~\ref{sec: contrast} this low S/N is linked to the intrinsic faintness of these disks relative to their central star.

\subsection{Disk polarized scattered light contrast} \label{sec: contrast}

The scattered-light brightness of a disk is not immediately assessed from the image. 
In fact, several elements contribute to the amount of detectable NIR flux (stellar luminosity, self-shadowing, disk geometry, scattering phase function). 
To alleviate this degeneracy, here we calculate the polarized contrast\footnote{Polarized contrast might be a somewhat misleading term. We use it here for consistency with previous studies. What we in fact calculate is effectively an albedo of the disk at the observed wavelength.} of our sources as in \cite{Garufi2014,Garufi2017}. 
This method consists of dividing the observed polarized flux at a certain location by the net stellar flux that is virtually incident on that disk region (that is the stellar flux diluted by the distance). 
Also, the measurement is performed along the major axis (approximately 90$^\circ$ scattering) to minimize the dependence on the scattering phase function. 
Finally, a unique number is obtained for each disk by averaging the values obtained over the separations with detected flux.

\begin{figure}
\centering
\includegraphics[width=0.495\textwidth]{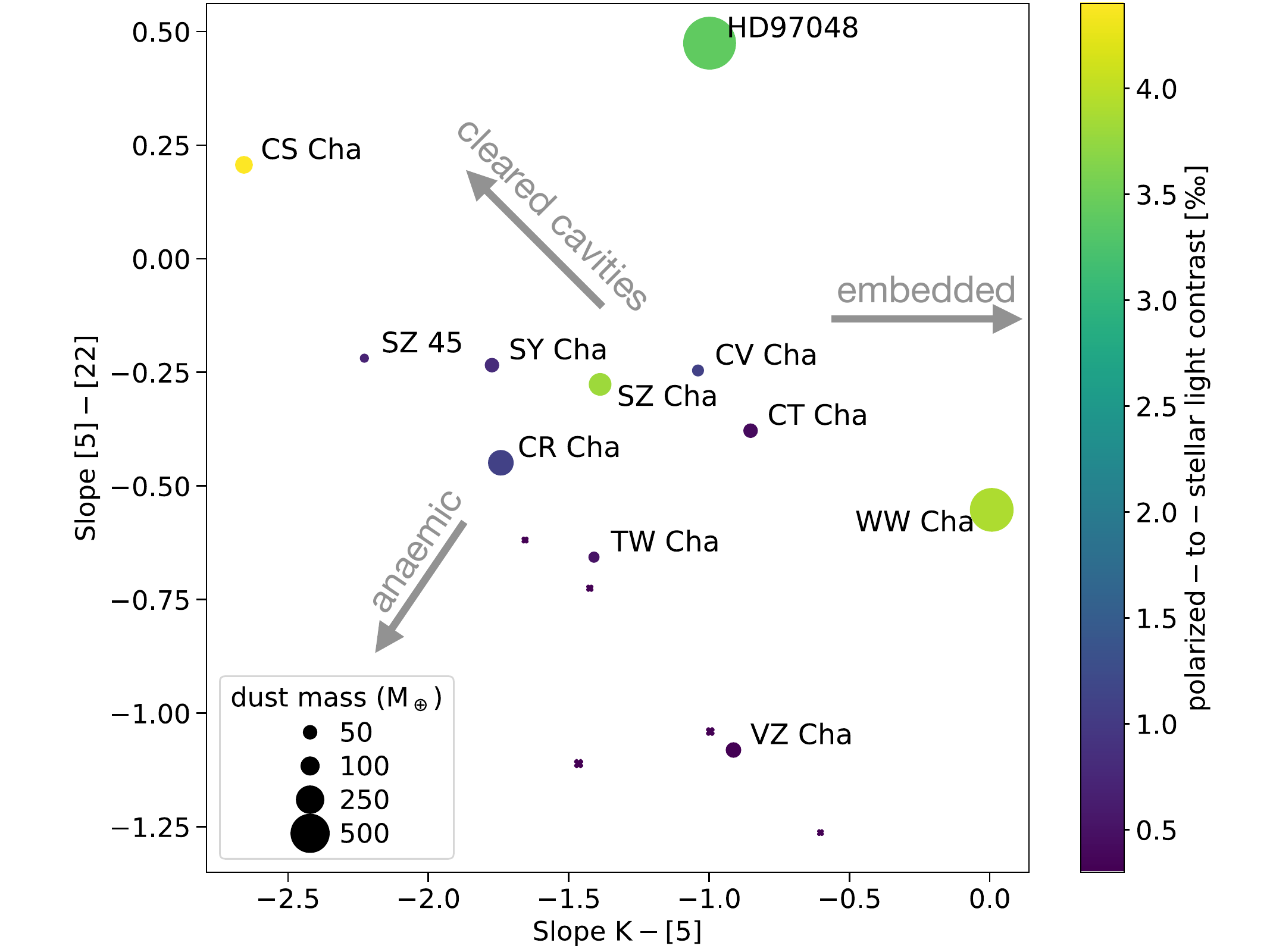}
\caption[]{Target colors vs disk brightness in scattered light. The slopes of the SEDs between K-band and 5\,$\mu$m and between 5\,$\mu$m and 22\,$\mu$m are compared with the polarized-to-stellar contrast, indicated by the symbol color. The disk dust mass is indicated by the symbol size.}
\label{fig: slope-contrast}
\end{figure}

In Figure~\ref{fig: slope-contrast} we plot the obtained polarized contrast versus the near (2.2\,$\mu m$ to 5\,$\mu m$) and mid-infrared (5\,$\mu m$ to 22\,$\mu m$) SED slopes for each system. A reversal in the sign from a negative near-infrared slope to a positive mid-infrared slope indicates classical transition disks, which show a lack of flux typically around 10\,$\mu m$, due to a cavity in the disk. In our target sample, only CS\,Cha and HD\,97048 are such classical transition disks. CS\,Cha shows by far the most prominent flip between near- and mid-infrared slopes. At the same time it is also clearly the brightest disk in scattered light. 
HD\,97048 equally appears bright in scattered light, though not as extreme as CS\,Cha. Its shallower near-infrared slope indicates the presence of some near-infrared excess emission and thus some disk material close to the central star. 

The only other comparably bright sources within our sample are WW\,Cha and SZ\,Cha. WW\,Cha shows a near flat near-infrared slope indicative of the star still being heavily embedded. In figure~\ref{fig: ww cha feature} we see that the disk is indeed still surrounded by either primordial cloud material or material ejected from the system. The position of SZ\,Cha in figure~\ref{fig: slope-contrast} is somewhat puzzling. Both its near- and mid-infrared slopes are average among our sample, indicating that dust is present in the inner and outer system. Although  it shows similar SED slopes to several other disks it is significantly brighter (i.e., more comparable to the transition disk HD\,97048). This is likely connected to the extreme flaring that we find for the disk in section~\ref{sec: ellipse-fitting}.

\begin{figure*}
\centering
\includegraphics[width=18.5cm]{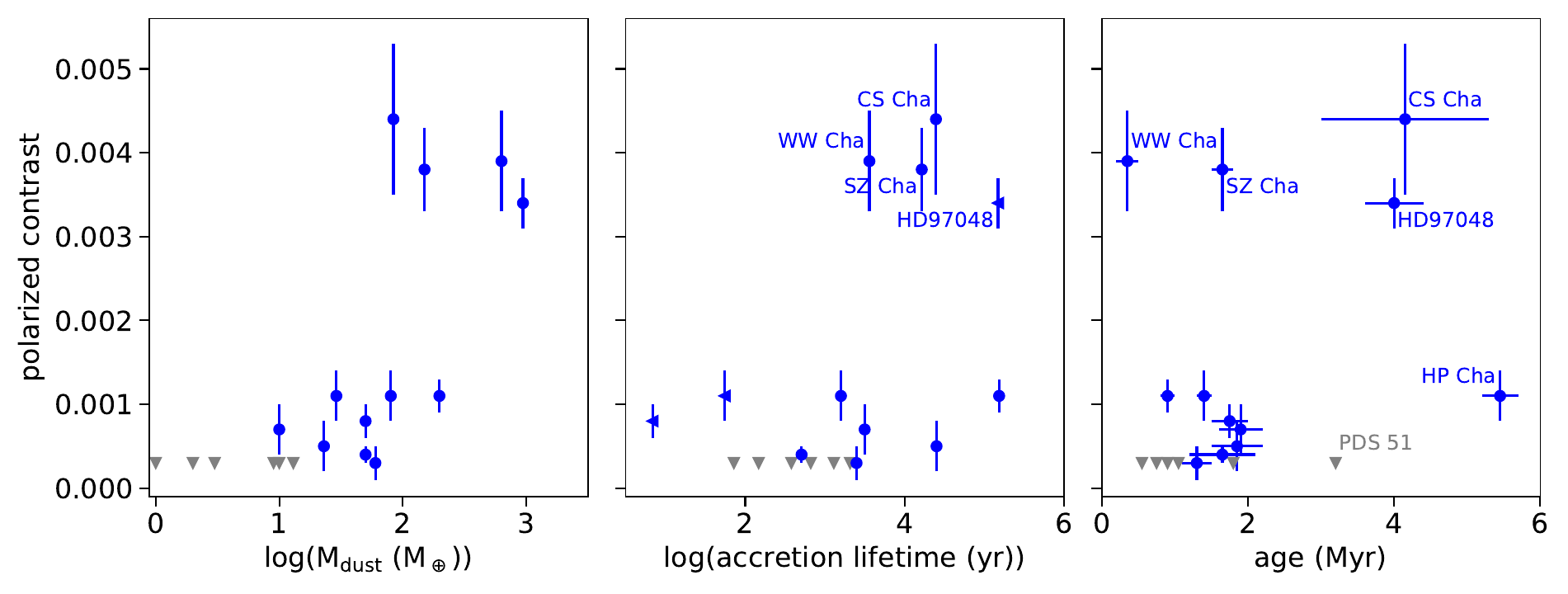}
\caption[]{Polarized contrast as a function of dust mass in the system, the accretion lifetime and the system age. \textit{Left:} Polarized contrast vs dust mass in the system as measured from ALMA continuum photometry. \textit{Middle:} Polarized contrast vs accretion lifetime of the systems as calculated from the accretion rate measured by \cite{Manara2019}, normalized by the dust mass in each system. \textit{Right:} Polarized contrast vs system age.}
\label{fig: system-contrast}
\end{figure*}


In general, Figure~\ref{fig: slope-contrast} shows a trend that disks appear fainter in scattered light if they have steep mid-infrared slopes ($<-0.25$). In particular all our faint nondetections have a mid-infrared SED slope smaller than $-0.5$. This is very similar in nature to the finding of \cite{Garufi2017} who studied disks around Herbig stars in scattered light. They found that there is a clear trend for Group I Herbig stars (as defined by \citealt{2001A&A...365..476M}) to be brighter than Group II Herbig stars. Group I is differentiated from Group II by the presence of a strong far-infrared excess in the SED, indicating the presence of a cold dust component. The mid-infrared slope of Figure~\ref{fig: slope-contrast} is by construction a proxy of the far-infrared excess. Therefore, the trend visible along the vertical direction of the diagram reflects this known behavior for which the removal of material close to the central star increases both the far-IR excess (and mid-IR slope) from the SED and the amount of detectable scattered light from the outer disk.

To understand how the scattered light brightness might depend on other system parameters, we show in Figure~\ref{fig: system-contrast} the polarized contrast plotted versus the system dust mass, the disk-mass normalized accretion rate\footnote{We are using this normalization since there is an observational correlation between dust mass and accretion rate in young systems (see, for example, \citealt{Manara2016}. We note that we are applying a linear scaling in accordance with the model of a pure Shakura \& Sunyayev-like disk. The empirical relation finds on the other hand a power law inex of 0.7 for the disk mass to accretion rate relation.). Since our systems have a variety of dust masses we rather compare the accretion rate in units of the total disk mass for individual systems.} (i.e., the disk accretion lifetime), and the system age. In particular the polarized contrast seems to increase with the logarithm of the dust mass. Conversely we did not detect scattered light signal from disks with dust masses lower than 10\,M$_\oplus$.
To quantify this trend we calculated the Kendall $\tau$-coefficient (\citealt{10.2307/2332226}). We find a correlation with $\tau_K$=0.65 and a probability of 0.03\,\% that both quantities are unrelated. We note that this correlation between the dust mass and the polarized contrast could be influenced by the fact that we used ALMA band 7 fluxes to obtain dust mass estimates. However, as recently discussed by \cite{Ballering2019} and \cite{Ribas2020} the disk may well be partially optically thick in band 7 and the measured fluxes may depend not just on dust mass and temperature but also on the viewing geometry. \\
For the disk accretion lifetime a possible weak correlation between long accretion lifetimes and high polarized contrast is visible. This may be explained if the accretion rate depends strongly on the gap opening in the systems, which in turn leads to stronger illumination of the outer part of the disk. However, CR\,Cha and SY\,Cha seem to not follow this trend (i.e., they have very low accretion rates compared to their dust mass); even so, they are significantly fainter in scattered light than CS\,Cha, SZ\,Cha and WW\,Cha, which have all higher accretion rates.\\
Finally we do not recover a clear correlation between system age and polarized contrast. However, it is noteworthy that out of the group of faint or undetected disks all but two are younger than 2\,Myr. The outliers are PDS\,51, which is a close binary and has additionally the lowest detected dust mass in our sample (2\,M$_\oplus$) and HP\,Cha which is an interacting triple systems (\citealt{Zhang2023}). Conversely the disks around CS\,Cha and HD\,97048 are the brightest objects in our sample and with the exception of HP\,Cha also the oldest. For the bright disks WW\,Cha and SZ\,Cha does not follow this potential age-polarized contrast relation. For WW\,Cha this may be related to the strong interaction of the system with the surrounding cloud. 

\section{Results on individual systems}
\label{sec: ind systems}
The thirteen detections of our sample are very diverse. CHX\,22 and HP\,Cha clearly reveal some evidence of interaction between disk and stellar companions. As described in \citet{Zhang2023}, the image of CHX\,22 reveals a tail-like structure that surrounds the close binary. The very low disk mass in dust revealed by the ALMA images ($<1 M_\oplus$, see section\,\ref{sec: stellar_properties}) points to a very small disk that is clearly shaped by the presence of the companion. Instead, the individual stars of the HP\,Cha are clearly more separated, and the primary disk is likely less disturbed by the companion even though a tenuous streamer indicates that an interaction between the two components is in place \citep{Zhang2023}.  

The disk of HD\,97048 is evidently the most extended disk of the sample, as well as one of the most prominent disk detections in the entire literature. As described by \citet{Ginski2016}, alternating rings and gaps are detected from 40\,au to 340\,au from the star. Conversely, the disks of VZ\,Cha, Sz\,45 and TW\,Cha are faint and apparently featureless in scattered light \citep[see][for an initial discussion of SZ\,45 and TW\,Cha]{Garufi2020, Garufi2022}. Their faintness is most likely due to a self-shadowed geometry where the disk's inner region intercepts and reprocesses a large fraction of the stellar photons \citep{Dullemond2004, Garufi2022}. 

Hereafter, we focus on the nine sources presenting features that are relevant to this work. These are CR, CS, CT, CV\,Cha, HD\,97048, SY, SZ, VZ, and WW\,Cha. Subsequently, we discuss the seven nondetections of the sample.    

\begin{figure}
\centering
\includegraphics[width=0.495\textwidth]{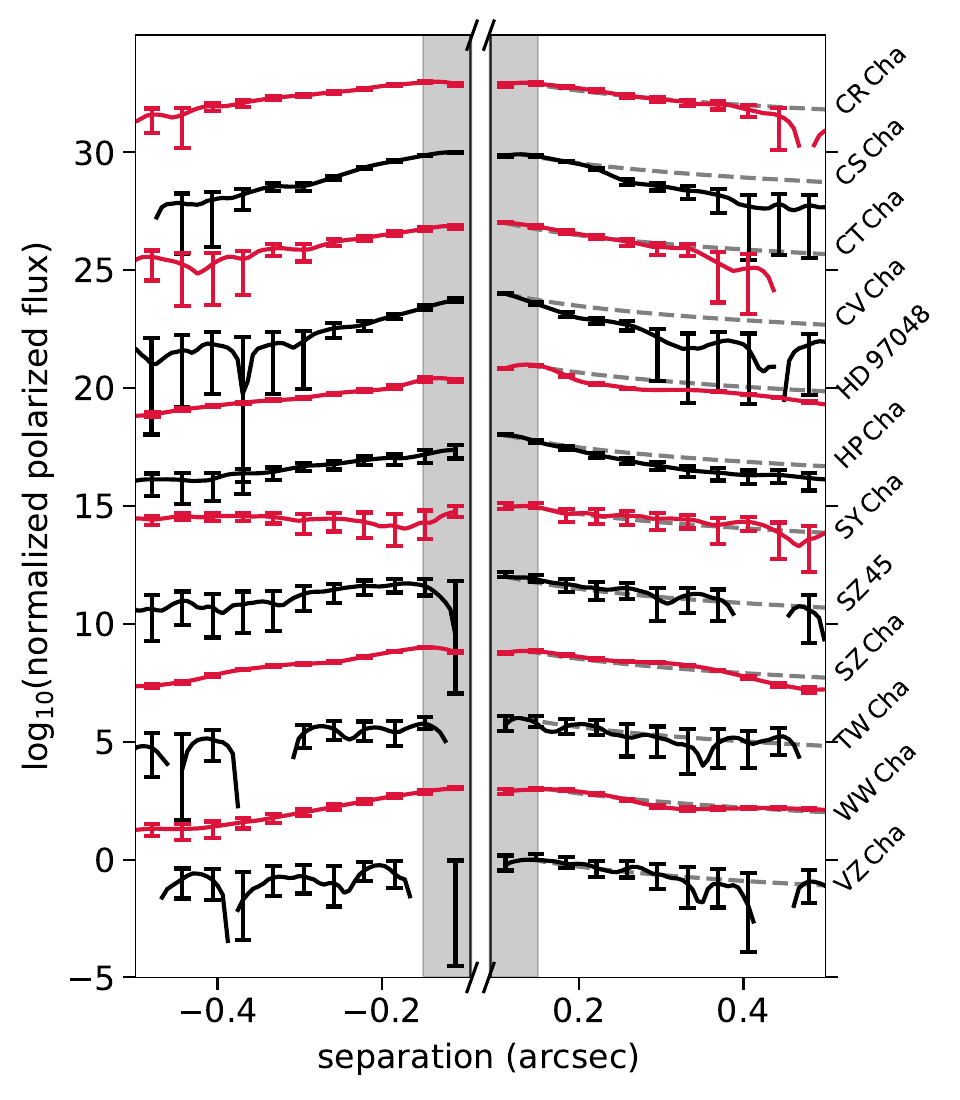}
\caption[]{Radial profiles of all detected disks along the major axis. The polarized flux is in all cases normalized to unity and then displayed on a logarithmic scale. The profiles of individual disks have been vertically offset from each other for better readability by 3 units of the y-axis (starting with an offset of 0 for VZ\,Cha). For the same reason the  individual profiles are alternated in red and black.
The region directly covered by the IRDIS coronagraph is excluded from the plot, while the region that is still influenced by coronagraph suppression is shaded light gray.
Gray dashed lines are added on the right side of the plot to indicate a $r^{-2}$ drop-off, expected from a scattered light signal.}
\label{fig: all disk profiles}
\end{figure}

\subsection{CR\,Cha}

We detect a smooth and featureless disk around CR\,Cha. In the noncoronagraphic images, we find a clear signal down to 85.6\,mas with no indication of a resolved inner cavity. This is consistent with recent ALMA observations by \cite{Kim2020}, who do likewise not recover a cavity in dust continuum emission or gas emission at similar spatial scales. 
In the deep coronagraphic images we find signal out to 0.6\arcsec{} (112.5\,au) along the disk major axis. This includes the region at 90\,au, where there is a clear detection of a gap and a narrow outer ring in the ALMA dust continuum. We do not find significant evidence for this gap in scattered light, but caution that the S/N of our data is low beyond $\sim$0.4\arcsec{} (79\,au). In figure~\ref{fig: all disk profiles} we show that the radial profile along the major axis is dropping consistent with the r$^{-2}$ illumination effect. It is thus possible that a longer observation will detect scattered light signal at larger separations. Indeed CO gas is detected with ALMA out to 240\,au (1.28\arcsec{}, \citealt{Kim2020}).\\
The disk shows a strong brightness asymmetry between the sorthwest and the southeast side, expected from phase function effects in scattered light. The northwest side is significantly brighter than the southwest, from which we conclude that the northwest side is the near side of the disk, showing strong forward scattering.

\begin{figure*}
\centering
\includegraphics[width=18.5cm]{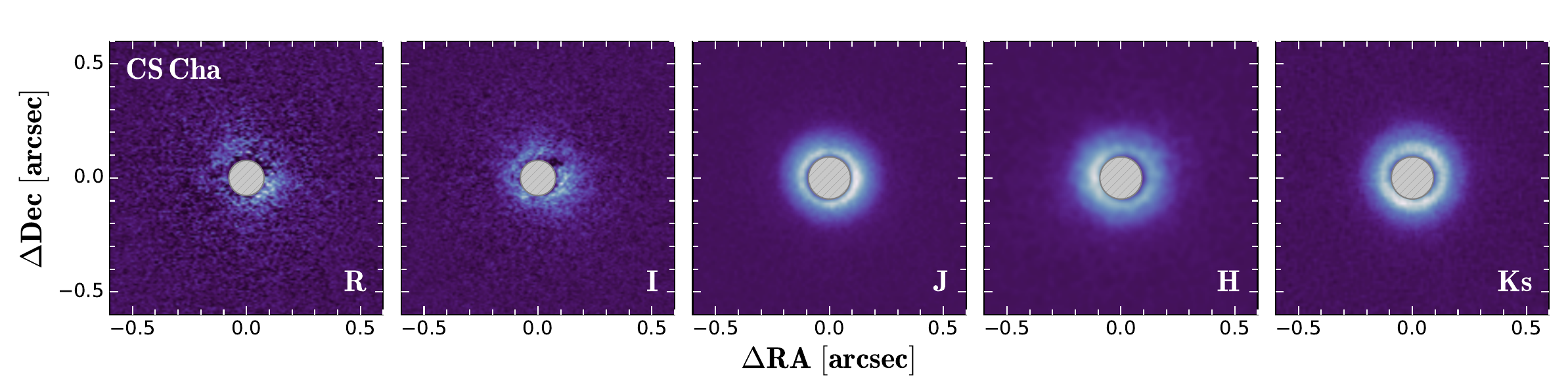}
\caption[]{Multi-band polarimetric observations of the disk around the CS\,Cha system with SPHERE/ZIMPOL and SPHERE/IRDIS. The filter band is indicated in each image. 
Displayed are the Q$_\phi$ images for all bands. The gray hashed disk in the center marks the size of the coronagraphic mask that was employed.}
\label{fig: cscha-multi-band}
\end{figure*}

\begin{figure*}
\centering
\includegraphics[width=18.5cm]{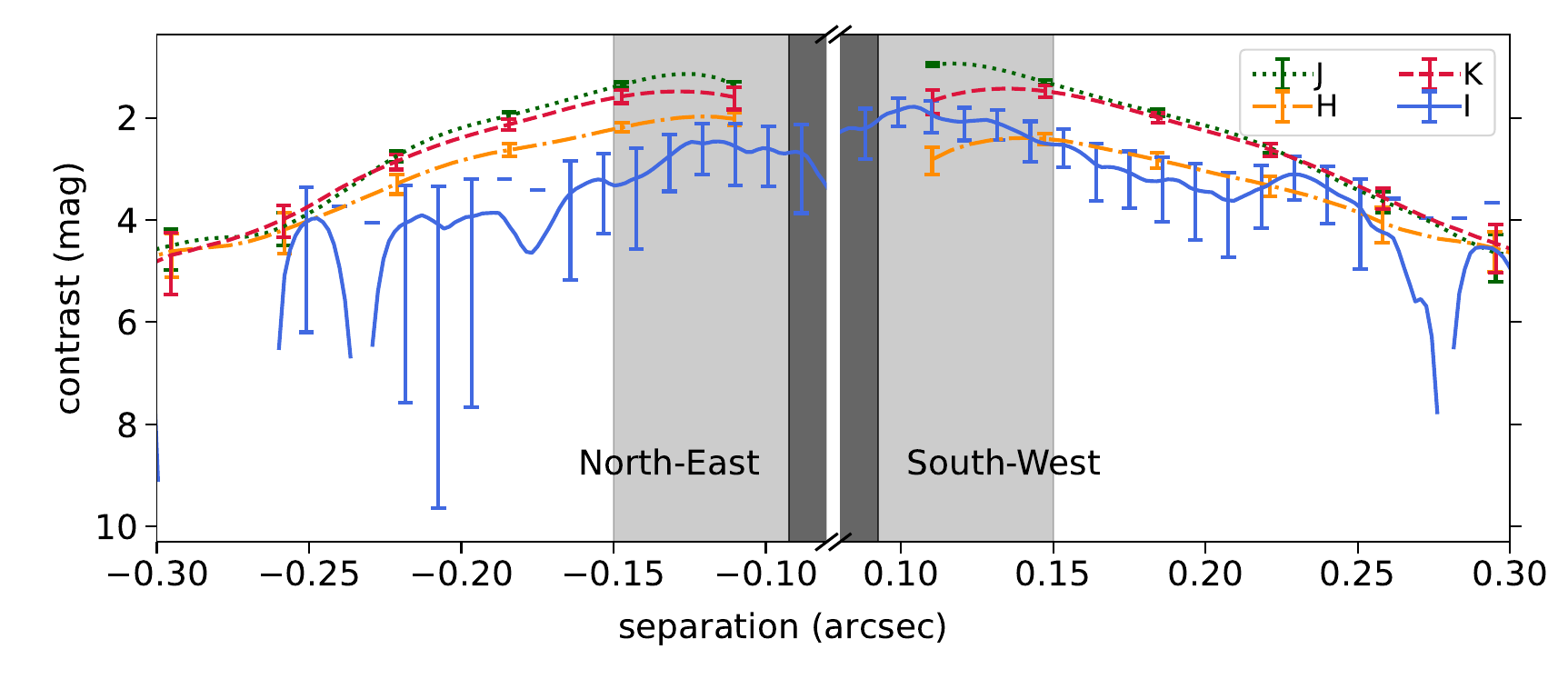}
\caption[]{Radial contrast profiles of the CS\,Cha multi-band observations along the major axis. The contrast is calculated relative to the total intensity of the central star.
The filter band is indicated by line style and color. The region covered by the utilized coronagraphs is shaded dark gray, while the region where coronagraph suppression is still significant is shaded light gray. 
The ZIMPOL coronagraph is smaller than the IRDIS coronagraph, and that the I-band data can be traced farther in. Due to the noisy nature of the R-band data, they are excluded here. }
\label{fig: cscha-multi-band-profiles}
\end{figure*}

\subsection{CS\,Cha}
\label{CSCha-discussion}

The circumstellar disk around CS\,Cha was first resolved in \cite{Ginski2018} in the J and H-band in polarized light. We have now added polarimetric observations with SPHERE/IRDIS in the K-band and with SPHERE/ZIMPOL in the R and I-band. The disk is detected in all bands, shown in figure~\ref{fig: cscha-multi-band}. We also show the extracted profiles along the major axis in figure~\ref{fig: cscha-multi-band-profiles}. \\
Based on ellipse fitting in scattered light the disk has an inclination of 21.6$\degree \pm$6.4$\degree$ and a position angle of $257.7\degree \pm 31.1\degree$. We used available high spatial resolution (beam size 0.09\arcsec\ $\times$ 0.06\arcsec) ALMA continuum Band 7 data \citep{Kurtovic2022} to test the inclination and position angle. An MCMC fit of a ring model to the visibilities finds broadly consistent values of 17.86$\degree $$^{+0.05}_{-0.01}$ for the inclination and $262.6\degree$ for the position angle. Since the ALMA data are not affected by illumination and phase function effects we adopt the ALMA values for further analysis.\\
As is visible in figure~\ref{fig: cscha-multi-band-profiles}, the disk around CS\,Cha appears smooth in all bands without distinct morphological features (we omit the R-band in this figure due to its low S/N). We do not resolve the inner cavity (seen in ALMA data and inferred from the SED) in scattered light down to the inner working angle of the employed coronagraphs (i.e., 92.5\,mas for IRDIS and 77.5\,mas for ZIMPOL). In figure~\ref{fig: all disk profiles} we show the radial profile of the disk along the major axis in J-band compared to a r$^{-2}$ illumination drop-off. Between 0.20\arcsec and 0.25\arcsec the profiles drop off more steeply than the expected illumination function. This is an indication that either the farther-out regions of the disk are strongly shadowed, or that the outer disk is sharply truncated.\\
From the different observation epochs of CS\,Cha we extracted flux-calibrated brightness profiles measured in contrast to the star in circular apertures along the major axis of the disk. Since the R-band data present a low S/N detection we excluded them from this analysis. \\
The disk profiles show a nearly identical slope and contrast in the J and K-band, while the disk appears slightly fainter in the H-band. In the optical I-band the disk is significantly fainter than in the near-infrared. The deviation of the H-band from the near gray scattering in the J and K-band is somewhat surprising. We however caution that the H-band data are of significantly worse quality than the J and K-band data due to the low atmosphere coherence time during the observation (see table~\ref{tab: observations}). In particular the coherence time degraded between the start and end of the observation, while flux calibration frames were only taken at the end of the observing sequence. This may have introduced a systematic effect in the photometric calibration of this data set, as was already discussed for the same data in \cite{Ginski2018}. However, we would expect such an effect to lead to an over-prediction of the disk brightness rather than an under-prediction.\footnote{The AO system produced a worse correction for the flux calibration frames than for the disk science data, making the star effectively appear fainter relative to the disk} 

\subsubsection{CS\,Cha: Scattered light color analysis}
\label{cscha: disk-color}
\begin{figure}
\centering
\includegraphics[width=9cm]{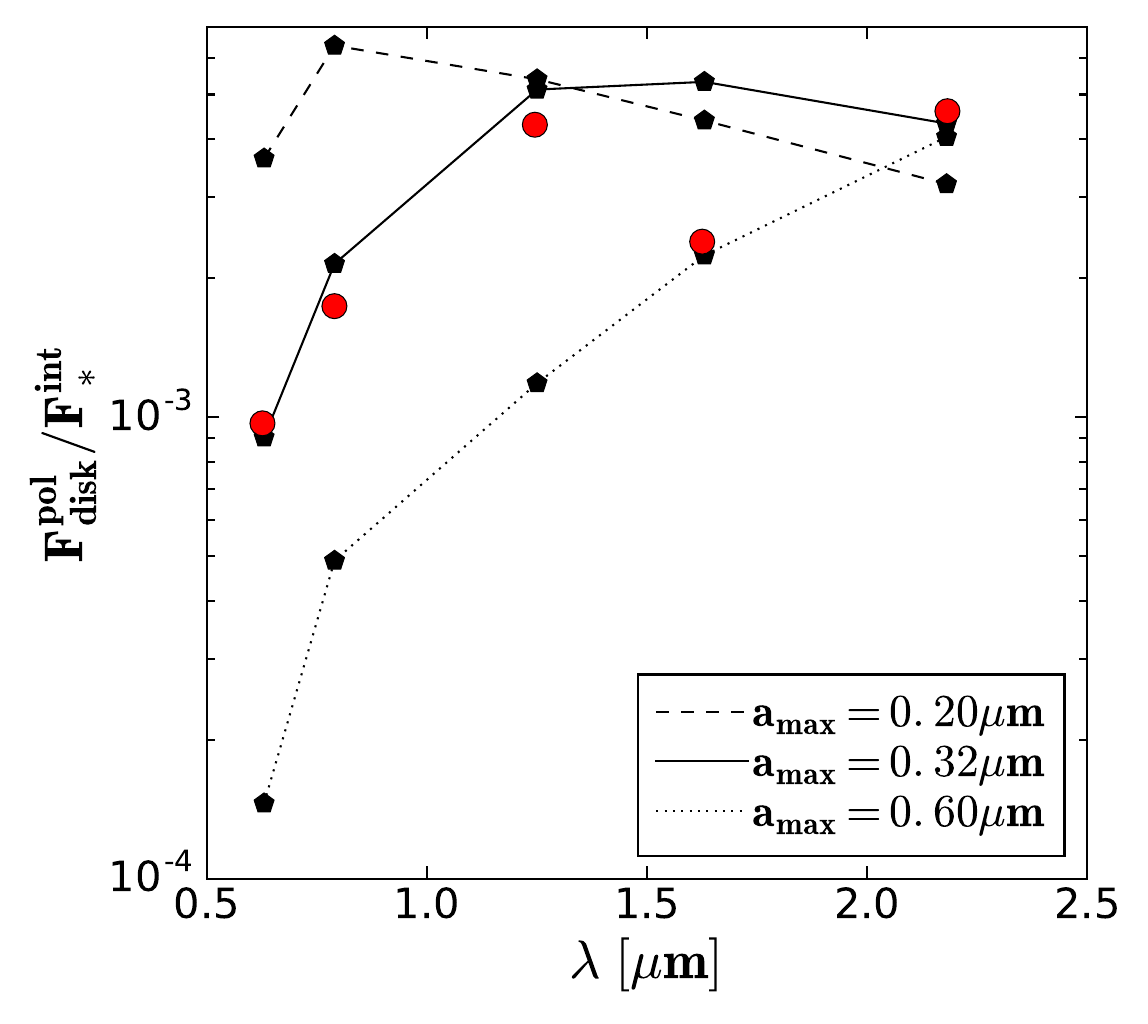}
\caption[]{Integrated polarized scattered light to total intensity stellar flux ratio for CS\,Cha (red dots). The H-band data might be affected by systematic calibration issues, as discussed in \cite{Ginski2018}.
Also shown are  the same for different compact dust aggregate models computed for the inclination of the CS\,Cha system, assuming a settled disk with a flaring exponent of 1.09.
Dust particles are spherical and a size distribution with minimum grain size of 0.1$\mu m$ is assumed. The maximum grain size is a free parameter and is indicated in the figure.
}
\label{fig: cscha-color-dust-model}
\end{figure}

To investigate the dust properties at the scattering surface of the disk we computed the integrated polarized scattered light in an annulus between 0.09" and 1.0" for the IRDIS, near-infrared images. The inner radius was chosen to exclude the region covered by the coronagraphic mask, while the outer region was selected such that all disk flux was included. We then computed the brightness ratio of polarized scattered light to stellar total intensity, as measured in the flux calibration frames of each epoch. We find very similar flux ratios of 4.3$\times 10^{-3}$ and 4.6$\times 10^{-3}$ for the J and K-band respectively. As already expected from the disk profiles, we find a lower ratio of 2.4$\times 10^{-3}$ for the H-band. For the ZIMPOL R and I-band data we measured the integrated polarized scattered light in an annulus between 0.07" and 1.0", taking into account the smaller coronagraph size. We find flux ratios of 1.0$\times 10^{-3}$ and 1.7$\times 10^{-3}$ for the two bands respectively. \\
Ignoring the H-band data, which may suffer from calibration issues discussed in detail in \cite{Ginski2018}, we find that the scattered light shows near gray color between the J and the K-band, while the polarized scattered light flux drops sharply between the J and the I-band. 
To model this we used the radiative transfer code RADMC3d (\citealt{2012ascl.soft02015D}). We assumed compact spherical grains made of pyroxene silicate (refractive index taken from \citealt{Dorschner1995}), with a size distribution with a minimum grain size of 0.1$\mu m$ and a power law index of -3.5. The maximum grain size $a_{max}$ was left as a free parameter. Optical properties were computed using the Mie theory by utilizing the public code \emph{Optool} (\citealt{Dominik2021}). We used the inclination of CS\,Cha of $\sim$22$^\circ$, as measured from the SPHERE data with a small flaring exponent of 1.09, to adjust the overall disk brightness to our observations. The remaining model parameters are identical to the ones presented in \cite{2019MNRAS.485.4951T}. We show the resulting integrated flux ratios in figure~\ref{fig: cscha-color-dust-model} and the corresponding synthetic images in figure~\ref{fig: CSCha-RT}.\\
For compact grains, the degree of linear polarization drops rapidly once the size parameter ($x = 2\pi a/ \lambda$, wherein $a$ is the grain radius) exceeds unity. This explains why the integrated polarized flux decreases for shorter wavelengths for all models. However, the wavelength at which the polarized flux drops depends on the grain radius. The observations suggest this happens between the I and J bands. From model calculations, we found that a model with $a_{max}=0.32 \mu m$ successfully reproduces the steep decline in polarized flux between I and J bands. The presence of such small grains is also in harmony with the lack of brightness asymmetry (i.e., forward scattering) in the observed images at near-IR wavelengths shown in figure~\ref{fig: cscha-multi-band}. However, we note that the model predicts a slightly blue J-K color, while our observations found a slightly red color. A large exploration of the parameter space may be necessary to fully reconcile our observations with the radiative transfer models, which is outside of the scope of this study. We note that new measurements of the polarization fraction in the disk can add additional strong constraints on the dust grain sizes as was recently demonstrated in the observational study by \cite{Ren2023}. Follow-up observations of the CS\,Cha system to detect the disk in total intensity scattered light should be undertaken in the future to compare to the results found here from the disk color.

\subsubsection{CS\,Cha: Stellar binary}
\label{sec: CSCha-orbit}
\begin{figure}
\centering
\includegraphics[width=9.5cm]{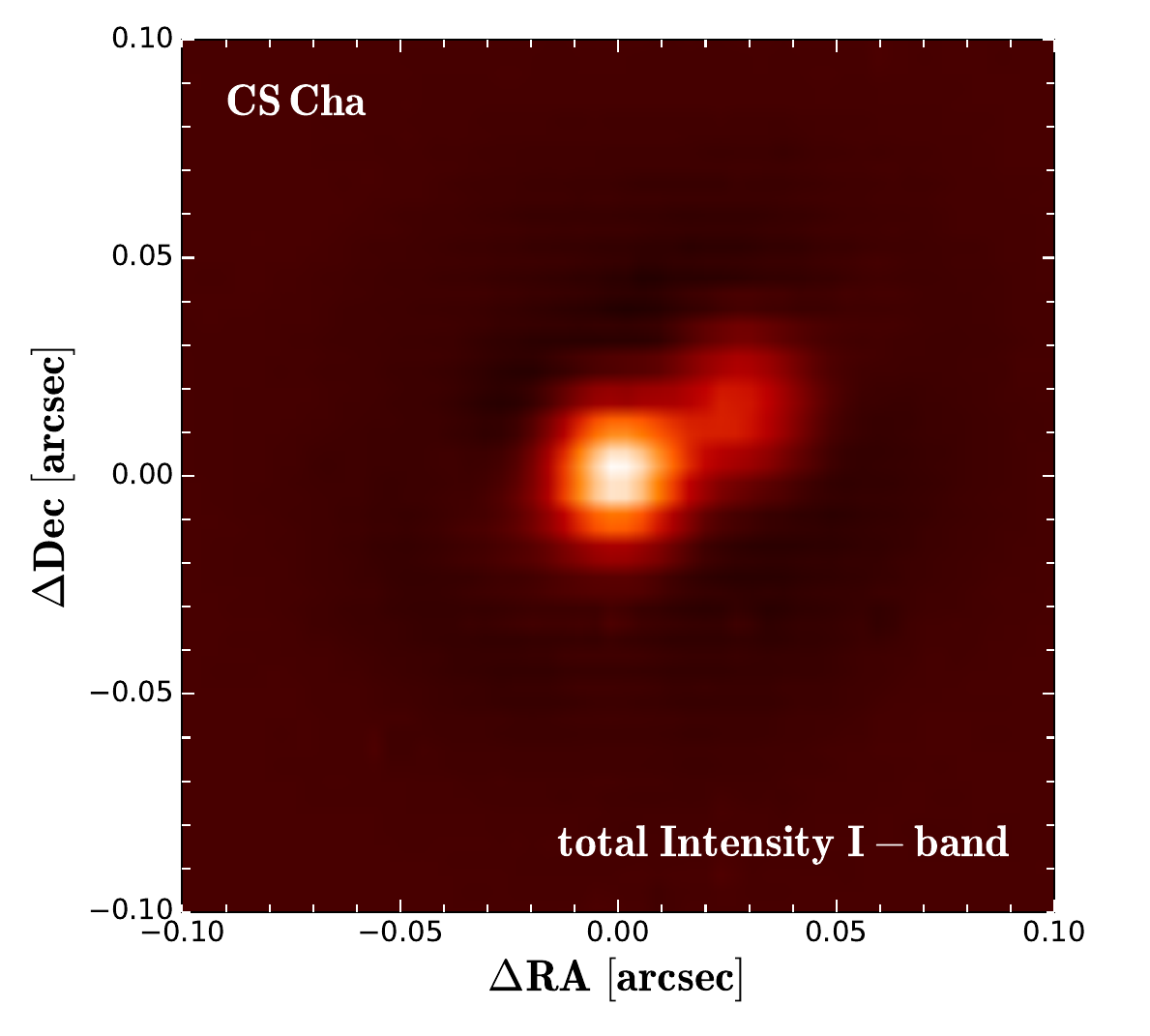}
\caption[]{SPHERE/ZIMPOL I-band observation of CS Cha A. The primary star is resolved in two components. The data are shown on a linear color scale.}
\label{fig: CSCha-binary}
\end{figure}

\begin{figure}
\centering
\includegraphics[width=9.0cm]{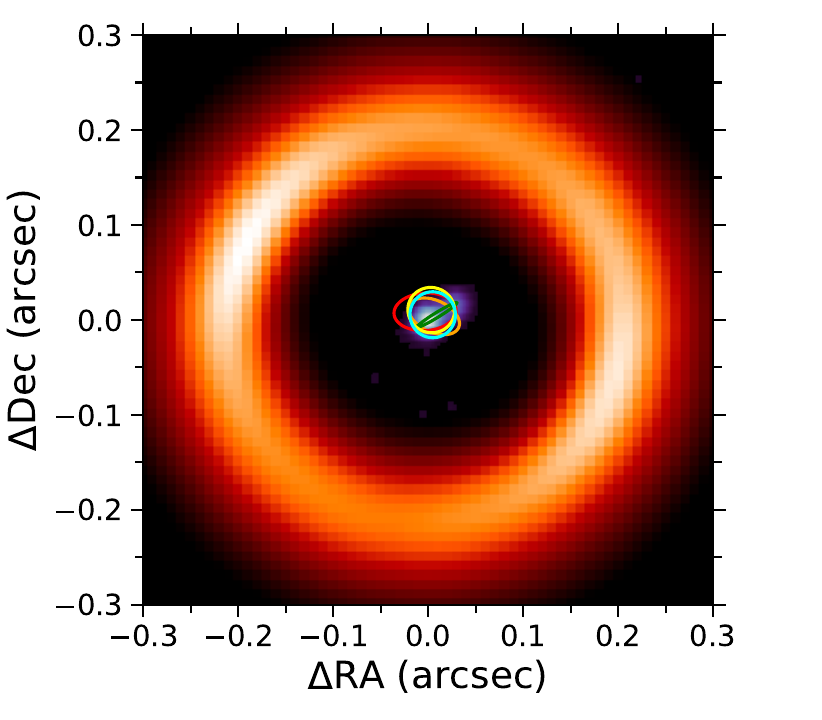}
\caption[]{Combined image showing the ALMA continuum observation at 887$\mu m$ of CS\,Cha, first published by \cite{2020ApJ...892..111F}, together with the new ZIMPOL data showing the resolved stellar binary. We overplot five random orbits from the resulting MCMC fit shown in figure~\ref{fig: CSCha-orbit} to illustrate the size of the binary orbit relative to the cavity size in millimeter continuum.}
\label{fig: CSCha-orbit-alma}
\end{figure}

The ZIMPOL observation sequences in R and I-band of the system included short noncoronagraphic sequences for flux calibration purposes. In these sequences, we found that the primary star in the system is for the first time resolved as a visual binary star (figure~\ref{fig: CSCha-binary}). CS\,Cha\,A was already known to be a spectroscopic binary from radial velocity observations taken by \cite{Guenther2007}. They found that the system may have a mass ratio close to 1 and an orbital period of or longer than 2482\,days. However, since they did not cover a full orbit they could not further constrain the orbital parameters. Our ZIMPOL observation finds a flux ratio of 0.31$\pm$0.03 in R-band and 0.41$\pm$0.02 in I-band. 
To convert the flux ratio to apparent magnitudes we use the cataloged I-band magnitude of 10.12$\pm$0.04\,mag (\citealt{2012AcA....62...67K}) and the R-band magnitude of 10.7$\pm$0.3\,mag (\citealt{2014A&A...570A..87S}). Since the binary was unresolved for these measurements we corrected the primary star magnitude following \cite{2020A&A...635A..73B}, using the flux ratio between the two components as input. This yielded corrected I-band and R-band magnitudes of the primary star of 10.49$\pm$0.04\,mag and 11.0$\pm$0.3\,mag respectively. Given our measured flux ratio, the secondary then has apparent I and R-band magnitudes of 11.46$\pm$0.06\,mag and 12.3$\pm$0.3\,mag, respectively. Assuming an average system age of 4.2\,Myr (see table~\ref{tab: sample}) and AMES-COND model isochrones (\citealt{Allard2001}), we find masses of 1.2\,M$_\odot$ and 0.8\,M$_\odot$ for Aa and Ab, with a typical uncertainty of 0.02\,M$_\odot$ based on the age and photometric uncertainty. This uncertainty may be underestimated, given that different isochrone models may well yield slightly different results. The total mass of the two components that we recover is well consistent with the total mass extracted from gas kinematics in the disk by \cite{Kurtovic2022}.\\
To extract the astrometry from the ZIMPOL images we simultaneously fitted two Moffat functions to the binary star. A Moffat function is typically a very good fit to the ZIMPOL point spread function (PSF), especially for fainter stars, since it still has a significant seeing limited halo surrounding the diffraction-limited PSF core. We find a separation of 31.6$\pm$1.3\,mas at a position angle of 297.7$^\circ \pm$2.1$^\circ$. The astrometric calibration of the detector was derived from several visual binary stars with well-known orbits as well as the internal pin-hole calibration grid (Ginski et al, in prep.). \\
We combined the radial velocity measurements given by \cite{Guenther2007} with our new astrometric measurement to derive the orbit of the CS\,Cha Aab pair. We use the \emph{orbitize!} python package for the fit with the included Markov-Chain Monte-Carlo sampler (\citealt{Blunt2020, Foreman-Mackey2013}). We set the prior of the combined system mass using a Gaussian distribution with the peak at 2.0\,M$_\odot$ and the conservative standard deviation of 0.2\,M$_\odot$ in line with the extracted masses of the stellar components from photometry. For the purpose of this orbit fit we adopted the
distance of 190\,pc (i.e., a parallax of 5.26\,mas). However we allow the fit to include the parallax and set a Gaussian prior with a standard deviation of 0.7\,mas, which then includes the nominal Gaia DR3 parallax of 5.93\,mas.
We ran $1.2 \times 10^{6}$ orbit solution with 50 walkers.  
The results for the semi-major axis, the eccentricity, and the inclination of the orbit are displayed in figure~\ref{fig: CSCha-orbit}.\\
We find that the binary orbit can be well constrained with the existing data points. The MCMC posterior distributions point to a most likely semi-major axis of 5.0$\pm$0.2\,au. The orbit shows an intermediate eccentricity of 0.40$\pm$0.04. The inclination distribution of the orbits shows a preference for prograde orbits peaking at $\sim$40$^\circ$. Somewhat unsurprisingly, given our limited astrometric data, we can not rule out retrograde orbits with an inclination of $\sim$140$^\circ$. Given the inclination of the circumstellar disk recovered from ALMA data of $\sim$18$^\circ$ (see table~\ref{tab: disk geometry}), it appears likely that the binary orbit may be somewhat misaligned with the disk plane. However, we note that we cannot rule out lower inclination orbits with our existing data and that indeed the orbital distributions shown in figure~\ref{fig: CSCha-orbit} still display a high density of solutions toward low inclinations. To illustrate the size of the orbit relative to the detected submillimeter continuum gap in the disk (\citealt{2020ApJ...892..111F}), we show several orbits over-plotted with the ZIMPOL and ALMA data in figure~\ref{fig: CSCha-orbit-alma}.\\
Given our orbit solutions it is interesting to investigate if the binary companion can be solely responsible for opening the gap in the disk traced by the ALMA data. The cavity radius from ALMA found by \cite{2020ApJ...892..111F} and scaled to our adopted distance of 190\,pc is 40\,au. The mass parameter $\mu =M_2/(M_1+M_2)$ for the CS\,Cha system is 0.4. The calculations by \cite{1994ApJ...421..651A} indicate that for an eccentric binary system with $e=0.4$ (identical to the peak we find in the orbit distribution for CS\,Cha) and a slightly lower mass companion with $\mu = 0.3$ as well as a disk viscosity of $\alpha_v = 10^{-3}$, the outer edge of the cavity should be located at 3.1$a$. This assumes that the binary orbit is co-planar with the circumbinary disk, which is consistent with at least some of the orbits that we find as discussed earlier. Given the peak of the semi-major axis distribution recovered by our orbit fit of 5\,au we would then expect for the inner cavity edge to be located at 15.5\,au. Given that in our case the binary component has possibly an even larger fraction of the total system mass, this value could still shift slightly to larger outer gap radii. The precise disk morphology will furthermore not only depend on the mass ratio of the binary and the eccentricity of the orbit, but also on the relative inclination of binary orbit to disk plane, as has been demonstrated (for example,  \citealt{Price2018, Ragusa2021}). 
In any case, our simple estimation indicates a significant discrepancy between the gap radius seen by ALMA and the expected gap radius based on the orbital configuration of the binary. 
In a recent study (based in part on our ZIMPOL detection of the stellar binary) \cite{Kurtovic2022} further argue that based on the brightness asymmetry observed in the ALMA data (not discussed in our study), an additional planet might be needed to explain the uneven distribution of material in the ring.
This would be well consistent with our result that the gap radius appears too large to be caused by the stellar binary alone. We refer to the hydrodynamic simulations by \cite{Kurtovic2022} for a deeper discussion of the star (and planet) disk interaction in the CS Cha system.\\
It is interesting that the large cavity seen in ALMA continuum is not detected in the scattered light observations of the system. This may indicate dust segregation as discussed by \cite{Ovelar2013}. While they do not discuss the case of a stellar companion they find that for various planetary mass objects that the disparity in cavity size between micron and millimeter sized dust grains increases with planet mass. Extrapolating these results to the case of CS\,Cha with a stellar companion (and possibly an additional planetary companion) might then explain the nondetection of the cavity in scattered light.

\subsection{CT\,Cha}

We find a compact smooth disk around CT\,Cha\,A. With the fitting of the scattered light data as described in section~\ref{sec: ellipse-fitting} we find that the disk major axis goes from the northeast to the southwest (see table~\ref{tab: disk geometry}). The disk shows an inclination of 45.7$^\circ \pm$5.0$^\circ$. The disk profile drops with the expected r$^{-2}$ illumination effect and is detected out to 0.35\arcsec{} (65\,au). Since the disk signal drops at the rate expected due to central illumination it is well possible that the disk extends beyond this range. Evidence for this is that we do not detect signal from the forward scattering peak of the backside of the disk, as is seen in other inclined systems (for example, IM\,Lup, DoAr\,25, IK\,Lup \citealt{Avenhaus2018, Garufi2020}). Such signal would be blocked from view if the disk indeed extends to larger radii. We however caution that the inclination of CT\,Cha is somewhat lower than these example disks (with IM\,Lup having the lowest inclination of 55$^\circ$). Thus it may be possible that the absence of the visible disk backside is simply a combination of viewing geometry and grain properties.\\
As expected from the polarized scattering phase function we receive strong signal from the disk ansae where scattering angles are close to 90$^\circ$. We detect the disk to slightly larger separation along the minor axis in the southeast than in the northwest. For a moderately to strongly inclined disk this is expected due to the bowl shape of the disk surface if the northwest side is the near side of the disk, which mainly shows a very narrow forward scattering rim. The peaks of the phase function are moved from the ansae in direction of the minor axis on the southeast side. This is again expected if we observe a flaring disk with the far side in the southeast and the near side in the northwest (for example, \citealt{deBoer2016}).\\
We note that we detect unresolved polarized light originating from the known substellar companion CT\,Cha\,b. This object is discussed in detail in Schmidt et al., in prep., including the polarimetric data presented here. 

\subsection{CV\,Cha}

\begin{figure}
\centering
\includegraphics[width=9cm]{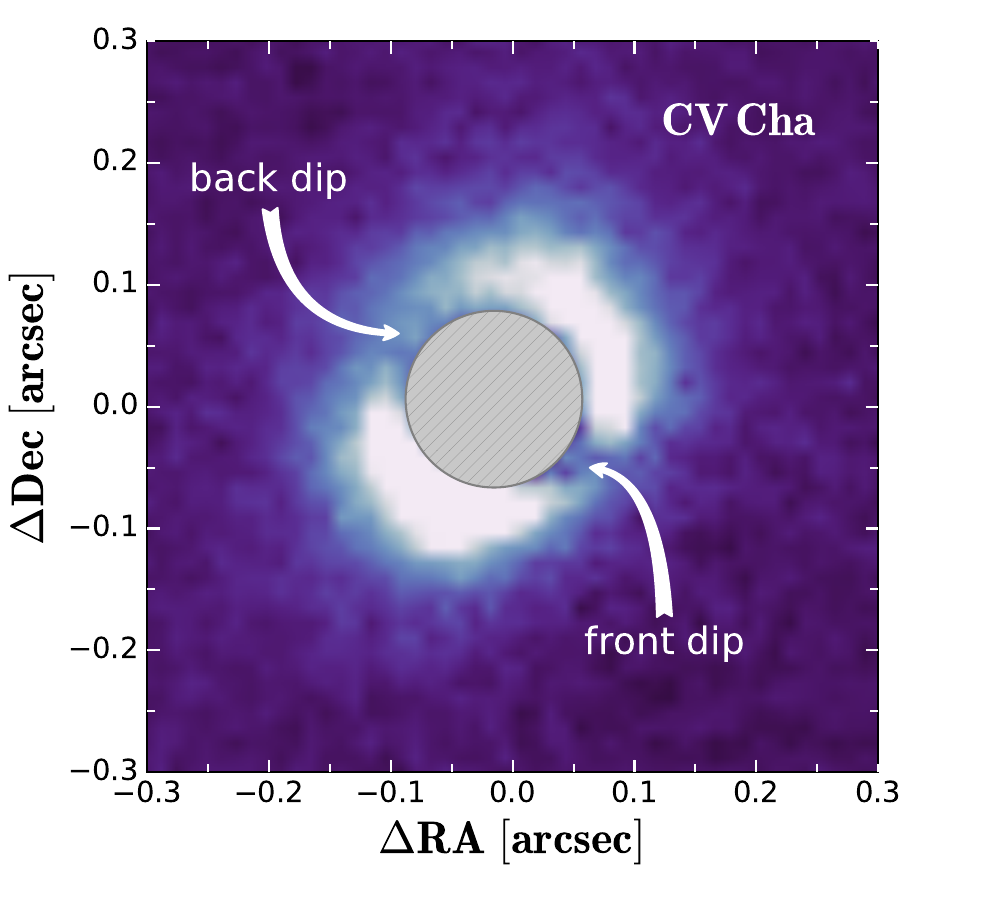}
\caption[]{Q$\phi$ image of the CV\,Cha system from J-band observations. Two steep azimuthal dips in brightness are detected.}
\label{fig: cv cha feature}
\end{figure}

We resolve a bright, compact disk around CV\,Cha. In the noncoronagraphic images of the system significant signal is detected down to 0.05\arcsec{} (9.5\,au). We do not resolve a cavity in scattered light. Ellipse fitting of the scattered light data finds the major axis from the southeast to the northwest. The disk appears similarly inclined as the CT\,Cha system with a recovered inclination of 43.0$^\circ \pm$5.3$^\circ$. However, different from the CT\,Cha system the radial profile along the major axis drops of much steeper than expected from illumination effects (see figure~\ref{fig: all disk profiles}).
This indicates that the disk is either indeed small or strongly self-shadowed beyond the point where we detect it. As was the case for CT\,Cha we do not detect the backside of the disk, which may indeed indicate that the disk extends farther and blocks our line of sight to the backside. This would then indicate that the region we observe is strongly puffed up and casts a shadow on the outer disk regions. The disk was not resolved in \cite{Pascucci2016} in ALMA millimeter continuum emission with a beam size of 0.7\arcsec{}$\times$0.5\arcsec{}, in agreement with our observations tracing smaller grains. However the gas could extend much farther and only small surface densities of $\mu$m dust grains are needed mixed in the gas to make the extended disk optically thick in the near-infrared.
The disk is detected at larger separations along the minor axis toward the northeast. Similar to CT\,Cha we interpret this as the far side of a flared, inclined disk. \\
We find two distinct dips in the azimuthal brightness distribution along the minor axis. We highlight these in figure~\ref{fig: cv cha feature}. In principle it is expected that the polarized scattering phase function peaks in the ansae and produces less signal toward forward and back-scattering sides of the disk. However, in practice the forward scattering side of the disk is rarely much fainter than the ansea when it is indeed resolved (see, for example, the polarized phase function measured for the HD\,97048 system, \citealt{Ginski2016}). Furthermore, in figure~\ref{fig: cv cha feature} it is visible that in particular the dip on the forward scattering side of the disk is very sharp. If this were a phase function effect we would expect a smoother transition from bright to dark azimuthal areas. Such azimuthal brightness dips have now been observed in a number of systems, for example HD\,142527 (\citealt{Marino2015}) and HD\,100453 (\citealt{Benisty2017}), and are usually attributed to a misaligned inner disk casting a shadow on the resolved outer disk. To produce narrow shadows the misalignment has to be significant. The position of the dips along the minor axis indicates not only a misalignment in inclination but also in position angle of the unresolved inner disk with respect to the outer disk.\footnote{Since the putative shadows line roughly up with the outer disk minor axis, the inner and outer disk would have to be misaligned in position angle by roughly 90$^\circ$.} 
With our data it is ultimately not possible to decide whether we see a steep phase function effect or shadows caused by disk misalignment. Optical interferometry or possibly the longest baseline ALMA observations may be able to resolve the inner disk zone and confirm or disproof the misalignment scenario. 

\subsection{HD\,97048}

\begin{figure}
\centering
\includegraphics[width=0.5\textwidth]{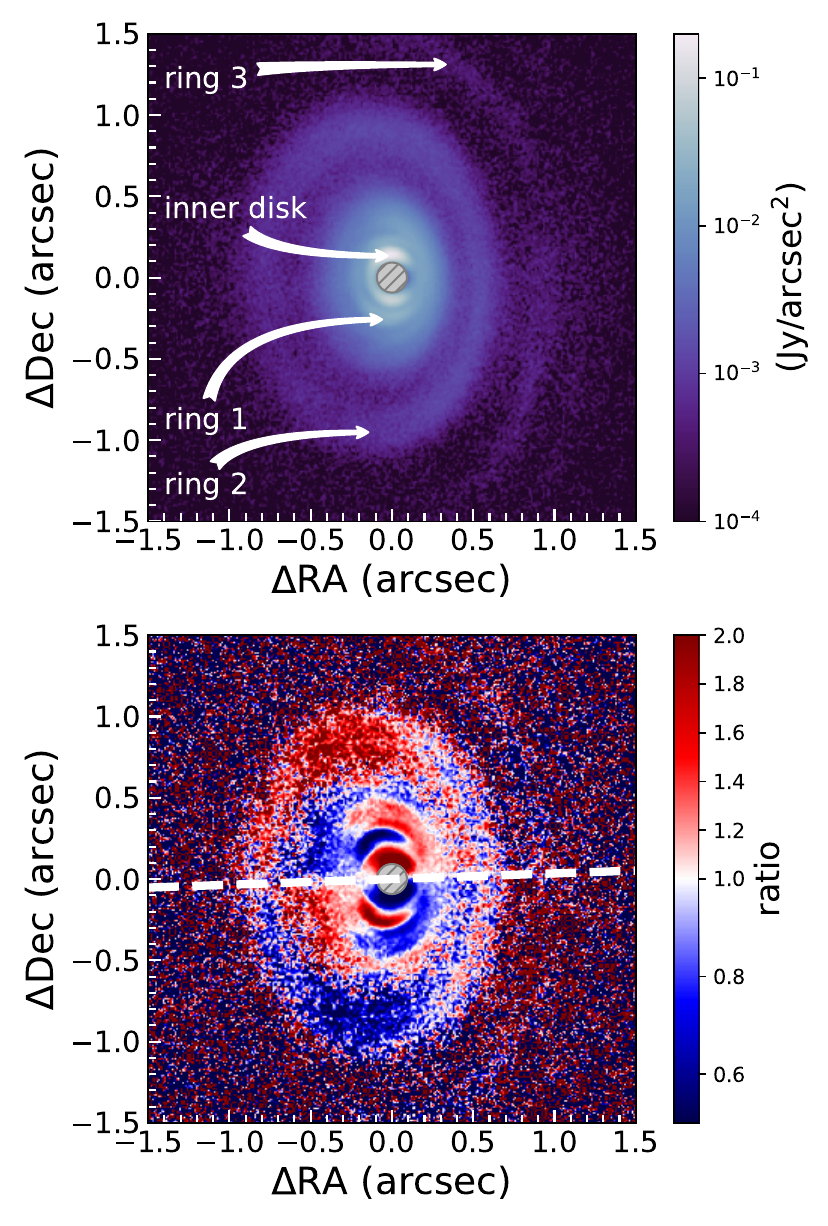}
\caption[]{Axial asymmetry of the disk surrounding HD\,97048 relative to the minor axis from SPHERE K-band observations. The disk was flipped around its minor axis and then  the original image was divided by the flipped image. The ratio of the two is indicated by the color map. The white dashed line indicates the disk minor axis.}
\label{fig: hd97 asymmetry}
\end{figure}

The HD\,97048 is a well-studied Herbig star. \cite{Lagage2006} resolved the disk around this system for the first time in mid-infrared polycyclic aromatic hydrocarbon (PAH) emission and found that it is extended and strongly flaring. Subsequent scattered light observations with HST/ACS traced disk structures out to $\sim$600\,au (\citealt{Doering2007}). More recently the planet-forming disk was resolved with extreme adaptive optics observations in the near-infrared with VLT/SPHERE and Gemini/GPI in the J and H-band, respectively (\citealt{Ginski2016, Rich2022}). These observations revealed multiple rings and gaps at radial separations between 40\,au and 340\,au. \cite{Ginski2016} extracted the height profile of the disk by tracing the center offset of rings and gaps along the disk minor axis and found that the disk scattered light surface height profile can be described by a single power law with a large flaring exponent of 1.73.\footnote{Recently \cite{Rich2021} found an even larger flaring exponent after suggesting a different fitting approach for one of the disk rings in HD\,97048.} Complementary submillimeter observations of the system found a central cavity in the dust continuum emission with a size of 40-46\,au (\citealt{vanderPlas2017, Walsh2016}). Using high spectral and spatial resolution ALMA gas line observations \cite{Pinte2019} reported the detection of dynamic signatures in the disk, consistent with the presence of an embedded gas-giant planet at an orbital separation of 130\,au. \\
Within our survey we present new VLT/SPHERE K-band polarimetric observations of the system. We show the corresponding Q$_\phi$ image in figure~\ref{fig: hd97 asymmetry}. The new K-band data are of significantly higher S/N compared to the J-band observations presented in \cite{Ginski2016}. In particular ring 3 (at $\sim280\,au$) from \cite{Ginski2016}, which they only detected in total intensity angular differential imaging reductions, is well recovered in the new K-band polarimetric image. As previously discussed in section~\ref{sec: ellipse-fitting}, we extracted the height profile of the disk from the K-band image and find a similar but overall lower profile as was recovered for the J-band data. This is consistent with the lower dust opacity in the K-band. \\
Due to the higher S/N of the new K-band data, we identify for the first time strong asymmetries in the disk scattered light image. We are highlighting these asymmetries in the bottom panel of figure~\ref{fig: hd97 asymmetry}. There we show the ratio of the disk surface brightness between the original image and an image flipped along the disk minor axis. For a perfectly symmetric disk, we would expect a ratio of 1 (i.e., axis symmetry with respect to the disk minor axis). Instead for HD\,97048 we find deviations of up to a factor of $\sim$2. Curiously we see a pattern along the disk major axis (from north to south in the image) where the disk surface brightness flips from rations smaller than 1 to ratios larger than 1. This brightness asymmetry can be well observed also in the top panel of figure~\ref{fig: hd97 asymmetry}. The inner disk appears brighter in the north, while ring 1 appears brighter in the south. A similar effect was observed for the multi-ringed disk around the T Tauri star RX J1615.3-3255 by \cite{deBoer2016}. They argued that this may be caused by oscillating shadowing of the outer rings by the next innermost ring. However, they suggest that such shadowing should only be expected for disks with low flaring exponents, whereas the disk around HD\,97048 is strongly flaring. \cite{Muro-Arena2020} showed that small (few degrees) relative misalignments within the disk around HD\,139614 can explain the large brightness asymmetries observed in this system. Following this logic the asymmetries observed in the HD\,97048 system may be interpreted as small-scale warps within the disk, which change alignment moving radially within the disk. This may be consistent with the evidence that at least one high-mass planet is embedded in the disk from \cite{Pinte2019}. If the orbital plane of this planet is slightly misaligned with the disk plane, then the planet could be responsible for warping the disk, similar as is suggested for the much more evolved $\beta$\,Pic system (\citealt{Dawson2011}). Given the multiple "flips" in brightness asymmetry that we observe it may also be possible that this disk asymmetry is not static, but that we observe signs of a dynamic warp, which produces time-dependent "ripples" within the disk.\\
As we show in Figure~\ref{fig: all disk profiles}, the extracted radial profile of HD\,97048 along the major axis does not follow the expected r$^2$ drop-off in illumination. Along with CV\,Cha, HD\,97048 is the only disk in our sample that shows a strong deviation from this expected behavior. In part this may be explained by the strong flaring of this particular disk, which is not taken into account for the simple computation of the distance in this plot (although  the even stronger flaring disk around SZ\,Cha appears consistent with the expected r$^2$ drop-off). However, this may be a further indicator that the disk is partially shadowed due to internal misalignment.  

\subsection{SY\,Cha}

\begin{figure*}
\centering
\includegraphics[width=18.5cm]{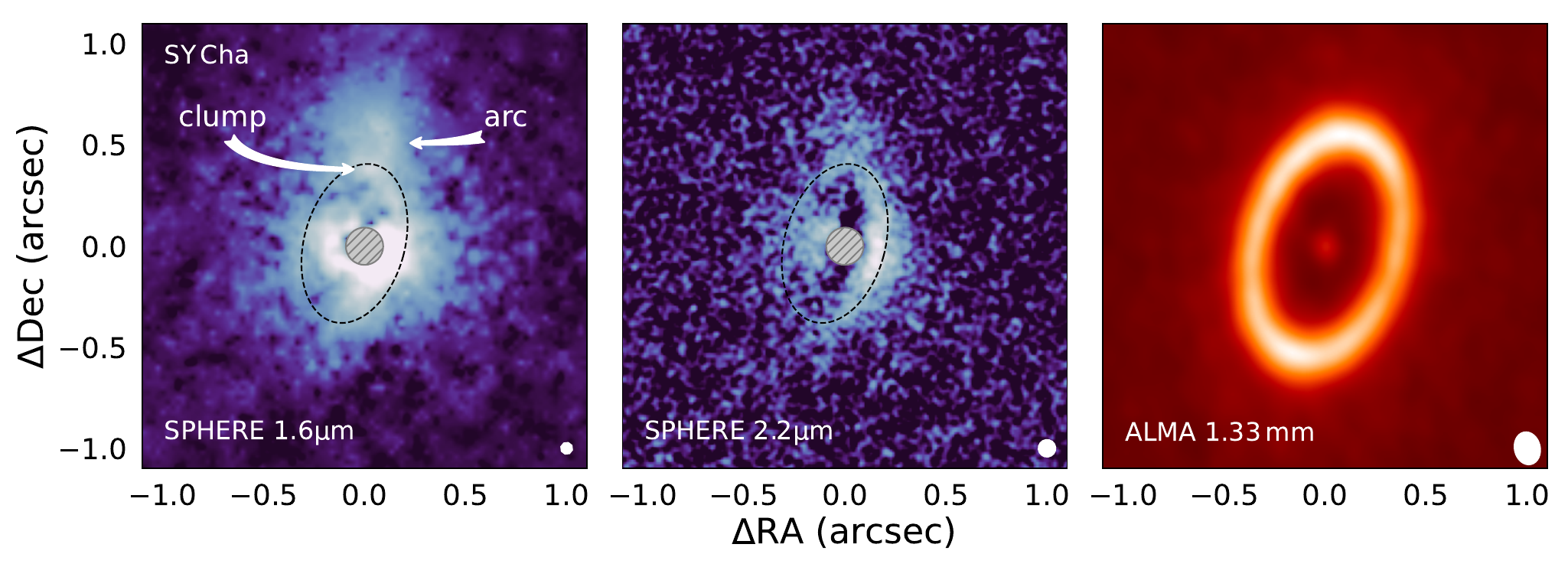}
\caption[]{Observational data of the SY\,Cha system. \textit{Left:} SPHERE/IRDIS polarized light H-band data of SY\,Cha. A variable Gaussian kernel was used to smooth the image. The FWHM of the kernel was set proportional to the inverse of the signal in the image (i.e., regions with strong signal were smoothed minimally), while regions that mostly contained background noise were smoothed strongly,  analogously to what was done for the UX\,Tau system in \citealt{Menard2020}). The data are shown on a logarithmic scale in order to highlight the disk morphology. The effective resolution element is shown on the lower right. The black dashed ellipse matches the ALMA inclination and position angle and highlights two features near the northern ansae. \textit{Middle:} SPHERE/IRIDS K-band data of SY\,Cha. The image is shown on a logarithmic scale. The data have been smoothed with a simple Gaussian to boost the signal.  \textit{Right:} ALMA dust continuum emission of SY\,Cha in Band 7 from \cite{Orihara2023}. The beam size is indicated by the white ellipse in the lower right corner.}
\label{fig: sycha-sphere-alma}
\end{figure*}

We detect clear scattered light signal around SY\,Cha in the H and K-band (see Figure~\ref{fig: sycha-sphere-alma}). However the shape of the signal appears complex and not easy to interpret. Some of the detected structures have low S/N, in part due to low data quality in the H-band and the intrinsically faint signal in the K-band. The primary star in the SY\,Cha system is faint in the optical (Gmag = 12.53$\pm$0.01, \citealt{2018A&A...616A...1G}) and thus a challenging target for the adaptive optics system of SPHERE (see \citealt{Jones2022}). During the H-band observation the tip-tilt stabilization had to be disabled, resulting in the star moving slightly behind the coronagraphic mask. While re-alignment of all frames was performed, using faint background stars as calibrators, this still resulted in degraded performance. \\
In figure~\ref{fig: sycha-sphere-alma} we see that signal was detected in the H-band right outside of the coronagraphic mask with a strong peak in the west. 
In the same figure we show the ALMA Band 6 continuum data from \cite{Orihara2023}. In the ALMA data the disk is resolved into an outer ring with an azimuthal asymmetry between the north-east and the south-west, an inner cavity and an unresolved inner disk component. The position angle of the major axis and the inclination were extracted from the ALMA data and are shown in table~\ref{tab: disk geometry}. The major axis is located at a position angle of $\sim$345$^\circ$. This fits well with the SPHERE data which show extended signal along the north--south direction.\\
The inclination of the disk is $\sim$52$^\circ$ as measured from ALMA (\citealt{Orihara2023}). In the left and middle panel of figure~\ref{fig: sycha-sphere-alma} we overlay the extracted ellipse from our fit of the scattered light data with inclination and position angle constraint to the ALMA values (see discussion in section~\ref{geometric-fitting-section}). The fit was performed on the K-band data only, as the faint ring structure appears more clear in this data set. The ellipse traces this low S/N structure which likely is the scattered light counterpart of the ALMA ring. 
Due to the higher polarized intensity signal in the west in both H- and K-band we assume that the western side is the near side of the disk.\\
The H-band data show an asymmetric structure to the north, outside of the fitted ellipse, which presents as a clump and a possible arc-like structure. Both of these asymmetric structures roughly coincide with the over-brightness in the ALMA Band 6 data which is visible in figure~\ref{fig: sycha-sphere-alma}. This asymmetry in ALMA continuum flux was already noted by \cite{Pascucci2016} in their lower resolution Band 7 data. In Band 7 the disk starts to become optically thick, thus the over-brightness could be either a temperature or a density effect. That the asymmetry is still present in the optically thinner Band 6 data may suggest that it is caused by an enhancement of material rather than an increase of temperature. Scattered light is typically most sensitive to the scale height of the disk which in turn determines the $\tau=1$ height of the scattering surface. Both, a dust (and gas) over-density or an increase in temperature may increase the scale height of the disk (see the case of SR\,21 where both effects are visible in scattered light, \citealt{Muro-Arena2020}). For a deep discussion of the ALMA Band 6 data we refer to \cite{Orihara2023}. \\
Given the low S/N of the the scattered light data, we can at this point only speculate on the origin of the described features. It may be possible that the clump feature in the north signifies the position of an accreting protoplanet with its own circumplanetary disk. If this is the case, the arc structure might be a spiral driven by the embedded planet in the outer disk (see, for example, \citealt{Muto2012, Dong2016}) and the over-brightness in the ALMA data may signify a density effect caused by a vortex also driven by the same planet. The correspondence between spirals in scattered light and asymmetries in dust continuum appears to be common among well-studied disks (\citealt{Garufi2018, Marel2021}). Such a scenario may be confirmed with deeper SPHERE observations in better weather conditions, or with high spectral resolution ALMA gas line observations which may trace the local deviation from Keplerian rotation induced by an embedded planet (\citealt{Pinte2018,Pinte2020,Teague2018}). \\
As we show in table~\ref{tab: disk geometry} and figure~\ref{fig: h_over_r}, we extract in principle an aspect ratio for SY\,Cha, which is well comparable to the strong flaring case of the HD\,97048 system (however with the caveat that the uncertainty of this measurement is large). If taken at face value this typically should correspond to a bright signal in scattered light, as the disk surface intercepts more stellar light. However, as we show in figure~\ref{fig: slope-contrast} it is among the faintest detected disks in our sample. Given the uncertainties of the fit the aspect ratio might be as small as 0.03 at 79\,au. That would be more in line with the faint disks in our sample. If the aspect ratio is indeed larger, as our fit hints, then the faint outer disk may be explained by the presence of the inner disk seen in the ALMA band 6 data in figure~\ref{fig: sycha-sphere-alma}. In the same figure it is also visible that we receive the strongest scattered light signal close to the coronagraphic mask in the SPHERE H-band scattered light observation. This signal may well originate from the inner disk. That the signal is more extended in scattered light than in ALMA millimeter continuum may be explained by radial drift of the larger dust particles (\citealt{Weidenschilling1977, Villenave2019}). However, it may also be possible that we trace an optical depth effect, with the optical depth to scattering at IR wavelengths remaining larger farther out in the disk than the optical depth of thermal emission at millimeter wavelengths.\\
If the bright signal around the coronagraph indeed originates from the inner disk, then this indicates, that, unlike the outer disk ring, it intercepts a lot of stellar light. The peak is particularly strong toward the west, which we identified as the forward scattering side of the disk. If this inner disk is slightly misaligned with the faint outer disk structure that we traced in the ellipse fitting, then shadowing might explain the large brightness contrast between these two disk zones (see, for example, \citealt{Muro-Arena2020a} for a detailed model of such an effect).
Such a disk misalignment might fit well with the speculative presence of a massive planet at the disk clump location if the orbit of such a planet was slightly misaligned with respect to the outer disk and thus is able to exert a torque on the disk warping or tilting the disk (\citealt{Xiang-Gruess2013}). 
Alternatively, the inner disk could simply be strongly puffed up due to stellar radiation, which may lead to full azimuthal shadowing of the outer disk, without the need to invoke a misalignment. 
High-resolution ALMA gas line observations in the future may enable us to detect a possible warp or tilt between inner and outer disk. Alternatively near-infrared interferometric observation might be able to measure the inclination of the inner disk (\citealt{Bohn2022}). 
Additionally, deep SPHERE observations in the H or K-band might be able to secure a higher S/N detection of the outer disk ring, which in turn would enable a  firmer conclusion on the disk aspect ratio.

\subsection{SZ\,Cha}

\begin{figure}
\centering
\includegraphics[width=0.5\textwidth]{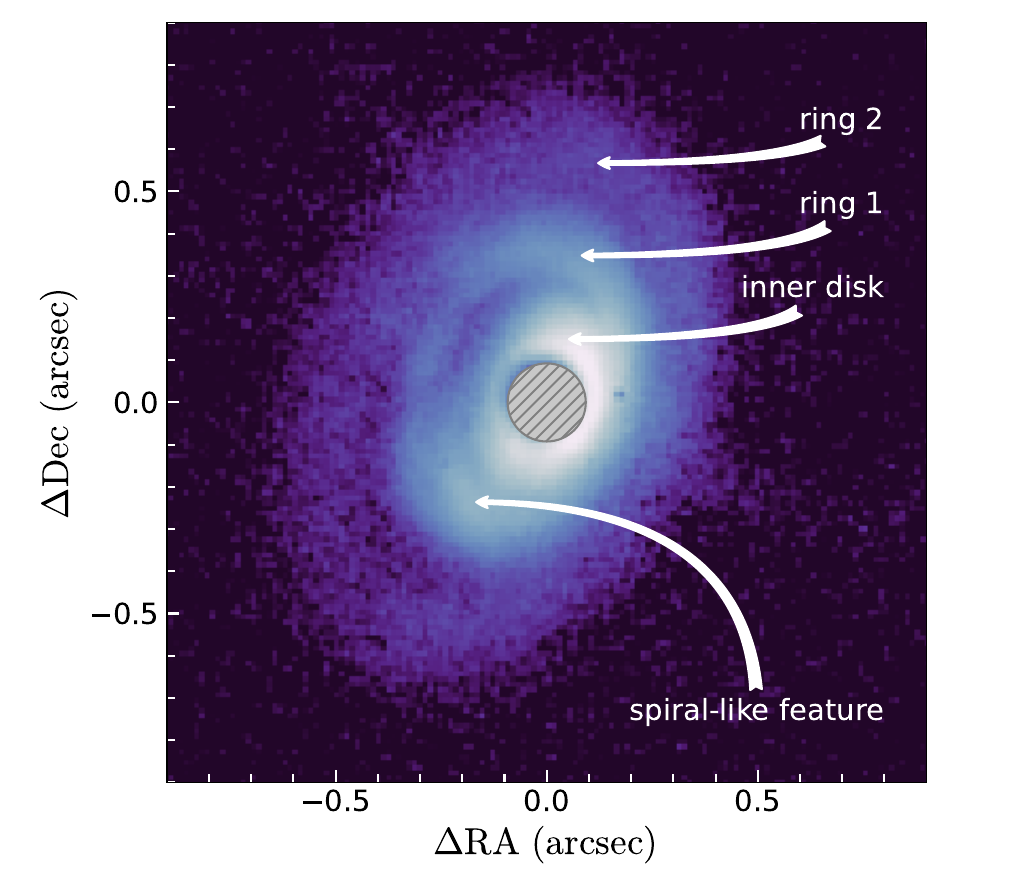}
\caption[]{Q$\phi$ image of the extended disk around the SZ\,Cha system. The color map is logarithmic to highlight the disk morphology. The hashed circle in the image center indicates the coronagraphic mask that was placed in front of the star.}
\label{fig: szcha annotated}
\end{figure}

SZ\,Cha possesses the second most extended disk of our target sample (0.86\,arcsec, i.e., 163\,au) with only HD\,97048 being more extended. The disk shows a multi-ringed substructure with two discernible rings with semi-major axes at 0.31\arcsec{} (59\,au) and 0.59\arcsec{} (112\,au). The near side of the disk is located in the west with the semi-major axis at a position angle between -12$^\circ$ and -23$^\circ$. Our fits of the ring-like features yield an intermediate disk inclination between 39$^\circ$ and 47$^\circ$. Ancillary ALMA data of the system, which will be presented in a forthcoming publication by Hagelberg et al., in prep. yield a consistent inclination of 42.1$^\circ$ and a position angle of -23$^\circ$.\\
Along the major axis there is some considerable brightness asymmetry in our scattered light image between the northern and the southern part of the disk with a possible spiral feature present in the south in between the inner disk right outside of the coronagraphic mask and partially overlapping with the first ring-like feature (see figure~\ref{fig: szcha annotated}). This geometry is reminiscent of the disk around the isolated Herbig star HD\,34282, for which \cite{2021A&A...649A..25D} suggested the presence of a possible one-armed spiral. Interestingly the aspect ratios recovered by \cite{2021A&A...649A..25D} for the ringed substructure in HD\,34282 (0.28 at 87.4\,au; 0.37 at 193.4\,au)
are well consistent with what we recover for SZ\,Cha. The similarities in morphology and height structure may indicate similar disk-shaping mechanisms in both systems. 
If we assume that the spiral-like feature in SZ\,Cha is driven by an embedded planet in the system, then visual comparison of the opening angle of the spiral arm with the model by \cite{Dong2016} suggest that the mass of the embedded planet should be low (i.e., likely $\leq3\,$M$_\mathrm{{Jup}}$) given that the opening angle of the spiral feature is small. 

\subsection{VZ\,Cha}

We find a low S/N detection of scattered light around VZ\,Cha. In figure~\ref{fig: coro_sample} it is visible that the Stokes Q and U images indeed show a butterfly pattern, with the majority of the signal located in the northeast lobe of the Stokes U image. Signal is detected out to a separation of 0.32\arcsec{} (61\,au). 
Transformation to Q$_\phi$ shows a faint disk with a position angle of 
$\sim$56$^\circ$. Ellipse fitting yields an inclination of 49.6$^\circ \pm$5.1$^\circ$ (see table~\ref{tab: disk geometry}). Due to the low S/N the fit of the disk height is highly uncertain (see figure~\ref{VZCha-ellipse}).\\
VZ\,Cha has a measured dust mass that is roughly a factor 2 higher than CV\,Cha which we detect at very high S/N (see table~\ref{tab: sample}). CT\,Cha and SY\,Cha have comparable dust masses, and both were likewise detected with higher S/N than VZ\,Cha. Thus the low polarization signal is likely not connected to low dust surface density but rather indicates that the disk around VZ\,Cha is either significantly vertically thinner than the other detected systems and thus intercepts less stellar light or is self-shadowed. It might of course well be possible that both effects are present in the system. 

\subsection{WW\,Cha}

\begin{figure}
\centering
\includegraphics[width=9cm]{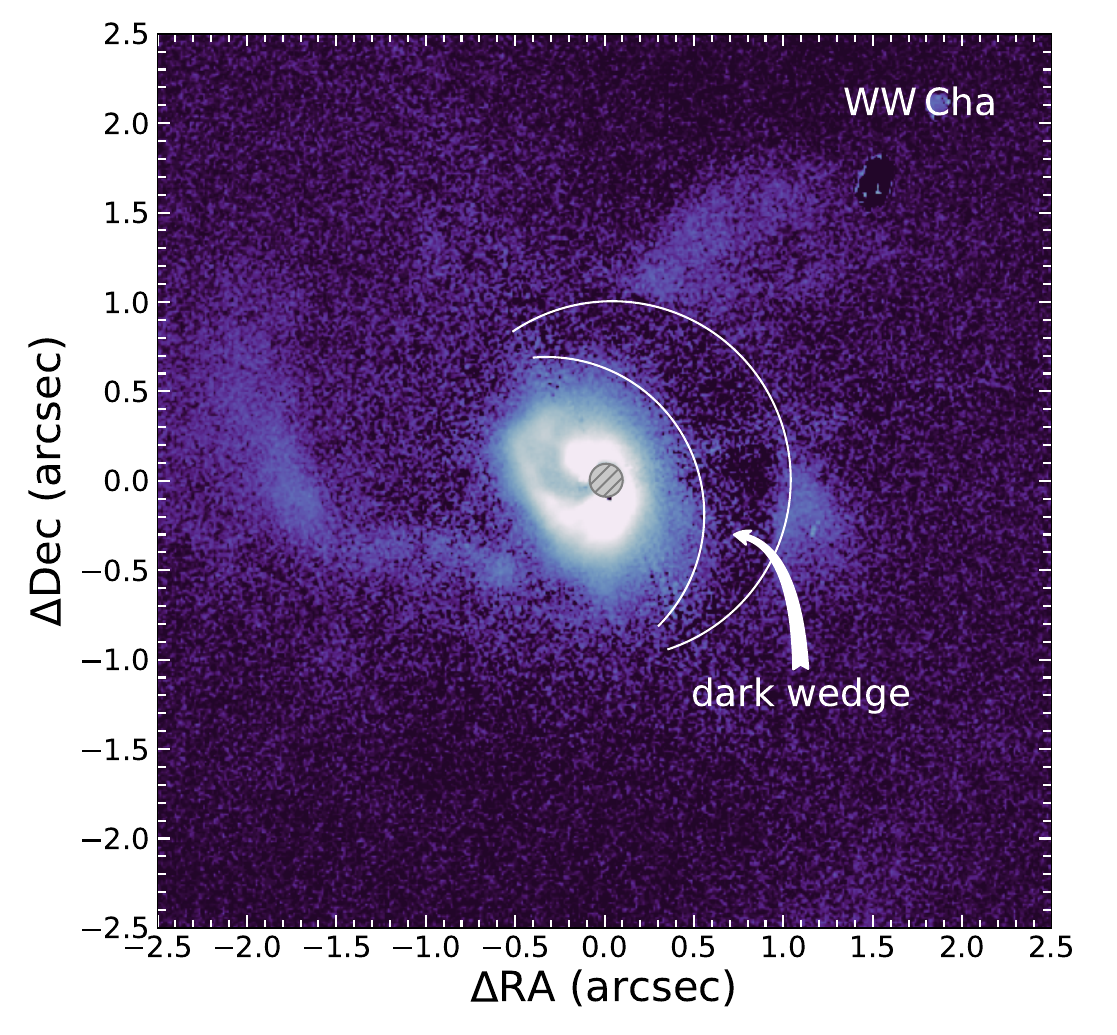}
\caption[]{Q$\phi$ image of the WW\,Cha system from H-band observations. An extended dust structure is visible in polarized light beyond the circumstellar disk. 
A dark  wedge  sharply separates the disk from the extended material in the northwest direction.}
\label{fig: ww cha feature}
\end{figure}

\begin{figure}
\centering
\includegraphics[width=9cm]{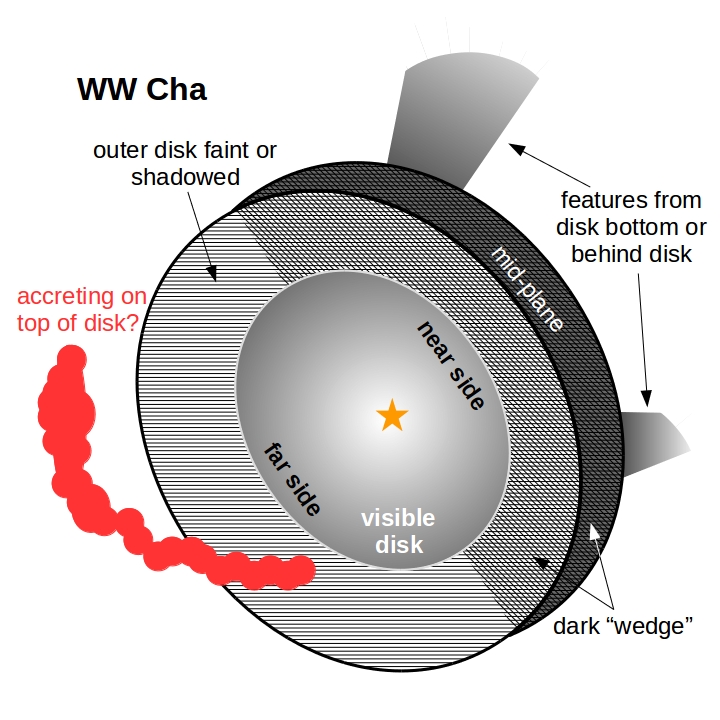}
\caption[]{Schematic of the WW\,Cha system geometry. The disk likely extends farther out in gas and small grains than   is detected in scattered light, either due to shadowing or due to the signal dropping below the noise threshold. The dark wedge in the data is caused by a combination of the unseen extended disk and the disk mid-plane.}
\label{fig: ww cha cartoon}
\end{figure}

We resolve a compact disk and extended emission around WW\,Cha using the same data set that was already presented in \cite{Garufi2020}. In particular, the disk is dominated by a bright spiral arm extending from the south of the disk and winding clockwise. The width of the spiral arm decreases smoothly from the launching point to the tip of the detected emission in the west of the disk. It is thus not trivial to measure an opening angle. \\
From the scattered light data we fit an inclination of 44.1$^\circ \pm$3.3$^\circ$ and a position angle of 29.8$^\circ \pm$4.8$^\circ$ (see table~\ref{tab: disk geometry}. High resolution (89.6 $\times$ 60\,mas beam size) ALMA band 7 data exist for the system, showing a non-axisymmetric, double-ringed disk in millimeter continuum emission, which were presented in \cite{Kanagawa2021}. They used the ALMA data to extract inclination and position angle and found 36.2$^\circ$ for the former and 34.0$^\circ$ for the latter.
The position angle is consistent with the scattered light data while the inclination is slightly smaller. This discrepancy may be caused by the prominent spiral structure which dominates the scattered light image.\\
In addition to the Keplerian disk we find a complex structure of extended material some of which was already mentioned in \cite{Garufi2020}. Most striking is the dark sickle or wedge-shaped region extending clockwise from the north to the south. We have marked this region in figure~\ref{fig: ww cha feature}.\\
Due to the asymmetry of the disk, it is not trivial to determine whether the northwest or the southeast side of the disk is the near side. However, the stellar position is clearly offset along the minor axis toward the northwest. Taking into account the height of the scattering surface this is a clear indication that the northwest is the near side of the disk and the southeast the far side (see, for example, \citealt{deBoer2016}).
If this is true, then the dark wedge can be explained by a combination of the (not illuminated) disk mid-plane and an outer area of the disk which is either self-shadowed or inherently faint. To clarify the geometry we show a sketch in figure~\ref{fig: ww cha cartoon}.\\
The wedge is likely too wide in the northwest direction to be just the disk mid-plane (see, for example, \citealt{Avenhaus2018, Garufi2020} for clear examples of the disk mid-plane). Furthermore, it does not appear that the wedge structure is converging to a point along the ansae as would be expected of the mid-plane and as shown in figure~\ref{fig: ww cha cartoon}, although we caution that there is little signal in the ansae to fully trace the wedge. However,  the sharp transition between the dark wedge and the outer structures to the northwest and west suggest a geometric origin. If we assume that the top-side of the disk around WW\,Cha extends farther than seen in scattered light, then a combination of this unseen outer part of the disk and the disk mid-plane can well explain the dark wedge region. This is displayed as the cross-hatched region in figure~\ref{fig: ww cha cartoon}. If this picture is correct, then the nebulous structures visible to the northwest and west beyond the dark wedge are either connected to the bottom side of the disk (the side facing away from the observer) or are part of the embedding cloud located behind the disk. Given that the structures are illuminated by WW\,Cha while other parts of the embedding cloud are not, it would follow that they can in any case not be arbitrarily far away from WW\,Cha. Indeed the SED of the system shown in figure~\ref{fig: sed}is more consistent with flat spectrum rather than a class II source, which suggests that WW Cha is still embedded in remaining cloud material.\\
The filamentary hook-shaped structure to the east of WW\,Cha was already noted by \cite{Garufi2020}. If the interpretation of the dark wedge that we present is correct then the filament must be placed above the disk (i.e.,  between the disk and the observer), otherwise we would not be able to trace it as close the visible disk of WW\,Cha without it being obscured. In that case the filament might trace material infalling onto the disk as was recently shown for SU\,Aur (\citealt{Ginski2021}). Based on the Herschel far-infrared data shown in figure~\ref{fig: app: pol_vetor}, WW\,Cha is located in one of the densest parts of the Cha\,I cloud, making an infall scenario likely. Deep ALMA gas emission line observations are needed to confirm the kinematics of the filament and other visible structures.

\subsection{Nondetections}

Of the 20 systems our sample seven yielded nondetections in polarized scattered light: either their disks are too small to be resolved with current instrumentation (radius $<92.5\,$mas, i.e., $\sim$18\,au at the average distance or Cha\,I), or too faint in scattered light. Of these seven systems, four are close visual binaries (DI\,Cha, PDS\,51, WX\,Cha, and WY\,Cha) with angular separations between 140\,mas (WY\,Cha) and 740\,mas (WX\,Cha). For these systems, it is plausible that the close stellar companions truncate the outer disk (see, for example, \citealt{Artymowicz1994}, \citealt{Miranda2015}), and thus they are almost certainly in the first category of nondetections (small disks). Of the remaining three systems Sz\,41 is a wide visual binary with an angular separation of $\sim$2\,arcsec ($\sim$380\,au). It also has an exceptionally low dust mass with an upper limit of less than 1\,M$_\oplus$, thus the disk in this system is likely small, but even if it were extended it would be faint with barely any dust entrained (assuming that the small $\mu$m dust population roughly correlates with the millimeter population detected at millimeter wavelength).

Finally CHX\,18N and RXJ1106.3-7721 are single stars.\footnote{No companions are reported in the literature and we do not detect any close stellar companion candidates within our SPHERE imaging data} CHX\,18N has a dust mass of 13\,M$_\oplus$, which is at the lower end for our sample. However, the even lower mass disk around Sz\,45 (10\,M$_\oplus$) yielded a scattered light detection. The SED of CHX\,18N does not show a significant dip close to 10$\mu$m, indicating that this is still a full disk without a large cavity. In comparison, Sz\,45, shows a significant dip in the SED. Thus while we can not rule out that the disk around CHX\,18N is simply compact it is possible that it is probable that it rather belongs to the second category of nondetections:  faint (self-shadowed) disks. For RXJ1106.3-7721 we do not have a measured dust mass, however, the system shows no significant mid- or far-infrared excess emission, indicating that indeed no extended disk is present.

In summary, of our nondetections we find that 5 are in all likelihood compact. One is faint and possibly self-shadowed and one might be either compact or faint. The majority of compact disks (4/5) is found (somewhat unsurprisingly) in close visual binary systems.



\section{Summary and discussion}
\label{sec: discussion}
We observed a sample of 20 systems in the nearby Chamaeleon I star-forming region with VLT/SPHERE in polarized scattered light. Our observations revealed a resolved circumstellar structure around 13 of these systems. Of these, the HD\,97048, SZ\,Cha, and the WW\,Cha systems showed clear morphological substructures, while we tentatively find shadow features in CV\,Cha and a possible dust clump and arc in the SY\,Cha system. The CR\,Cha, CS\,Cha, CT\,Cha, SZ\,45, TW\,Cha, and VZ\,Cha systems all show relatively small disks that appear  smooth at our spatial resolution and sensitivity, but that vary by a factor of $\sim$4 in polarized brightness contrast relative to the central star. Of our seven nondetections, five  are located in binary systems, four  of which are closer than 1\arcsec{}.
We find a tentative trend between the "transitional" nature of the system SED (the presence of a dip at $\sim$10 $\mu$m) and the brightness in scattered light, with the two classical transitional disks (based on their SED) around CS\,Cha and HD\,97048 being the brightest disks in our sample. We also find a possible correlation of this trend with the dust mass in the system. However, the dust masses of CR\,Cha and VZ\,Cha are similar or even slightly higher than for the bright disk around CS\,Cha. Both of these former systems are significantly fainter in scattered light, with VZ\,Cha being the faintest detected disk in our sample. This suggests that the brightness in scattered light is at least in some cases dominated by geometric shadowing effects by inner-system material rather than overall depletion. This conclusion is in line with the recent findings of \cite{Garufi2022}, who studied a sample of 15 disks that are faint in scattered light and found an anti-correlation between the amount of near-infrared excess and the brightness in near-infrared scattered light.\\
Since the scattered light brightness of the disks appears to be linked to the opening of a cavity in the inner disk region, it is interesting to study its age dependence. 
From our sample we do not recover a clear trend between system age and scattered light brightness. However, 12 of the 14 systems that are younger than 2\,Myr are all faint in scattered light. The exceptions are the WW\,Cha system, which appears to be still interacting with surrounding cloud material, and the extremely flaring SZ\,Cha system. The two bright transition disks within our sample, CS\,Cha and HD\,97048, are among the oldest systems within our sample. 
These general findings are consistent with the findings of \cite{Garufi2018}, who looked at a larger sample of scattered light disks inhomogeneously drawn from multiple star-forming regions and isolated stars. However, their sample was, by construction, biased toward known bright transition disks. 
A larger unbiased sample size is needed in order to draw a clear conclusion on the connection between disk brightness and stellar age. 
A possible correlation between scattered light brightness and system age paints an interesting picture. Young disks at the start of planet formation may still be faint in scattered light because the forming protoplanets have not yet opened a gap or cavity in the inner disk. When planetary cores have formed, the disk becomes increasingly bright in scattered light due to the lack of shadowing material in the inner disk regions. The increased illumination of the outer disk may in turn also lead   to an increase in temperature and disk flaring. A similar scenario was speculated upon in the context of mid-infrared interferometric observations by \cite{Menu2015}.
One of the oldest systems in our study, the CS\,Cha system, might be an extreme case of such an evolution, due to the circumbinary nature of the disk. We note that  known circumbinary disks in the literature indeed tend to be bright in scattered light, for example the disks around HD\,34700 (\citealt{Monnier2019}), GG\,Tau (\citealt{Krist2002,Itoh2014,Keppler2020}), and HD\,142527 (\citealt{Fukagawa2006,Avenhaus2014,Hunziker2021}).\\
Given the orbit constraints we can put on the inner stellar companion in the CS\,Cha system, it appears questionable whether the stellar binary companion alone can be responsible for opening the cavity within the disk, seen in ALMA millimeter continuum emission and also evident from the system SED. An additional planet,   proposed by \cite{Kurtovic2022} to explain the disk asymmetry seen in ALMA dust continuum, may also explain the large inner cavity. Our polarized scattered light color analysis suggests that the disk around CS\,Cha consists of small compact aggregates which may be relatively settled. This is  consistent with numerical simulations that suggest that the binary companion will excite eccentric dust particle orbits, which lead to orbit crossing and collisions with speeds above fragmentation velocity (\citealt{Meschiari2012, Paardekooper2012, Pierens2021}). In a recent study, using three-dimensional hydrodynamic simulations, \cite{Pierens2021} also found that the dust scale height near the tidally truncated cavity edge is low, especially for high solid-to-gas ratios.  
That the dust grains in the disk are already at an advanced stage of settling fits well with the very low aspect ratio of $\sim$0.03 that we find at a radius of 85\,au; although we note that this value comes with a large uncertainty (see discussion in section~\ref{sec: ellipse-fitting}). This is in turn also   consistent with the picture of a rather old disk (3-5.3\,Myr). 
The fact that we observe grain segregation in the CS\,Cha disk, with a narrow ring and large cavity in millimeter emission, and no resolved cavity in scattered light, might be explained by the circumbinary nature of this system (see, for example, \citealt{Ovelar2013}). We note that the {accretion lifetime} of CS\,Cha is among the longest in the sample, which may indicate that the disk is somewhat depleted in gas compared to the younger systems. \\
Considering the young age and bright disk in the WW\,Cha system, as well as the old age and faint disk in the HP\,Cha system, both appear as outliers in a possible age-disk brightness correlation. The WW\,Cha system seems to still be interacting with surrounding cloud material, seen in scattered light. This was already pointed out by \cite{Garufi2020}, who note that the system is located in a known network of clumpy filaments (\citealt{Haikala2005}).  The morphology, with possible streamers connected to the disk, as seen in scattered light, is somewhat similar to  the case of SU\,Aur, where infall of material was shown by \cite{Ginski2021}. Deep, high spatial resolution molecular line observations with ALMA could shed light on the nature of the interaction of WW\,Cha with the surrounding cloud.
Conversely, the HP\,Cha system is a complex triple system with signs of ongoing interaction between primary and circumsecondary disks (\citealt{Zhang2023}). 
This may in turn strongly influence the disk evolution around the primary star.
\\
Our high spatial resolution, near-infrared study of Cha I systems has shown some interesting trends that connect the scattered light appearance of the circumstellar disks with the disk evolution. It will be most interesting to compare these findings with similar studies of other nearby star-forming regions.
Companion papers to this study, focusing on the Taurus star-forming region as well as the Orion star-forming region, are currently under review (Garufi et al., submitted; Valegard et al., submitted).
Furthermore, it will be critical in the future to enable similar observations of optically fainter low-mass stars (i.e., the bulk of the stellar content of Cha I and other nearby star-forming regions). This may be enabled by the SPHERE+ initiative (\citealt{Boccaletti2020}), which aims to upgrade the SPHERE AO system to observe fainter and redder objects.  

\begin{acknowledgements}
We would like to thank an anonymous referee for an excellent report that greatly improved the clarity of this article.
      Based on observations made with ESO Telescopes at the La Silla Paranal Observatory under programme ID XX.XXXX
      This work has made use of data from the European Space Agency (ESA) mission
{\it Gaia} (\url{https://www.cosmos.esa.int/gaia}), processed by the {\it Gaia}
Data Processing and Analysis Consortium (DPAC,
\url{https://www.cosmos.esa.int/web/gaia/dpac/consortium}). Funding for the DPAC
has been provided by national institutions, in particular, the institutions
participating in the {\it Gaia} Multilateral Agreement.
SPHERE was designed and built by a consortium made of IPAG (Grenoble, France), MPIA (Heidelberg, Germany), LAM (Marseille, France), LESIA (Paris, France), Laboratoire Lagrange (Nice, France), INAF–Osservatorio di Padova (Italy), Observatoire de Genève (Switzerland), ETH Zurich (Switzerland), NOVA (Netherlands), ONERA (France) and ASTRON (Netherlands) in collaboration with ESO.  SPHERE was funded by ESO, with additional contributions from CNRS (France), MPIA (Germany), INAF (Italy), FINES (Switzerland) and NOVA (Netherlands).  
Additional funding from EC's 6th and 7th Framework Programmes as part of OPTICON was received (grant number RII3-Ct-2004-001566 for FP6 (2004–2008); 226604 for FP7 (2009–2012); 312430 for FP7 (2013–2016)).
N.K. and P.P. acknowledge the support provided by the Alexander von Humboldt Foundation in the framework of the Sofja Kovalevskaja Award endowed by the Federal Ministry of Education and Research.
CH.R. is grateful for support from the Max Planck Society and acknowledges funding by the Deutsche Forschungsgemeinschaft (DFG, German Research Foundation) - 325594231. 
This paper makes use of the following ALMA data: ADS/JAO.ALMA\#2018.1.00689.S. ALMA is a partnership of ESO (representing its member states), NSF (USA) and NINS (Japan), together with NRC (Canada), MOST and ASIAA (Taiwan), and KASI (Republic of Korea), in cooperation with the Republic of Chile. The Joint ALMA Observatory is operated by ESO, AUI/NRAO and NAOJ. A.Z. acknowledges support from ANID -- Millennium Science Initiative Program -- Center Code NCN2021\_080.
This project has received funding from the European Research Council (ERC) under the European Union's Horizon 2020 research and innovation programme (PROTOPLANETS, grant agreement No.~101002188). 
\end{acknowledgements}


\bibliographystyle{aa}
\bibliography{myBib}

\begin{thebibliography}{119}
\expandafter\ifx\csname natexlab\endcsname\relax\def\natexlab#1{#1}\fi

\bibitem[{{Allard} {et~al.}(2001){Allard}, {Hauschildt}, {Alexander}, {Tamanai}, \& {Schweitzer}}]{Allard2001}
{Allard}, F., {Hauschildt}, P.~H., {Alexander}, D.~R., {Tamanai}, A., \& {Schweitzer}, A. 2001, \apj, 556, 357

\bibitem[{{Andrews} {et~al.}(2011){Andrews}, {Wilner}, {Espaillat}, {Hughes}, {Dullemond}, {McClure}, {Qi}, \& {Brown}}]{Andrews2011}
{Andrews}, S.~M., {Wilner}, D.~J., {Espaillat}, C., {et~al.} 2011, \apj, 732, 42

\bibitem[{{Ansdell} {et~al.}(2017){Ansdell}, {Williams}, {Manara}, {Miotello}, {Facchini}, {van der Marel}, {Testi}, \& {van Dishoeck}}]{Ansdell2017}
{Ansdell}, M., {Williams}, J.~P., {Manara}, C.~F., {et~al.} 2017, \aj, 153, 240

\bibitem[{{Ansdell} {et~al.}(2016){Ansdell}, {Williams}, {van der Marel}, {Carpenter}, {Guidi}, {Hogerheijde}, {Mathews}, {Manara}, {Miotello}, {Natta}, {Oliveira}, {Tazzari}, {Testi}, {van Dishoeck}, \& {van Terwisga}}]{Ansdell2016}
{Ansdell}, M., {Williams}, J.~P., {van der Marel}, N., {et~al.} 2016, \apj, 828, 46

\bibitem[{{Artymowicz} \& {Lubow}(1994{\natexlab{a}})}]{1994ApJ...421..651A}
{Artymowicz}, P. \& {Lubow}, S.~H. 1994{\natexlab{a}}, \apj, 421, 651

\bibitem[{{Artymowicz} \& {Lubow}(1994{\natexlab{b}})}]{Artymowicz1994}
{Artymowicz}, P. \& {Lubow}, S.~H. 1994{\natexlab{b}}, \apj, 421, 651

\bibitem[{{Avenhaus} {et~al.}(2018){Avenhaus}, {Quanz}, {Garufi}, {Perez}, {Casassus}, {Pinte}, {Bertrang}, {Caceres}, {Benisty}, \& {Dominik}}]{Avenhaus2018}
{Avenhaus}, H., {Quanz}, S.~P., {Garufi}, A., {et~al.} 2018, \apj, 863, 44

\bibitem[{{Avenhaus} {et~al.}(2014){Avenhaus}, {Quanz}, {Schmid}, {Meyer}, {Garufi}, {Wolf}, \& {Dominik}}]{Avenhaus2014}
{Avenhaus}, H., {Quanz}, S.~P., {Schmid}, H.~M., {et~al.} 2014, \apj, 781, 87

\bibitem[{{Ballering} \& {Eisner}(2019)}]{Ballering2019}
{Ballering}, N.~P. \& {Eisner}, J.~A. 2019, \aj, 157, 144

\bibitem[{{Baraffe} {et~al.}(2015){Baraffe}, {Homeier}, {Allard}, \& {Chabrier}}]{Baraffe2015}
{Baraffe}, I., {Homeier}, D., {Allard}, F., \& {Chabrier}, G. 2015, \aap, 577, A42

\bibitem[{{Beckwith} \& {Sargent}(1991)}]{Beckwith1991}
{Beckwith}, S.~V.~W. \& {Sargent}, A.~I. 1991, \apj, 381, 250

\bibitem[{{Benisty} {et~al.}(2017){Benisty}, {Stolker}, {Pohl}, {de Boer}, {Lesur}, {Dominik}, {Dullemond}, {Langlois}, {Min}, {Wagner}, {Henning}, {Juhasz}, {Pinilla}, {Facchini}, {Apai}, {van Boekel}, {Garufi}, {Ginski}, {M{\'e}nard}, {Pinte}, {Quanz}, {Zurlo}, {Boccaletti}, {Bonnefoy}, {Beuzit}, {Chauvin}, {Cudel}, {Desidera}, {Feldt}, {Fontanive}, {Gratton}, {Kasper}, {Lagrange}, {LeCoroller}, {Mouillet}, {Mesa}, {Sissa}, {Vigan}, {Antichi}, {Buey}, {Fusco}, {Gisler}, {Llored}, {Magnard}, {Moeller-Nilsson}, {Pragt}, {Roelfsema}, {Sauvage}, \& {Wildi}}]{Benisty2017}
{Benisty}, M., {Stolker}, T., {Pohl}, A., {et~al.} 2017, \aap, 597, A42

\bibitem[{{Beuzit} {et~al.}(2019){Beuzit}, {Vigan}, {Mouillet}, {Dohlen}, {Gratton}, {Boccaletti}, {Sauvage}, {Schmid}, {Langlois}, {Petit}, {Baruffolo}, {Feldt}, {Milli}, {Wahhaj}, {Abe}, {Anselmi}, {Antichi}, {Barette}, {Baudrand}, {Baudoz}, {Bazzon}, {Bernardi}, {Blanchard}, {Brast}, {Bruno}, {Buey}, {Carbillet}, {Carle}, {Cascone}, {Chapron}, {Charton}, {Chauvin}, {Claudi}, {Costille}, {De Caprio}, {de Boer}, {Delboulb{\'e}}, {Desidera}, {Dominik}, {Downing}, {Dupuis}, {Fabron}, {Fantinel}, {Farisato}, {Feautrier}, {Fedrigo}, {Fusco}, {Gigan}, {Ginski}, {Girard}, {Giro}, {Gisler}, {Gluck}, {Gry}, {Henning}, {Hubin}, {Hugot}, {Incorvaia}, {Jaquet}, {Kasper}, {Lagadec}, {Lagrange}, {Le Coroller}, {Le Mignant}, {Le Ruyet}, {Lessio}, {Lizon}, {Llored}, {Lundin}, {Madec}, {Magnard}, {Marteaud}, {Martinez}, {Maurel}, {M{\'e}nard}, {Mesa}, {M{\"o}ller-Nilsson}, {Moulin}, {Moutou}, {Orign{\'e}}, {Parisot}, {Pavlov}, {Perret}, {Pragt}, {Puget}, {Rabou}, {Ramos}, {Reess}, {Rigal}, {Rochat}, {Roelfsema}, {Rousset},
  {Roux}, {Saisse}, {Salasnich}, {Santambrogio}, {Scuderi}, {Segransan}, {Sevin}, {Siebenmorgen}, {Soenke}, {Stadler}, {Suarez}, {Tiph{\`e}ne}, {Turatto}, {Udry}, {Vakili}, {Waters}, {Weber}, {Wildi}, {Zins}, \& {Zurlo}}]{Beuzit2019}
{Beuzit}, J.~L., {Vigan}, A., {Mouillet}, D., {et~al.} 2019, \aap, 631, A155

\bibitem[{{Blunt} {et~al.}(2020){Blunt}, {Wang}, {Angelo}, {Ngo}, {Cody}, {De Rosa}, {Graham}, {Hirsch}, {Nagpal}, {Nielsen}, {Pearce}, {Rice}, \& {Tejada}}]{Blunt2020}
{Blunt}, S., {Wang}, J.~J., {Angelo}, I., {et~al.} 2020, \aj, 159, 89

\bibitem[{{Boccaletti} {et~al.}(2020){Boccaletti}, {Chauvin}, {Mouillet}, {Absil}, {Allard}, {Antoniucci}, {Augereau}, {Barge}, {Baruffolo}, {Baudino}, {Baudoz}, {Beaulieu}, {Benisty}, {Beuzit}, {Bianco}, {Biller}, {Bonavita}, {Bonnefoy}, {Bos}, {Bouret}, {Brandner}, {Buchschache}, {Carry}, {Cantalloube}, {Cascone}, {Carlotti}, {Charnay}, {Chiavassa}, {Choquet}, {Clenet}, {Crida}, {De Boer}, {De Caprio}, {Desidera}, {Desert}, {Delisle}, {Delorme}, {Dohlen}, {Doelman}, {Dominik}, {Orazi}, {Dougados}, {Doute}, {Fedele}, {Feldt}, {Ferreira}, {Fontanive}, {Fusco}, {Galicher}, {Garufi}, {Gendron}, {Ghedina}, {Ginski}, {Gonzalez}, {Gratadour}, {Gratton}, {Guillot}, {Haffert}, {Hagelberg}, {Henning}, {Huby}, {Janson}, {Kamp}, {Keller}, {Kenworthy}, {Kervella}, {Kral}, {Kuhn}, {Lagadec}, {Laibe}, {Langlois}, {Lagrange}, {Launhardt}, {Leboulleux}, {Le Coroller}, {Li Causi}, {Loupias}, {Maire}, {Marleau}, {Martinache}, {Martinez}, {Mary}, {Mattioli}, {Mazoyer}, {Meheut}, {Menard}, {Mesa}, {Meunier}, {Miguel}, {Milli},
  {Min}, {Molliere}, {Mordasini}, {Moretto}, {Mugnier}, {Muro Arena}, {Nardetto}, {Diaye}, {Nesvadba}, {Pedichini}, {Pinilla}, {Por}, {Potier}, {Quanz}, {Rameau}, {Roelfsema}, {Rouan}, {Rigliaco}, {Salasnich}, {Samland}, {Sauvage}, {Schmid}, {Segransan}, {Snellen}, {Snik}, {Soulez}, {Stadler}, {Stam}, {Tallon}, {Thebault}, {Thiebaut}, {Tschudi}, {Udry}, {van Holstein}, {Vernazza}, {Vidal}, {Vigan}, {Waters}, {Wildi}, {Willson}, {Zanutta}, {Zavagno}, \& {Zurlo}}]{Boccaletti2020}
{Boccaletti}, A., {Chauvin}, G., {Mouillet}, D., {et~al.} 2020, arXiv e-prints, arXiv:2003.05714

\bibitem[{{Bohn} {et~al.}(2022){Bohn}, {Benisty}, {Perraut}, {van der Marel}, {W{\"o}lfer}, {van Dishoeck}, {Facchini}, {Manara}, {Teague}, {Francis}, {Berger}, {Garcia-Lopez}, {Ginski}, {Henning}, {Kenworthy}, {Kraus}, {M{\'e}nard}, {M{\'e}rand}, \& {P{\'e}rez}}]{Bohn2022}
{Bohn}, A.~J., {Benisty}, M., {Perraut}, K., {et~al.} 2022, \aap, 658, A183

\bibitem[{{Bohn} {et~al.}(2020){Bohn}, {Southworth}, {Ginski}, {Kenworthy}, {Maxted}, \& {Evans}}]{2020A&A...635A..73B}
{Bohn}, A.~J., {Southworth}, J., {Ginski}, C., {et~al.} 2020, \aap, 635, A73

\bibitem[{{Bressan} {et~al.}(2012){Bressan}, {Marigo}, {Girardi}, {Salasnich}, {Dal Cero}, {Rubele}, \& {Nanni}}]{Bressan2012}
{Bressan}, A., {Marigo}, P., {Girardi}, L., {et~al.} 2012, \mnras, 427, 127

\bibitem[{{Canovas} {et~al.}(2011){Canovas}, {Rodenhuis}, {Jeffers}, {Min}, \& {Keller}}]{2011A&A...531A.102C}
{Canovas}, H., {Rodenhuis}, M., {Jeffers}, S.~V., {Min}, M., \& {Keller}, C.~U. 2011, \aap, 531, A102

\bibitem[{{Carbillet} {et~al.}(2011){Carbillet}, {Bendjoya}, {Abe}, {Guerri}, {Boccaletti}, {Daban}, {Dohlen}, {Ferrari}, {Robbe-Dubois}, {Douet}, \& {Vakili}}]{2011ExA....30...39C}
{Carbillet}, M., {Bendjoya}, P., {Abe}, L., {et~al.} 2011, Experimental Astronomy, 30, 39

\bibitem[{{Choi} {et~al.}(2016){Choi}, {Dotter}, {Conroy}, {Cantiello}, {Paxton}, \& {Johnson}}]{Choi2016}
{Choi}, J., {Dotter}, A., {Conroy}, C., {et~al.} 2016, \apj, 823, 102

\bibitem[{{Cieza} {et~al.}(2019){Cieza}, {Ru{\'\i}z-Rodr{\'\i}guez}, {Hales}, {Casassus}, {P{\'e}rez}, {Gonzalez-Ruilova}, {C{\'a}novas}, {Williams}, {Zurlo}, {Ansdell}, {Avenhaus}, {Bayo}, {Bertrang}, {Christiaens}, {Dent}, {Ferrero}, {Gamen}, {Olofsson}, {Orcajo}, {Pe{\~n}a Ram{\'\i}rez}, {Principe}, {Schreiber}, \& {van der Plas}}]{Cieza2019}
{Cieza}, L.~A., {Ru{\'\i}z-Rodr{\'\i}guez}, D., {Hales}, A., {et~al.} 2019, \mnras, 482, 698

\bibitem[{{Covino} {et~al.}(1997){Covino}, {Palazzi}, {Penprase}, {Schwarz}, \& {Terranegra}}]{Covino1997}
{Covino}, E., {Palazzi}, E., {Penprase}, B.~E., {Schwarz}, H.~E., \& {Terranegra}, L. 1997, \aaps, 122, 95

\bibitem[{{Dawson} {et~al.}(2011){Dawson}, {Murray-Clay}, \& {Fabrycky}}]{Dawson2011}
{Dawson}, R.~I., {Murray-Clay}, R.~A., \& {Fabrycky}, D.~C. 2011, \apjl, 743, L17

\bibitem[{{de Boer} {et~al.}(2021){de Boer}, {Ginski}, {Chauvin}, {M{\'e}nard}, {Benisty}, {Dominik}, {Maaskant}, {Girard}, {van der Plas}, {Garufi}, {Perrot}, {Stolker}, {Avenhaus}, {Bohn}, {Delboulb{\'e}}, {Jaquet}, {Buey}, {M{\"o}ller-Nilsson}, {Pragt}, \& {Fusco}}]{2021A&A...649A..25D}
{de Boer}, J., {Ginski}, C., {Chauvin}, G., {et~al.} 2021, \aap, 649, A25

\bibitem[{{de Boer} {et~al.}(2020){de Boer}, {Langlois}, {van Holstein}, {Girard}, {Mouillet}, {Vigan}, {Dohlen}, {Snik}, {Keller}, {Ginski}, {Stam}, {Milli}, {Wahhaj}, {Kasper}, {Schmid}, {Rabou}, {Gluck}, {Hugot}, {Perret}, {Martinez}, {Weber}, {Pragt}, {Sauvage}, {Boccaletti}, {Le Coroller}, {Dominik}, {Henning}, {Lagadec}, {M{\'e}nard}, {Turatto}, {Udry}, {Chauvin}, {Feldt}, \& {Beuzit}}]{2020A&A...633A..63D}
{de Boer}, J., {Langlois}, M., {van Holstein}, R.~G., {et~al.} 2020, \aap, 633, A63

\bibitem[{{de Boer} {et~al.}(2016){de Boer}, {Salter}, {Benisty}, {Vigan}, {Boccaletti}, {Pinilla}, {Ginski}, {Juhasz}, {Maire}, {Messina}, {Desidera}, {Cheetham}, {Girard}, {Wahhaj}, {Langlois}, {Bonnefoy}, {Beuzit}, {Buenzli}, {Chauvin}, {Dominik}, {Feldt}, {Gratton}, {Hagelberg}, {Isella}, {Janson}, {Keller}, {Lagrange}, {Lannier}, {Menard}, {Mesa}, {Mouillet}, {Mugrauer}, {Peretti}, {Perrot}, {Sissa}, {Snik}, {Vogt}, {Zurlo}, \& {SPHERE Consortium}}]{deBoer2016}
{de Boer}, J., {Salter}, G., {Benisty}, M., {et~al.} 2016, \aap, 595, A114

\bibitem[{{de Juan Ovelar} {et~al.}(2013){de Juan Ovelar}, {Min}, {Dominik}, {Thalmann}, {Pinilla}, {Benisty}, \& {Birnstiel}}]{Ovelar2013}
{de Juan Ovelar}, M., {Min}, M., {Dominik}, C., {et~al.} 2013, \aap, 560, A111

\bibitem[{{Doering} {et~al.}(2007){Doering}, {Meixner}, {Holfeltz}, {Krist}, {Ardila}, {Kamp}, {Clampin}, \& {Lubow}}]{Doering2007}
{Doering}, R.~L., {Meixner}, M., {Holfeltz}, S.~T., {et~al.} 2007, \aj, 133, 2122

\bibitem[{{Dohlen} {et~al.}(2008){Dohlen}, {Langlois}, {Saisse}, {Hill}, {Origne}, {Jacquet}, {Fabron}, {Blanc}, {Llored}, {Carle}, {Moutou}, {Vigan}, {Boccaletti}, {Carbillet}, {Mouillet}, \& {Beuzit}}]{2008SPIE.7014E..3LD}
{Dohlen}, K., {Langlois}, M., {Saisse}, M., {et~al.} 2008, in \procspie, Vol. 7014, Ground-based and Airborne Instrumentation for Astronomy II, 70143L

\bibitem[{{Dominik} {et~al.}(2021){Dominik}, {Min}, \& {Tazaki}}]{Dominik2021}
{Dominik}, C., {Min}, M., \& {Tazaki}, R. 2021, {OpTool: Command-line driven tool for creating complex dust opacities}

\bibitem[{{Dong} {et~al.}(2016){Dong}, {Fung}, \& {Chiang}}]{Dong2016}
{Dong}, R., {Fung}, J., \& {Chiang}, E. 2016, \apj, 826, 75

\bibitem[{{Dorschner} {et~al.}(1995){Dorschner}, {Begemann}, {Henning}, {Jaeger}, \& {Mutschke}}]{Dorschner1995}
{Dorschner}, J., {Begemann}, B., {Henning}, T., {Jaeger}, C., \& {Mutschke}, H. 1995, \aap, 300, 503

\bibitem[{{Dullemond} \& {Dominik}(2004)}]{Dullemond2004}
{Dullemond}, C.~P. \& {Dominik}, C. 2004, \aap, 417, 159

\bibitem[{{Dullemond} {et~al.}(2012){Dullemond}, {Juhasz}, {Pohl}, {Sereshti}, {Shetty}, {Peters}, {Commercon}, \& {Flock}}]{2012ascl.soft02015D}
{Dullemond}, C.~P., {Juhasz}, A., {Pohl}, A., {et~al.} 2012, {RADMC-3D: A multi-purpose radiative transfer tool}

\bibitem[{{Engler} {et~al.}(2017){Engler}, {Schmid}, {Thalmann}, {Boccaletti}, {Bazzon}, {Baruffolo}, {Beuzit}, {Claudi}, {Costille}, {Desidera}, {Dohlen}, {Dominik}, {Feldt}, {Fusco}, {Ginski}, {Gisler}, {Girard}, {Gratton}, {Henning}, {Hubin}, {Janson}, {Kasper}, {Kral}, {Langlois}, {Lagadec}, {M{\'e}nard}, {Meyer}, {Milli}, {Mouillet}, {Olofsson}, {Pavlov}, {Pragt}, {Puget}, {Quanz}, {Roelfsema}, {Salasnich}, {Siebenmorgen}, {Sissa}, {Suarez}, {Szulagyi}, {Turatto}, {Udry}, \& {Wildi}}]{Engler2017}
{Engler}, N., {Schmid}, H.~M., {Thalmann}, C., {et~al.} 2017, \aap, 607, A90

\bibitem[{{Foreman-Mackey} {et~al.}(2013){Foreman-Mackey}, {Hogg}, {Lang}, \& {Goodman}}]{Foreman-Mackey2013}
{Foreman-Mackey}, D., {Hogg}, D.~W., {Lang}, D., \& {Goodman}, J. 2013, \pasp, 125, 306

\bibitem[{{Francis} \& {van der Marel}(2020)}]{2020ApJ...892..111F}
{Francis}, L. \& {van der Marel}, N. 2020, \apj, 892, 111

\bibitem[{{Fukagawa} {et~al.}(2006){Fukagawa}, {Tamura}, {Itoh}, {Kudo}, {Imaeda}, {Oasa}, {Hayashi}, \& {Hayashi}}]{Fukagawa2006}
{Fukagawa}, M., {Tamura}, M., {Itoh}, Y., {et~al.} 2006, \apjl, 636, L153

\bibitem[{{Gaia Collaboration} {et~al.}(2018){Gaia Collaboration}, {Brown}, {Vallenari}, {Prusti}, {de Bruijne}, {Babusiaux}, {Bailer-Jones}, {Biermann}, {Evans}, {Eyer}, {Jansen}, {Jordi}, {Klioner}, {Lammers}, {Lindegren}, {Luri}, {Mignard}, {Panem}, {Pourbaix}, {Randich}, {Sartoretti}, {Siddiqui}, {Soubiran}, {van Leeuwen}, {Walton}, {Arenou}, {Bastian}, {Cropper}, {Drimmel}, {Katz}, {Lattanzi}, {Bakker}, {Cacciari}, {Casta{\~n}eda}, {Chaoul}, {Cheek}, {De Angeli}, {Fabricius}, {Guerra}, {Holl}, {Masana}, {Messineo}, {Mowlavi}, {Nienartowicz}, {Panuzzo}, {Portell}, {Riello}, {Seabroke}, {Tanga}, {Th{\'e}venin}, {Gracia-Abril}, {Comoretto}, {Garcia-Reinaldos}, {Teyssier}, {Altmann}, {Andrae}, {Audard}, {Bellas-Velidis}, {Benson}, {Berthier}, {Blomme}, {Burgess}, {Busso}, {Carry}, {Cellino}, {Clementini}, {Clotet}, {Creevey}, {Davidson}, {De Ridder}, {Delchambre}, {Dell'Oro}, {Ducourant}, {Fern{\'a}ndez-Hern{\'a}ndez}, {Fouesneau}, {Fr{\'e}mat}, {Galluccio}, {Garc{\'\i}a-Torres},
  {Gonz{\'a}lez-N{\'u}{\~n}ez}, {Gonz{\'a}lez-Vidal}, {Gosset}, {Guy}, {Halbwachs}, {Hambly}, {Harrison}, {Hern{\'a}ndez}, {Hestroffer}, {Hodgkin}, {Hutton}, {Jasniewicz}, {Jean-Antoine-Piccolo}, {Jordan}, {Korn}, {Krone-Martins}, {Lanzafame}, {Lebzelter}, {L{\"o}ffler}, {Manteiga}, {Marrese}, {Mart{\'\i}n-Fleitas}, {Moitinho}, {Mora}, {Muinonen}, {Osinde}, {Pancino}, {Pauwels}, {Petit}, {Recio-Blanco}, {Richards}, {Rimoldini}, {Robin}, {Sarro}, {Siopis}, {Smith}, {Sozzetti}, {S{\"u}veges}, {Torra}, {van Reeven}, {Abbas}, {Abreu Aramburu}, {Accart}, {Aerts}, {Altavilla}, {{\'A}lvarez}, {Alvarez}, {Alves}, {Anderson}, {Andrei}, {Anglada Varela}, {Antiche}, {Antoja}, {Arcay}, {Astraatmadja}, {Bach}, {Baker}, {Balaguer-N{\'u}{\~n}ez}, {Balm}, {Barache}, {Barata}, {Barbato}, {Barblan}, {Barklem}, {Barrado}, {Barros}, {Barstow}, {Bartholom{\'e} Mu{\~n}oz}, {Bassilana}, {Becciani}, {Bellazzini}, {Berihuete}, {Bertone}, {Bianchi}, {Bienaym{\'e}}, {Blanco-Cuaresma}, {Boch}, {Boeche}, {Bombrun}, {Borrachero},
  {Bossini}, {Bouquillon}, {Bourda}, {Bragaglia}, {Bramante}, {Breddels}, {Bressan}, {Brouillet}, {Br{\"u}semeister}, {Brugaletta}, {Bucciarelli}, {Burlacu}, {Busonero}, {Butkevich}, {Buzzi}, {Caffau}, {Cancelliere}, {Cannizzaro}, {Cantat-Gaudin}, {Carballo}, {Carlucci}, {Carrasco}, {Casamiquela}, {Castellani}, {Castro-Ginard}, {Charlot}, {Chemin}, {Chiavassa}, {Cocozza}, {Costigan}, {Cowell}, {Crifo}, {Crosta}, {Crowley}, {Cuypers}, {Dafonte}, {Damerdji}, {Dapergolas}, {David}, {David}, {de Laverny}, {De Luise}, {De March}, {de Martino}, {de Souza}, {de Torres}, {Debosscher}, {del Pozo}, {Delbo}, {Delgado}, {Delgado}, {Di Matteo}, {Diakite}, {Diener}, {Distefano}, {Dolding}, {Drazinos}, {Dur{\'a}n}, {Edvardsson}, {Enke}, {Eriksson}, {Esquej}, {Eynard Bontemps}, {Fabre}, {Fabrizio}, {Faigler}, {Falc{\~a}o}, {Farr{\`a}s Casas}, {Federici}, {Fedorets}, {Fernique}, {Figueras}, {Filippi}, {Findeisen}, {Fonti}, {Fraile}, {Fraser}, {Fr{\'e}zouls}, {Gai}, {Galleti}, {Garabato}, {Garc{\'\i}a-Sedano}, {Garofalo},
  {Garralda}, {Gavel}, {Gavras}, {Gerssen}, {Geyer}, {Giacobbe}, {Gilmore}, {Girona}, {Giuffrida}, {Glass}, {Gomes}, {Granvik}, {Gueguen}, {Guerrier}, {Guiraud}, {Guti{\'e}rrez-S{\'a}nchez}, {Haigron}, {Hatzidimitriou}, {Hauser}, {Haywood}, {Heiter}, {Helmi}, {Heu}, {Hilger}, {Hobbs}, {Hofmann}, {Holland}, {Huckle}, {Hypki}, {Icardi}, {Jan{\ss}en}, {Jevardat de Fombelle}, {Jonker}, {Juh{\'a}sz}, {Julbe}, {Karampelas}, {Kewley}, {Klar}, {Kochoska}, {Kohley}, {Kolenberg}, {Kontizas}, {Kontizas}, {Koposov}, {Kordopatis}, {Kostrzewa-Rutkowska}, {Koubsky}, {Lambert}, {Lanza}, {Lasne}, {Lavigne}, {Le Fustec}, {Le Poncin-Lafitte}, {Lebreton}, {Leccia}, {Leclerc}, {Lecoeur-Taibi}, {Lenhardt}, {Leroux}, {Liao}, {Licata}, {Lindstr{\o}m}, {Lister}, {Livanou}, {Lobel}, {L{\'o}pez}, {Managau}, {Mann}, {Mantelet}, {Marchal}, {Marchant}, {Marconi}, {Marinoni}, {Marschalk{\'o}}, {Marshall}, {Martino}, {Marton}, {Mary}, {Massari}, {Matijevi{\v{c}}}, {Mazeh}, {McMillan}, {Messina}, {Michalik}, {Millar}, {Molina}, {Molinaro},
  {Moln{\'a}r}, {Montegriffo}, {Mor}, {Morbidelli}, {Morel}, {Morris}, {Mulone}, {Muraveva}, {Musella}, {Nelemans}, {Nicastro}, {Noval}, {O'Mullane}, {Ord{\'e}novic}, {Ord{\'o}{\~n}ez-Blanco}, {Osborne}, {Pagani}, {Pagano}, {Pailler}, {Palacin}, {Palaversa}, {Panahi}, {Pawlak}, {Piersimoni}, {Pineau}, {Plachy}, {Plum}, {Poggio}, {Poujoulet}, {Pr{\v{s}}a}, {Pulone}, {Racero}, {Ragaini}, {Rambaux}, {Ramos-Lerate}, {Regibo}, {Reyl{\'e}}, {Riclet}, {Ripepi}, {Riva}, {Rivard}, {Rixon}, {Roegiers}, {Roelens}, {Romero-G{\'o}mez}, {Rowell}, {Royer}, {Ruiz-Dern}, {Sadowski}, {Sagrist{\`a} Sell{\'e}s}, {Sahlmann}, {Salgado}, {Salguero}, {Sanna}, {Santana-Ros}, {Sarasso}, {Savietto}, {Schultheis}, {Sciacca}, {Segol}, {Segovia}, {S{\'e}gransan}, {Shih}, {Siltala}, {Silva}, {Smart}, {Smith}, {Solano}, {Solitro}, {Sordo}, {Soria Nieto}, {Souchay}, {Spagna}, {Spoto}, {Stampa}, {Steele}, {Steidelm{\"u}ller}, {Stephenson}, {Stoev}, {Suess}, {Surdej}, {Szabados}, {Szegedi-Elek}, {Tapiador}, {Taris}, {Tauran}, {Taylor},
  {Teixeira}, {Terrett}, {Teyssandier}, {Thuillot}, {Titarenko}, {Torra Clotet}, {Turon}, {Ulla}, {Utrilla}, {Uzzi}, {Vaillant}, {Valentini}, {Valette}, {van Elteren}, {Van Hemelryck}, {van Leeuwen}, {Vaschetto}, {Vecchiato}, {Veljanoski}, {Viala}, {Vicente}, {Vogt}, {von Essen}, {Voss}, {Votruba}, {Voutsinas}, {Walmsley}, {Weiler}, {Wertz}, {Wevers}, {Wyrzykowski}, {Yoldas}, {{\v{Z}}erjal}, {Ziaeepour}, {Zorec}, {Zschocke}, {Zucker}, {Zurbach}, \& {Zwitter}}]{2018A&A...616A...1G}
{Gaia Collaboration}, {Brown}, A.~G.~A., {Vallenari}, A., {et~al.} 2018, \aap, 616, A1

\bibitem[{{Gaia Collaboration} {et~al.}(2022){Gaia Collaboration}, {Vallenari}, {Brown}, {Prusti}, {de Bruijne}, {Arenou}, {Babusiaux}, {Biermann}, {Creevey}, {Ducourant}, {Evans}, {Eyer}, {Guerra}, {Hutton}, {Jordi}, {Klioner}, {Lammers}, {Lindegren}, {Luri}, {Mignard}, {Panem}, {Pourbaix}, {Randich}, {Sartoretti}, {Soubiran}, {Tanga}, {Walton}, {Bailer-Jones}, {Bastian}, {Drimmel}, {Jansen}, {Katz}, {Lattanzi}, {van Leeuwen}, {Bakker}, {Cacciari}, {Casta{\~n}eda}, {De Angeli}, {Fabricius}, {Fouesneau}, {Fr{\'e}mat}, {Galluccio}, {Guerrier}, {Heiter}, {Masana}, {Messineo}, {Mowlavi}, {Nicolas}, {Nienartowicz}, {Pailler}, {Panuzzo}, {Riclet}, {Roux}, {Seabroke}, {Sordo{\o}rcit}, {Th{\'e}venin}, {Gracia-Abril}, {Portell}, {Teyssier}, {Altmann}, {Andrae}, {Audard}, {Bellas-Velidis}, {Benson}, {Berthier}, {Blomme}, {Burgess}, {Busonero}, {Busso}, {C{\'a}novas}, {Carry}, {Cellino}, {Cheek}, {Clementini}, {Damerdji}, {Davidson}, {de Teodoro}, {Nu{\~n}ez Campos}, {Delchambre}, {Dell'Oro}, {Esquej},
  {Fern{\'a}ndez-Hern{\'a}ndez}, {Fraile}, {Garabato}, {Garc{\'\i}a-Lario}, {Gosset}, {Haigron}, {Halbwachs}, {Hambly}, {Harrison}, {Hern{\'a}ndez}, {Hestroffer}, {Hodgkin}, {Holl}, {Jan{\ss}en}, {Jevardat de Fombelle}, {Jordan}, {Krone-Martins}, {Lanzafame}, {L{\"o}ffler}, {Marchal}, {Marrese}, {Moitinho}, {Muinonen}, {Osborne}, {Pancino}, {Pauwels}, {Recio-Blanco}, {Reyl{\'e}}, {Riello}, {Rimoldini}, {Roegiers}, {Rybizki}, {Sarro}, {Siopis}, {Smith}, {Sozzetti}, {Utrilla}, {van Leeuwen}, {Abbas}, {{\'A}brah{\'a}m}, {Abreu Aramburu}, {Aerts}, {Aguado}, {Ajaj}, {Aldea-Montero}, {Altavilla}, {{\'A}lvarez}, {Alves}, {Anders}, {Anderson}, {Anglada Varela}, {Antoja}, {Baines}, {Baker}, {Balaguer-N{\'u}{\~n}ez}, {Balbinot}, {Balog}, {Barache}, {Barbato}, {Barros}, {Barstow}, {Bartolom{\'e}}, {Bassilana}, {Bauchet}, {Becciani}, {Bellazzini}, {Berihuete}, {Bernet}, {Bertone}, {Bianchi}, {Binnenfeld}, {Blanco-Cuaresma}, {Blazere}, {Boch}, {Bombrun}, {Bossini}, {Bouquillon}, {Bragaglia}, {Bramante}, {Breedt},
  {Bressan}, {Brouillet}, {Brugaletta}, {Bucciarelli}, {Burlacu}, {Butkevich}, {Buzzi}, {Caffau}, {Cancelliere}, {Cantat-Gaudin}, {Carballo}, {Carlucci}, {Carnerero}, {Carrasco}, {Casamiquela}, {Castellani}, {Castro-Ginard}, {Chaoul}, {Charlot}, {Chemin}, {Chiaramida}, {Chiavassa}, {Chornay}, {Comoretto}, {Contursi}, {Cooper}, {Cornez}, {Cowell}, {Crifo}, {Cropper}, {Crosta}, {Crowley}, {Dafonte}, {Dapergolas}, {David}, {David}, {de Laverny}, {De Luise}, {De March}, {De Ridder}, {de Souza}, {de Torres}, {del Peloso}, {del Pozo}, {Delbo}, {Delgado}, {Delisle}, {Demouchy}, {Dharmawardena}, {Di Matteo}, {Diakite}, {Diener}, {Distefano}, {Dolding}, {Edvardsson}, {Enke}, {Fabre}, {Fabrizio}, {Faigler}, {Fedorets}, {Fernique}, {Fienga}, {Figueras}, {Fournier}, {Fouron}, {Fragkoudi}, {Gai}, {Garcia-Gutierrez}, {Garcia-Reinaldos}, {Garc{\'\i}a-Torres}, {Garofalo}, {Gavel}, {Gavras}, {Gerlach}, {Geyer}, {Giacobbe}, {Gilmore}, {Girona}, {Giuffrida}, {Gomel}, {Gomez}, {Gonz{\'a}lez-N{\'u}{\~n}ez},
  {Gonz{\'a}lez-Santamar{\'\i}a}, {Gonz{\'a}lez-Vidal}, {Granvik}, {Guillout}, {Guiraud}, {Guti{\'e}rrez-S{\'a}nchez}, {Guy}, {Hatzidimitriou}, {Hauser}, {Haywood}, {Helmer}, {Helmi}, {Sarmiento}, {Hidalgo}, {Hilger}, {H{\l}adczuk}, {Hobbs}, {Holland}, {Huckle}, {Jardine}, {Jasniewicz}, {Jean-Antoine Piccolo}, {Jim{\'e}nez-Arranz}, {Jorissen}, {Juaristi Campillo}, {Julbe}, {Karbevska}, {Kervella}, {Khanna}, {Kontizas}, {Kordopatis}, {Korn}, {K{\'o}sp{\'a}l}, {Kostrzewa-Rutkowska}, {Kruszy{\'n}ska}, {Kun}, {Laizeau}, {Lambert}, {Lanza}, {Lasne}, {Le Campion}, {Lebreton}, {Lebzelter}, {Leccia}, {Leclerc}, {Lecoeur-Taibi}, {Liao}, {Licata}, {Lindstr{\o}m}, {Lister}, {Livanou}, {Lobel}, {Lorca}, {Loup}, {Madrero Pardo}, {Magdaleno Romeo}, {Managau}, {Mann}, {Manteiga}, {Marchant}, {Marconi}, {Marcos}, {Marcos Santos}, {Mar{\'\i}n Pina}, {Marinoni}, {Marocco}, {Marshall}, {Polo}, {Mart{\'\i}n-Fleitas}, {Marton}, {Mary}, {Masip}, {Massari}, {Mastrobuono-Battisti}, {Mazeh}, {McMillan}, {Messina}, {Michalik},
  {Millar}, {Mints}, {Molina}, {Molinaro}, {Moln{\'a}r}, {Monari}, {Mongui{\'o}}, {Montegriffo}, {Montero}, {Mor}, {Mora}, {Morbidelli}, {Morel}, {Morris}, {Muraveva}, {Murphy}, {Musella}, {Nagy}, {Noval}, {Oca{\~n}a}, {Ogden}, {Ordenovic}, {Osinde}, {Pagani}, {Pagano}, {Palaversa}, {Palicio}, {Pallas-Quintela}, {Panahi}, {Payne-Wardenaar}, {Pe{\~n}alosa Esteller}, {Penttil{\"a}}, {Pichon}, {Piersimoni}, {Pineau}, {Plachy}, {Plum}, {Poggio}, {Pr{\v{s}}a}, {Pulone}, {Racero}, {Ragaini}, {Rainer}, {Raiteri}, {Rambaux}, {Ramos}, {Ramos-Lerate}, {Re Fiorentin}, {Regibo}, {Richards}, {Rios Diaz}, {Ripepi}, {Riva}, {Rix}, {Rixon}, {Robichon}, {Robin}, {Robin}, {Roelens}, {Rogues}, {Rohrbasser}, {Romero-G{\'o}mez}, {Rowell}, {Royer}, {Ruz Mieres}, {Rybicki}, {Sadowski}, {S{\'a}ez N{\'u}{\~n}ez}, {Sagrist{\`a} Sell{\'e}s}, {Sahlmann}, {Salguero}, {Samaras}, {Sanchez Gimenez}, {Sanna}, {Santove{\~n}a}, {Sarasso}, {Schultheis}, {Sciacca}, {Segol}, {Segovia}, {S{\'e}gransan}, {Semeux}, {Shahaf}, {Siddiqui}, {Siebert},
  {Siltala}, {Silvelo}, {Slezak}, {Slezak}, {Smart}, {Snaith}, {Solano}, {Solitro}, {Souami}, {Souchay}, {Spagna}, {Spina}, {Spoto}, {Steele}, {Steidelm{\"u}ller}, {Stephenson}, {S{\"u}veges}, {Surdej}, {Szabados}, {Szegedi-Elek}, {Taris}, {Taylo}, {Teixeira}, {Tolomei}, {Tonello}, {Torra}, {Torra}, {Torralba Elipe}, {Trabucchi}, {Tsounis}, {Turon}, {Ulla}, {Unger}, {Vaillant}, {van Dillen}, {van Reeven}, {Vanel}, {Vecchiato}, {Viala}, {Vicente}, {Voutsinas}, {Weiler}, {Wevers}, {Wyrzykowski}, {Yoldas}, {Yvard}, {Zhao}, {Zorec}, {Zucker}, \& {Zwitter}}]{Gaia2023}
{Gaia Collaboration}, {Vallenari}, A., {Brown}, A.~G.~A., {et~al.} 2022, arXiv e-prints, arXiv:2208.00211

\bibitem[{{Garufi} {et~al.}(2020){Garufi}, {Avenhaus}, {P{\'e}rez}, {Quanz}, {van Holstein}, {Bertrang}, {Casassus}, {Cieza}, {Principe}, {van der Plas}, \& {Zurlo}}]{Garufi2020}
{Garufi}, A., {Avenhaus}, H., {P{\'e}rez}, S., {et~al.} 2020, \aap, 633, A82

\bibitem[{{Garufi} {et~al.}(2018){Garufi}, {Benisty}, {Pinilla}, {Tazzari}, {Dominik}, {Ginski}, {Henning}, {Kral}, {Langlois}, {M{\'e}nard}, {Stolker}, {Szulagyi}, {Villenave}, \& {van der Plas}}]{Garufi2018}
{Garufi}, A., {Benisty}, M., {Pinilla}, P., {et~al.} 2018, \aap, 620, A94

\bibitem[{{Garufi} {et~al.}(2022){Garufi}, {Dominik}, {Ginski}, {Benisty}, {van Holstein}, {Henning}, {Pawellek}, {Pinte}, {Avenhaus}, {Facchini}, {Galicher}, {Gratton}, {M{\'e}nard}, {Muro-Arena}, {Milli}, {Stolker}, {Vigan}, {Villenave}, {Moulin}, {Origne}, {Rigal}, {Sauvage}, \& {Weber}}]{Garufi2022}
{Garufi}, A., {Dominik}, C., {Ginski}, C., {et~al.} 2022, \aap, 658, A137

\bibitem[{{Garufi} {et~al.}(2017){Garufi}, {Meeus}, {Benisty}, {Quanz}, {Banzatti}, {Kama}, {Canovas}, {Eiroa}, {Schmid}, {Stolker}, {Pohl}, {Rigliaco}, {M{\'e}nard}, {Meyer}, {van Boekel}, \& {Dominik}}]{Garufi2017}
{Garufi}, A., {Meeus}, G., {Benisty}, M., {et~al.} 2017, \aap, 603, A21

\bibitem[{{Garufi} {et~al.}(2014){Garufi}, {Quanz}, {Schmid}, {Avenhaus}, {Buenzli}, \& {Wolf}}]{Garufi2014}
{Garufi}, A., {Quanz}, S.~P., {Schmid}, H.~M., {et~al.} 2014, \aap, 568, A40

\bibitem[{{Garufi} {et~al.}(2016){Garufi}, {Quanz}, {Schmid}, {Mulders}, {Avenhaus}, {Boccaletti}, {Ginski}, {Langlois}, {Stolker}, {Augereau}, {Benisty}, {Lopez}, {Dominik}, {Gratton}, {Henning}, {Janson}, {M{\'e}nard}, {Meyer}, {Pinte}, {Sissa}, {Vigan}, {Zurlo}, {Bazzon}, {Buenzli}, {Bonnefoy}, {Brandner}, {Chauvin}, {Cheetham}, {Cudel}, {Desidera}, {Feldt}, {Galicher}, {Kasper}, {Lagrange}, {Lannier}, {Maire}, {Mesa}, {Mouillet}, {Peretti}, {Perrot}, {Salter}, \& {Wildi}}]{Garufi2016}
{Garufi}, A., {Quanz}, S.~P., {Schmid}, H.~M., {et~al.} 2016, \aap, 588, A8

\bibitem[{{Ginski} {et~al.}(2018){Ginski}, {Benisty}, {van Holstein}, {Juh{\'a}sz}, {Schmidt}, {Chauvin}, {de Boer}, {Wilby}, {Manara}, {Delorme}, {M{\'e}nard}, {Pinilla}, {Birnstiel}, {Flock}, {Keller}, {Kenworthy}, {Milli}, {Olofsson}, {P{\'e}rez}, {Snik}, \& {Vogt}}]{Ginski2018}
{Ginski}, C., {Benisty}, M., {van Holstein}, R.~G., {et~al.} 2018, \aap, 616, A79

\bibitem[{{Ginski} {et~al.}(2021){Ginski}, {Facchini}, {Huang}, {Benisty}, {Vaendel}, {Stapper}, {Dominik}, {Bae}, {M{\'e}nard}, {Muro-Arena}, {Hogerheijde}, {McClure}, {van Holstein}, {Birnstiel}, {Boehler}, {Bohn}, {Flock}, {Mamajek}, {Manara}, {Pinilla}, {Pinte}, \& {Ribas}}]{Ginski2021}
{Ginski}, C., {Facchini}, S., {Huang}, J., {et~al.} 2021, \apjl, 908, L25

\bibitem[{{Ginski} {et~al.}(2016){Ginski}, {Stolker}, {Pinilla}, {Dominik}, {Boccaletti}, {de Boer}, {Benisty}, {Biller}, {Feldt}, {Garufi}, {Keller}, {Kenworthy}, {Maire}, {M{\'e}nard}, {Mesa}, {Milli}, {Min}, {Pinte}, {Quanz}, {van Boekel}, {Bonnefoy}, {Chauvin}, {Desidera}, {Gratton}, {Girard}, {Keppler}, {Kopytova}, {Lagrange}, {Langlois}, {Rouan}, \& {Vigan}}]{Ginski2016}
{Ginski}, C., {Stolker}, T., {Pinilla}, P., {et~al.} 2016, \aap, 595, A112

\bibitem[{{Grady} {et~al.}(2001){Grady}, {Polomski}, {Henning}, {Stecklum}, {Woodgate}, {Telesco}, {Pi{\~n}a}, {Gull}, {Boggess}, {Bowers}, {Bruhweiler}, {Clampin}, {Danks}, {Green}, {Heap}, {Hutchings}, {Jenkins}, {Joseph}, {Kaiser}, {Kimble}, {Kraemer}, {Lindler}, {Linsky}, {Maran}, {Moos}, {Plait}, {Roesler}, {Timothy}, \& {Weistrop}}]{Grady2001}
{Grady}, C.~A., {Polomski}, E.~F., {Henning}, T., {et~al.} 2001, \aj, 122, 3396

\bibitem[{{Guenther} {et~al.}(2007){Guenther}, {Esposito}, {Mundt}, {Covino}, {Alcal{\'a}}, {Cusano}, \& {Stecklum}}]{Guenther2007}
{Guenther}, E.~W., {Esposito}, M., {Mundt}, R., {et~al.} 2007, \aap, 467, 1147

\bibitem[{{Haikala} {et~al.}(2005){Haikala}, {Harju}, {Mattila}, \& {Toriseva}}]{Haikala2005}
{Haikala}, L.~K., {Harju}, J., {Mattila}, K., \& {Toriseva}, M. 2005, \aap, 431, 149

\bibitem[{Hal\'{i}\v{r} \& Flusser(1998)}]{oy1998NUMERICALLYSD}
Hal\'{i}\v{r}, R. \& Flusser, J. 1998

\bibitem[{{Hartmann} {et~al.}(1998){Hartmann}, {Calvet}, {Gullbring}, \& {D'Alessio}}]{1998ApJ...495..385H}
{Hartmann}, L., {Calvet}, N., {Gullbring}, E., \& {D'Alessio}, P. 1998, \apj, 495, 385

\bibitem[{{Hauschildt} {et~al.}(1999){Hauschildt}, {Allard}, \& {Baron}}]{Hauschildt1999}
{Hauschildt}, P.~H., {Allard}, F., \& {Baron}, E. 1999, \apj, 512, 377

\bibitem[{{Hunziker} {et~al.}(2021){Hunziker}, {Schmid}, {Ma}, {Menard}, {Avenhaus}, {Boccaletti}, {Beuzit}, {Chauvin}, {Dohlen}, {Dominik}, {Engler}, {Ginski}, {Gratton}, {Henning}, {Langlois}, {Milli}, {Mouillet}, {Tschudi}, {van Holstein}, \& {Vigan}}]{Hunziker2021}
{Hunziker}, S., {Schmid}, H.~M., {Ma}, J., {et~al.} 2021, \aap, 648, A110

\bibitem[{{Itoh} {et~al.}(2014){Itoh}, {Oasa}, {Kudo}, {Kusakabe}, {Hashimoto}, {Abe}, {Brandner}, {Brandt}, {Carson}, {Egner}, {Feldt}, {Grady}, {Guyon}, {Hayano}, {Hayashi}, {Hayashi}, {Henning}, {Hodapp}, {Ishii}, {Iye}, {Janson}, {Kandori}, {Knapp}, {Kuzuhara}, {Kwon}, {Matsuo}, {McElwain}, {Miyama}, {Morino}, {Moro-Martin}, {Nishimura}, {Pyo}, {Serabyn}, {Suenaga}, {Suto}, {Suzuki}, {Takahashi}, {Takato}, {Terada}, {Thalmann}, {Tomono}, {Turner}, {Watanabe}, {Wisniewski}, {Yamada}, {Mayama}, {Currie}, {Takami}, {Usuda}, \& {Tamura}}]{Itoh2014}
{Itoh}, Y., {Oasa}, Y., {Kudo}, T., {et~al.} 2014, Research in Astronomy and Astrophysics, 14, 1438

\bibitem[{{Jones} {et~al.}(2022){Jones}, {Milli}, {Blanchard}, {Wahhaj}, {de Rosa}, \& {Romero}}]{Jones2022}
{Jones}, M.~I., {Milli}, J., {Blanchard}, I., {et~al.} 2022, arXiv e-prints, arXiv:2204.11746

\bibitem[{{Kanagawa} {et~al.}(2021){Kanagawa}, {Hashimoto}, {Muto}, {Tsukagoshi}, {Takahashi}, {Hasegawa}, {Konishi}, {Nomura}, {Liu}, {Dong}, {Kataoka}, {Momose}, {Ono}, {Sitko}, {Takami}, \& {Tomida}}]{Kanagawa2021}
{Kanagawa}, K.~D., {Hashimoto}, J., {Muto}, T., {et~al.} 2021, \apj, 909, 212

\bibitem[{Kendall(1938)}]{10.2307/2332226}
Kendall, M.~G. 1938, Biometrika, 30, 81

\bibitem[{{Keppler} {et~al.}(2020){Keppler}, {Penzlin}, {Benisty}, {van Boekel}, {Henning}, {van Holstein}, {Kley}, {Garufi}, {Ginski}, {Brandner}, {Bertrang}, {Boccaletti}, {de Boer}, {Bonavita}, {Brown Sevilla}, {Chauvin}, {Dominik}, {Janson}, {Langlois}, {Lodato}, {Maire}, {M{\'e}nard}, {Pantin}, {Pinte}, {Stolker}, {Szul{\'a}gyi}, {Thebault}, {Villenave}, {Zurlo}, {Rabou}, {Feautrier}, {Feldt}, {Madec}, \& {Wildi}}]{Keppler2020}
{Keppler}, M., {Penzlin}, A., {Benisty}, M., {et~al.} 2020, \aap, 639, A62

\bibitem[{{Kim} {et~al.}(2020){Kim}, {Takahashi}, {Nomura}, {Tsukagoshi}, {Lee}, {Muto}, {Dong}, {Hasegawa}, {Hashimoto}, {Kanagawa}, {Kataoka}, {Konishi}, {Liu}, {Momose}, {Sitko}, \& {Tomida}}]{Kim2020}
{Kim}, S., {Takahashi}, S., {Nomura}, H., {et~al.} 2020, \apj, 888, 72

\bibitem[{{Kiraga}(2012)}]{2012AcA....62...67K}
{Kiraga}, M. 2012, \actaa, 62, 67

\bibitem[{{Krist} {et~al.}(2002){Krist}, {Stapelfeldt}, \& {Watson}}]{Krist2002}
{Krist}, J.~E., {Stapelfeldt}, K.~R., \& {Watson}, A.~M. 2002, \apj, 570, 785

\bibitem[{{Kurtovic} {et~al.}(2022){Kurtovic}, {Pinilla}, {Penzlin}, {Benisty}, {P{\'e}rez}, {Ginski}, {Isella}, {Kley}, {Menard}, {P{\'e}rez}, \& {Bayo}}]{Kurtovic2022}
{Kurtovic}, N.~T., {Pinilla}, P., {Penzlin}, A. B.~T., {et~al.} 2022, \aap, 664, A151

\bibitem[{{Lagage} {et~al.}(2006){Lagage}, {Doucet}, {Pantin}, {Habart}, {Duch{\^e}ne}, {M{\'e}nard}, {Pinte}, {Charnoz}, \& {Pel}}]{Lagage2006}
{Lagage}, P.-O., {Doucet}, C., {Pantin}, E., {et~al.} 2006, Science, 314, 621

\bibitem[{{Langlois} {et~al.}(2014){Langlois}, {Dohlen}, {Vigan}, {Zurlo}, {Moutou}, {Schmid}, {Mili}, {Beuzit}, {Boccaletti}, {Carle}, {Costille}, {Dorn}, {Gluck}, {Hubin}, {Feldt}, {Kasper}, {Lizon}, {Madec}, {Le Mignant}, {Mouillet}, {Puget}, {Sauvage}, \& {Wildi}}]{2014SPIE.9147E..1RL}
{Langlois}, M., {Dohlen}, K., {Vigan}, A., {et~al.} 2014, in \procspie, Vol. 9147, Ground-based and Airborne Instrumentation for Astronomy V, 91471R

\bibitem[{{Long} {et~al.}(2019){Long}, {Herczeg}, {Harsono}, {Pinilla}, {Tazzari}, {Manara}, {Pascucci}, {Cabrit}, {Nisini}, {Johnstone}, {Edwards}, {Salyk}, {Menard}, {Lodato}, {Boehler}, {Mace}, {Liu}, {Mulders}, {Hendler}, {Ragusa}, {Fischer}, {Banzatti}, {Rigliaco}, {van de Plas}, {Dipierro}, {Gully-Santiago}, \& {Lopez-Valdivia}}]{Long2019}
{Long}, F., {Herczeg}, G.~J., {Harsono}, D., {et~al.} 2019, \apj, 882, 49

\bibitem[{{Manara} {et~al.}(2016{\natexlab{a}}){Manara}, {Fedele}, {Herczeg}, \& {Teixeira}}]{2016A&A...585A.136M}
{Manara}, C.~F., {Fedele}, D., {Herczeg}, G.~J., \& {Teixeira}, P.~S. 2016{\natexlab{a}}, \aap, 585, A136

\bibitem[{{Manara} {et~al.}(2019){Manara}, {Mordasini}, {Testi}, {Williams}, {Miotello}, {Lodato}, \& {Emsenhuber}}]{Manara2019}
{Manara}, C.~F., {Mordasini}, C., {Testi}, L., {et~al.} 2019, \aap, 631, L2

\bibitem[{{Manara} {et~al.}(2016{\natexlab{b}}){Manara}, {Rosotti}, {Testi}, {Natta}, {Alcal{\'a}}, {Williams}, {Ansdell}, {Miotello}, {van der Marel}, {Tazzari}, {Carpenter}, {Guidi}, {Mathews}, {Oliveira}, {Prusti}, \& {van Dishoeck}}]{Manara2016}
{Manara}, C.~F., {Rosotti}, G., {Testi}, L., {et~al.} 2016{\natexlab{b}}, \aap, 591, L3

\bibitem[{{Marino} {et~al.}(2015){Marino}, {Perez}, \& {Casassus}}]{Marino2015}
{Marino}, S., {Perez}, S., \& {Casassus}, S. 2015, \apjl, 798, L44

\bibitem[{{Martinez} {et~al.}(2009){Martinez}, {Dorrer}, {Aller Carpentier}, {Kasper}, {Boccaletti}, {Dohlen}, \& {Yaitskova}}]{2009A&A...495..363M}
{Martinez}, P., {Dorrer}, C., {Aller Carpentier}, E., {et~al.} 2009, \aap, 495, 363

\bibitem[{{Meeus} {et~al.}(2001){Meeus}, {Waters}, {Bouwman}, {van den Ancker}, {Waelkens}, \& {Malfait}}]{2001A&A...365..476M}
{Meeus}, G., {Waters}, L.~B.~F.~M., {Bouwman}, J., {et~al.} 2001, \aap, 365, 476

\bibitem[{{M{\'e}nard} {et~al.}(2020){M{\'e}nard}, {Cuello}, {Ginski}, {van der Plas}, {Villenave}, {Gonzalez}, {Pinte}, {Benisty}, {Boccaletti}, {Price}, {Boehler}, {Chripko}, {de Boer}, {Dominik}, {Garufi}, {Gratton}, {Hagelberg}, {Henning}, {Langlois}, {Maire}, {Pinilla}, {Ruane}, {Schmid}, {van Holstein}, {Vigan}, {Zurlo}, {Hubin}, {Pavlov}, {Rochat}, {Sauvage}, \& {Stadler}}]{Menard2020}
{M{\'e}nard}, F., {Cuello}, N., {Ginski}, C., {et~al.} 2020, \aap, 639, L1

\bibitem[{{Menu} {et~al.}(2015){Menu}, {van Boekel}, {Henning}, {Leinert}, {Waelkens}, \& {Waters}}]{Menu2015}
{Menu}, J., {van Boekel}, R., {Henning}, T., {et~al.} 2015, \aap, 581, A107

\bibitem[{{Meschiari}(2012)}]{Meschiari2012}
{Meschiari}, S. 2012, \apjl, 761, L7

\bibitem[{{Miranda} \& {Lai}(2015)}]{Miranda2015}
{Miranda}, R. \& {Lai}, D. 2015, \mnras, 452, 2396

\bibitem[{{Monnier} {et~al.}(2019){Monnier}, {Harries}, {Bae}, {Setterholm}, {Laws}, {Aarnio}, {Adams}, {Andrews}, {Calvet}, {Espaillat}, {Hartmann}, {Kraus}, {McClure}, {Miller}, {Oppenheimer}, {Wilner}, \& {Zhu}}]{Monnier2019}
{Monnier}, J.~D., {Harries}, T.~J., {Bae}, J., {et~al.} 2019, \apj, 872, 122

\bibitem[{{Muro-Arena} {et~al.}(2020{\natexlab{a}}){Muro-Arena}, {Benisty}, {Ginski}, {Dominik}, {Facchini}, {Villenave}, {van Boekel}, {Chauvin}, {Garufi}, {Henning}, {Janson}, {Keppler}, {Matter}, {M{\'e}nard}, {Stolker}, {Zurlo}, {Blanchard}, {Maurel}, {Moeller-Nilsson}, {Petit}, {Roux}, {Sevin}, \& {Wildi}}]{Muro-Arena2020a}
{Muro-Arena}, G.~A., {Benisty}, M., {Ginski}, C., {et~al.} 2020{\natexlab{a}}, \aap, 635, A121

\bibitem[{{Muro-Arena} {et~al.}(2020{\natexlab{b}}){Muro-Arena}, {Ginski}, {Dominik}, {Benisty}, {Pinilla}, {Bohn}, {Moldenhauer}, {Kley}, {Harsono}, {Henning}, {van Holstein}, {Janson}, {Keppler}, {M{\'e}nard}, {P{\'e}rez}, {Stolker}, {Tazzari}, {Villenave}, {Zurlo}, {Petit}, {Rigal}, {M{\"o}ller-Nilsson}, {Llored}, {Moulin}, \& {Rabou}}]{Muro-Arena2020}
{Muro-Arena}, G.~A., {Ginski}, C., {Dominik}, C., {et~al.} 2020{\natexlab{b}}, \aap, 636, L4

\bibitem[{{Muto} {et~al.}(2012){Muto}, {Grady}, {Hashimoto}, {Fukagawa}, {Hornbeck}, {Sitko}, {Russell}, {Werren}, {Cur{\'e}}, {Currie}, {Ohashi}, {Okamoto}, {Momose}, {Honda}, {Inutsuka}, {Takeuchi}, {Dong}, {Abe}, {Brandner}, {Brandt}, {Carson}, {Egner}, {Feldt}, {Fukue}, {Goto}, {Guyon}, {Hayano}, {Hayashi}, {Hayashi}, {Henning}, {Hodapp}, {Ishii}, {Iye}, {Janson}, {Kandori}, {Knapp}, {Kudo}, {Kusakabe}, {Kuzuhara}, {Matsuo}, {Mayama}, {McElwain}, {Miyama}, {Morino}, {Moro-Martin}, {Nishimura}, {Pyo}, {Serabyn}, {Suto}, {Suzuki}, {Takami}, {Takato}, {Terada}, {Thalmann}, {Tomono}, {Turner}, {Watanabe}, {Wisniewski}, {Yamada}, {Takami}, {Usuda}, \& {Tamura}}]{Muto2012}
{Muto}, T., {Grady}, C.~A., {Hashimoto}, J., {et~al.} 2012, \apjl, 748, L22

\bibitem[{{Orihara} {et~al.}(2023){Orihara}, {Momose}, {Muto}, {Hashimoto}, {Liu}, {Tsukagoshi}, {Kudo}, {Takahashi}, {Yang}, {Hasegawa}, {Dong}, {Konishi}, \& {Akiyama}}]{Orihara2023}
{Orihara}, R., {Momose}, M., {Muto}, T., {et~al.} 2023, \pasj, 75, 424

\bibitem[{{Paardekooper} {et~al.}(2012){Paardekooper}, {Leinhardt}, {Th{\'e}bault}, \& {Baruteau}}]{Paardekooper2012}
{Paardekooper}, S.-J., {Leinhardt}, Z.~M., {Th{\'e}bault}, P., \& {Baruteau}, C. 2012, \apjl, 754, L16

\bibitem[{{Pascucci} {et~al.}(2016){Pascucci}, {Testi}, {Herczeg}, {Long}, {Manara}, {Hendler}, {Mulders}, {Krijt}, {Ciesla}, {Henning}, {Mohanty}, {Drabek-Maunder}, {Apai}, {Sz{\H{u}}cs}, {Sacco}, \& {Olofsson}}]{Pascucci2016}
{Pascucci}, I., {Testi}, L., {Herczeg}, G.~J., {et~al.} 2016, \apj, 831, 125

\bibitem[{{Pierens} {et~al.}(2021){Pierens}, {Nelson}, \& {McNally}}]{Pierens2021}
{Pierens}, A., {Nelson}, R.~P., \& {McNally}, C.~P. 2021, \mnras

\bibitem[{{Pineda} {et~al.}(2014){Pineda}, {Quanz}, {Meru}, {Mulders}, {Meyer}, {Pani{\'c}}, \& {Avenhaus}}]{Pineda2014}
{Pineda}, J.~E., {Quanz}, S.~P., {Meru}, F., {et~al.} 2014, \apjl, 788, L34

\bibitem[{{Pinte} {et~al.}(2020){Pinte}, {Price}, {M{\'e}nard}, {Duch{\^e}ne}, {Christiaens}, {Andrews}, {Huang}, {Hill}, {van der Plas}, {Perez}, {Isella}, {Boehler}, {Dent}, {Mentiplay}, \& {Loomis}}]{Pinte2020}
{Pinte}, C., {Price}, D.~J., {M{\'e}nard}, F., {et~al.} 2020, \apjl, 890, L9

\bibitem[{{Pinte} {et~al.}(2018){Pinte}, {Price}, {M{\'e}nard}, {Duch{\^e}ne}, {Dent}, {Hill}, {de Gregorio-Monsalvo}, {Hales}, \& {Mentiplay}}]{Pinte2018}
{Pinte}, C., {Price}, D.~J., {M{\'e}nard}, F., {et~al.} 2018, \apjl, 860, L13

\bibitem[{{Pinte} {et~al.}(2019){Pinte}, {van der Plas}, {M{\'e}nard}, {Price}, {Christiaens}, {Hill}, {Mentiplay}, {Ginski}, {Choquet}, {Boehler}, {Duch{\^e}ne}, {Perez}, \& {Casassus}}]{Pinte2019}
{Pinte}, C., {van der Plas}, G., {M{\'e}nard}, F., {et~al.} 2019, Nature Astronomy, 3, 1109

\bibitem[{{Price} {et~al.}(2018){Price}, {Cuello}, {Pinte}, {Mentiplay}, {Casassus}, {Christiaens}, {Kennedy}, {Cuadra}, {Sebastian Perez}, {Marino}, {Armitage}, {Zurlo}, {Juhasz}, {Ragusa}, {Laibe}, \& {Lodato}}]{Price2018}
{Price}, D.~J., {Cuello}, N., {Pinte}, C., {et~al.} 2018, \mnras, 477, 1270

\bibitem[{{Ragusa} {et~al.}(2021){Ragusa}, {Fasano}, {Toci}, {Duch{\^e}ne}, {Cuello}, {Villenave}, {van der Plas}, {Lodato}, {M{\'e}nard}, {Price}, {Pinte}, {Stapelfeldt}, \& {Wolff}}]{Ragusa2021}
{Ragusa}, E., {Fasano}, D., {Toci}, C., {et~al.} 2021, \mnras, 507, 1157

\bibitem[{{Ren} {et~al.}(2023){Ren}, {Benisty}, {Ginski}, {Tazaki}, {Wallack}, {Milli}, {Garufi}, {Bae}, {Facchini}, {M{\'e}nard}, {Pinilla}, {Swastik}, {Teague}, \& {Wahhaj}}]{Ren2023}
{Ren}, B.~B., {Benisty}, M., {Ginski}, C., {et~al.} 2023, arXiv e-prints, arXiv:2310.08589

\bibitem[{{Ribas} {et~al.}(2017){Ribas}, {Espaillat}, {Mac{\'\i}as}, {Bouy}, {Andrews}, {Calvet}, {Naylor}, {Riviere-Marichalar}, {van der Wiel}, \& {Wilner}}]{Ribas2017}
{Ribas}, {\'A}., {Espaillat}, C.~C., {Mac{\'\i}as}, E., {et~al.} 2017, \apj, 849, 63

\bibitem[{{Ribas} {et~al.}(2020){Ribas}, {Espaillat}, {Mac{\'\i}as}, \& {Sarro}}]{Ribas2020}
{Ribas}, {\'A}., {Espaillat}, C.~C., {Mac{\'\i}as}, E., \& {Sarro}, L.~M. 2020, \aap, 642, A171

\bibitem[{{Rich} {et~al.}(2022){Rich}, {Monnier}, {Aarnio}, {Laws}, {Setterholm}, {Wilner}, {Calvet}, {Harries}, {Miller}, {Davies}, {Adams}, {Andrews}, {Bae}, {Espaillat}, {Greenbaum}, {Hinkley}, {Kraus}, {Hartmann}, {Isella}, {McClure}, {Oppenheimer}, {P{\'e}rez}, \& {Zhu}}]{Rich2022}
{Rich}, E.~A., {Monnier}, J.~D., {Aarnio}, A., {et~al.} 2022, \aj, 164, 109

\bibitem[{{Rich} {et~al.}(2021){Rich}, {Teague}, {Monnier}, {Davies}, {Bosman}, {Harries}, {Calvet}, {Adams}, {Wilner}, \& {Zhu}}]{Rich2021}
{Rich}, E.~A., {Teague}, R., {Monnier}, J.~D., {et~al.} 2021, \apj, 913, 138

\bibitem[{{Schmid} {et~al.}(2018){Schmid}, {Bazzon}, {Roelfsema}, {Mouillet}, {Milli}, {Menard}, {Gisler}, {Hunziker}, {Pragt}, {Dominik}, {Boccaletti}, {Ginski}, {Abe}, {Antoniucci}, {Avenhaus}, {Baruffolo}, {Baudoz}, {Beuzit}, {Carbillet}, {Chauvin}, {Claudi}, {Costille}, {Daban}, {de Haan}, {Desidera}, {Dohlen}, {Downing}, {Elswijk}, {Engler}, {Feldt}, {Fusco}, {Girard}, {Gratton}, {Hanenburg}, {Henning}, {Hubin}, {Joos}, {Kasper}, {Keller}, {Langlois}, {Lagadec}, {Martinez}, {Mulder}, {Pavlov}, {Podio}, {Puget}, {Quanz}, {Rigal}, {Salasnich}, {Sauvage}, {Schuil}, {Siebenmorgen}, {Sissa}, {Snik}, {Suarez}, {Thalmann}, {Turatto}, {Udry}, {van Duin}, {van Holstein}, {Vigan}, \& {Wildi}}]{Schmid2018}
{Schmid}, H.~M., {Bazzon}, A., {Roelfsema}, R., {et~al.} 2018, \aap, 619, A9

\bibitem[{{Schmid} {et~al.}(2006){Schmid}, {Joos}, \& {Tschan}}]{2006A&A...452..657S}
{Schmid}, H.~M., {Joos}, F., \& {Tschan}, D. 2006, \aap, 452, 657

\bibitem[{{Serkowski} {et~al.}(1975){Serkowski}, {Mathewson}, \& {Ford}}]{1975ApJ...196..261S}
{Serkowski}, K., {Mathewson}, D.~S., \& {Ford}, V.~L. 1975, \apj, 196, 261

\bibitem[{{Siess} {et~al.}(2000){Siess}, {Dufour}, \& {Forestini}}]{Siess2000}
{Siess}, L., {Dufour}, E., \& {Forestini}, M. 2000, \aap, 358, 593

\bibitem[{{Smart} \& {Nicastro}(2014)}]{2014A&A...570A..87S}
{Smart}, R.~L. \& {Nicastro}, L. 2014, \aap, 570, A87

\bibitem[{{Stolker} {et~al.}(2016){Stolker}, {Dominik}, {Avenhaus}, {Min}, {de Boer}, {Ginski}, {Schmid}, {Juhasz}, {Bazzon}, {Waters}, {Garufi}, {Augereau}, {Benisty}, {Boccaletti}, {Henning}, {Langlois}, {Maire}, {M{\'e}nard}, {Meyer}, {Pinte}, {Quanz}, {Thalmann}, {Beuzit}, {Carbillet}, {Costille}, {Dohlen}, {Feldt}, {Gisler}, {Mouillet}, {Pavlov}, {Perret}, {Petit}, {Pragt}, {Rochat}, {Roelfsema}, {Salasnich}, {Soenke}, \& {Wildi}}]{Stolker2016}
{Stolker}, T., {Dominik}, C., {Avenhaus}, H., {et~al.} 2016, \aap, 595, A113

\bibitem[{{Tazaki} {et~al.}(2019){Tazaki}, {Tanaka}, {Muto}, {Kataoka}, \& {Okuzumi}}]{2019MNRAS.485.4951T}
{Tazaki}, R., {Tanaka}, H., {Muto}, T., {Kataoka}, A., \& {Okuzumi}, S. 2019, \mnras, 485, 4951

\bibitem[{{Teague} {et~al.}(2018){Teague}, {Bae}, {Bergin}, {Birnstiel}, \& {Foreman-Mackey}}]{Teague2018}
{Teague}, R., {Bae}, J., {Bergin}, E.~A., {Birnstiel}, T., \& {Foreman-Mackey}, D. 2018, \apjl, 860, L12

\bibitem[{{van der Marel} {et~al.}(2021){van der Marel}, {Birnstiel}, {Garufi}, {Ragusa}, {Christiaens}, {Price}, {Sallum}, {Muley}, {Francis}, \& {Dong}}]{Marel2021}
{van der Marel}, N., {Birnstiel}, T., {Garufi}, A., {et~al.} 2021, \aj, 161, 33

\bibitem[{{van der Marel} {et~al.}(2016){van der Marel}, {Cazzoletti}, {Pinilla}, \& {Garufi}}]{Marel2016}
{van der Marel}, N., {Cazzoletti}, P., {Pinilla}, P., \& {Garufi}, A. 2016, \apj, 832, 178

\bibitem[{{van der Plas} {et~al.}(2017{\natexlab{a}}){van der Plas}, {Wright}, {M{\'e}nard}, {Casassus}, {Canovas}, {Pinte}, {Maddison}, {Maaskant}, {Avenhaus}, {Cieza}, {Perez}, \& {Ubach}}]{Plas2017}
{van der Plas}, G., {Wright}, C.~M., {M{\'e}nard}, F., {et~al.} 2017{\natexlab{a}}, \aap, 597, A32

\bibitem[{{van der Plas} {et~al.}(2017{\natexlab{b}}){van der Plas}, {Wright}, {M{\'e}nard}, {Casassus}, {Canovas}, {Pinte}, {Maddison}, {Maaskant}, {Avenhaus}, {Cieza}, {Perez}, \& {Ubach}}]{vanderPlas2017}
{van der Plas}, G., {Wright}, C.~M., {M{\'e}nard}, F., {et~al.} 2017{\natexlab{b}}, \aap, 597, A32

\bibitem[{{van Holstein} {et~al.}(2020){van Holstein}, {Girard}, {de Boer}, {Snik}, {Milli}, {Stam}, {Ginski}, {Mouillet}, {Wahhaj}, {Schmid}, {Keller}, {Langlois}, {Dohlen}, {Vigan}, {Pohl}, {Carbillet}, {Fantinel}, {Maurel}, {Orign{\'e}}, {Petit}, {Ramos}, {Rigal}, {Sevin}, {Boccaletti}, {Le Coroller}, {Dominik}, {Henning}, {Lagadec}, {M{\'e}nard}, {Turatto}, {Udry}, {Chauvin}, {Feldt}, \& {Beuzit}}]{2020A&A...633A..64V}
{van Holstein}, R.~G., {Girard}, J.~H., {de Boer}, J., {et~al.} 2020, \aap, 633, A64

\bibitem[{{van Holstein} {et~al.}(2021){van Holstein}, {Stolker}, {Jensen-Clem}, {Ginski}, {Milli}, {de Boer}, {Girard}, {Wahhaj}, {Bohn}, {Millar-Blanchaer}, {Benisty}, {Bonnefoy}, {Chauvin}, {Dominik}, {Hinkley}, {Keller}, {Keppler}, {Langlois}, {Marino}, {M{\'e}nard}, {Perrot}, {Schmidt}, {Vigan}, {Zurlo}, \& {Snik}}]{2021A&A...647A..21V}
{van Holstein}, R.~G., {Stolker}, T., {Jensen-Clem}, R., {et~al.} 2021, \aap, 647, A21

\bibitem[{{Villenave} {et~al.}(2019){Villenave}, {Benisty}, {Dent}, {M{\'e}nard}, {Garufi}, {Ginski}, {Pinilla}, {Pinte}, {Williams}, {de Boer}, {Morino}, {Fukagawa}, {Dominik}, {Flock}, {Henning}, {Juh{\'a}sz}, {Keppler}, {Muro-Arena}, {Olofsson}, {P{\'e}rez}, {van der Plas}, {Zurlo}, {Carle}, {Feautrier}, {Pavlov}, {Pragt}, {Ramos}, {Sauvage}, {Stadler}, \& {Weber}}]{Villenave2019}
{Villenave}, M., {Benisty}, M., {Dent}, W.~R.~F., {et~al.} 2019, \aap, 624, A7

\bibitem[{{Voirin} {et~al.}(2018){Voirin}, {Manara}, \& {Prusti}}]{Voirin2018}
{Voirin}, J., {Manara}, C.~F., \& {Prusti}, T. 2018, \aap, 610, A64

\bibitem[{{Walsh} {et~al.}(2016){Walsh}, {Juh{\'a}sz}, {Meeus}, {Dent}, {Maud}, {Aikawa}, {Millar}, \& {Nomura}}]{Walsh2016}
{Walsh}, C., {Juh{\'a}sz}, A., {Meeus}, G., {et~al.} 2016, \apj, 831, 200

\bibitem[{{Walsh} {et~al.}(2014){Walsh}, {Juh{\'a}sz}, {Pinilla}, {Harsono}, {Mathews}, {Dent}, {Hogerheijde}, {Birnstiel}, {Meeus}, {Nomura}, {Aikawa}, {Millar}, \& {Sandell}}]{Walsh2014}
{Walsh}, C., {Juh{\'a}sz}, A., {Pinilla}, P., {et~al.} 2014, \apjl, 791, L6

\bibitem[{{Weidenschilling}(1977)}]{Weidenschilling1977}
{Weidenschilling}, S.~J. 1977, \apss, 51, 153

\bibitem[{{Xiang-Gruess} \& {Papaloizou}(2013)}]{Xiang-Gruess2013}
{Xiang-Gruess}, M. \& {Papaloizou}, J.~C.~B. 2013, \mnras, 431, 1320

\bibitem[{{Zhang} {et~al.}(2023){Zhang}, {Ginski}, {Huang}, {Zurlo}, {Beust}, {Bae}, {Benisty}, {Garufi}, {Hogerheijde}, {van Holstein}, {Kenworthy}, {Langlois}, {Manara}, {Pinilla}, {Rab}, {Ribas}, {Rosotti}, \& {Williams}}]{Zhang2023}
{Zhang}, Y., {Ginski}, C., {Huang}, J., {et~al.} 2023, \aap, 672, A145

\end{thebibliography}

\begin{appendix} 

\section{Observing condition and instrument setup} \label{appendix: observations}
The detailed dates and weather conditions of our observations as well as the instrument setup are listed in Table \ref{tab: observations}. 

\begin{table*}[!h]
 \centering
 \caption{Observing dates, instrument setup, and weather conditions for all systems in our study. The average values for the Seeing and the coherence time $\tau_0$ of the atmosphere during the observation blocks are listed.}
  \begin{tabular}{@{}lccccccc@{}}
  \hline 
 Target         & Date          & Filter                & Coronagraph           & DIT\,[s]        &  \# frames    & Seeing\,[arcsec]      & $\tau_0$\,[ms]         \\
 \hline
 CHX 18 N           & 30-01-2020 & BB\_H & N\_ALC\_YJH\_S &  64  & 56 & 0.77 & 4.4 \\
 CHX 22             & 29-02-2020 & BB\_H & N\_ALC\_YJH\_S &   8  & 416 & 0.68 & 11.6 \\
 CR\,Cha                   & 15-01-2019 & BB\_H & N\_ALC\_YJH\_S & 32   & 16              & 0.65                  & 5.4                   \\
                              & 15-01-2019 & BB\_H & none                       & 0.83 & 480              & 0.65                  & 5.4                   \\
 CS\,Cha                   & 17-02-2017 & BB\_J & N\_ALC\_YJH\_S & 96 & 40              &       0.64            &       -               \\
                              & 17-06-2017 & BB\_H & N\_ALC\_YJH\_S     & 64 & 28         &       0.66            & 2.4                   \\
                              & 23-12-2018 & I\_PRIM/R\_PRIM & V\_CLC\_MT\_WF & 10 & 120 & 0.65 & 5.9 \\
                              & 20-01-2019 & I\_PRIM/R\_PRIM & V\_CLC\_MT\_WF & 58 & 32 & 0.55 & 10.5\\
                              & 24-03-2019 & BB\_K & N\_ALC\_YJH\_S     & 64 & 28         & 0.70                  & 8.8                   \\
 CT\,Cha            & 25-02-2018        & BB\_H & N\_ALC\_YJH\_S & 64 & 108     &       0.44            &       10.9            \\
 CV\,Cha            & 26-03-2016        & BB\_J & N\_ALC\_YJ\_S  & 64 & 48              &       1.02            &               3.1     \\
                              & 26-03-2016 & BB\_J + ND\_1.0& none      & 2        & 240          &       1.08            & 2.4                   \\
 DI Cha             & 21-03-2017 & BB\_H & N\_ALC\_YJH\_S & 32 & 50 & 0.54 & 7.5 \\
 HD97048            & 24-05-2019 & BB\_K & N\_ALC\_KS     & 64 & 152 & 0.69 & 4.9\\
 HP Cha             & 13-01-2020 & BB\_H & N\_ALC\_YJH\_S & 64 & 56 & 0.42 & 9.0\\
 PDS 51             & 21-01-2020 & BB\_H & none & 0.83 & 2240 & 0.38 & 17.9 \\
 RX J1106.3-7721    & 24-02-2020 & BB\_H & N\_ALC\_YJH\_S & 64 & 56 & 0.56 & 9.2 \\
 SY\,Cha               & 16-05-2017     & BB\_H & N\_ALC\_YJH\_S & 64 & 32              & 0.75                    & -                     \\
                        & 02-01-2021 & BB\_K & N\_ALC\_KS     & 16 & 156                & 0.40                    & 13.3                  \\
 Sz 41              & 18-02-2020 & BB\_H & N\_ALC\_YJH\_S & 2 & 1056 & 0.53 & 11.6 \\
 Sz 45              & 21-03-2020 & BB\_H & N\_ALC\_YJH\_S & 64 & 34 & 0.75 & 5.4 \\
 SZ\,Cha               & 21-03-2017     & BB\_H & N\_ALC\_YJH\_S & 96 & 20              & 0.70                    & -                     \\
 TW Cha             & 09-01-2020 & BB\_H & N\_ALC\_YJH\_S & 64 & 56 & 0.57 & 5.2 \\
 VZ\,Cha               & 28-02-2018     & BB\_H & N\_ALC\_YJH\_S & 64           & 12              & 0.66                  & 3.5                   \\
 WW\,Cha               & 12-03-2017     & BB\_H & N\_ALC\_YJH\_S & 32           & 48              & 0.50                  & -                     \\
 WX\,Cha               & 01-04-2019     & BB\_H & none                   & 0.83          & 480           & 0.85                  & 3.0                   \\
 WY\,Cha                   & 18-02-2019 & BB\_H & none                   & 0.83          & 480           & 0.65                  & 15                    \\
 
\hline\end{tabular}
\label{tab: observations}
\end{table*}

\clearpage

\section{Detailed Stokes images}

\begin{figure*}
\centering
\includegraphics[width=0.995\textwidth]{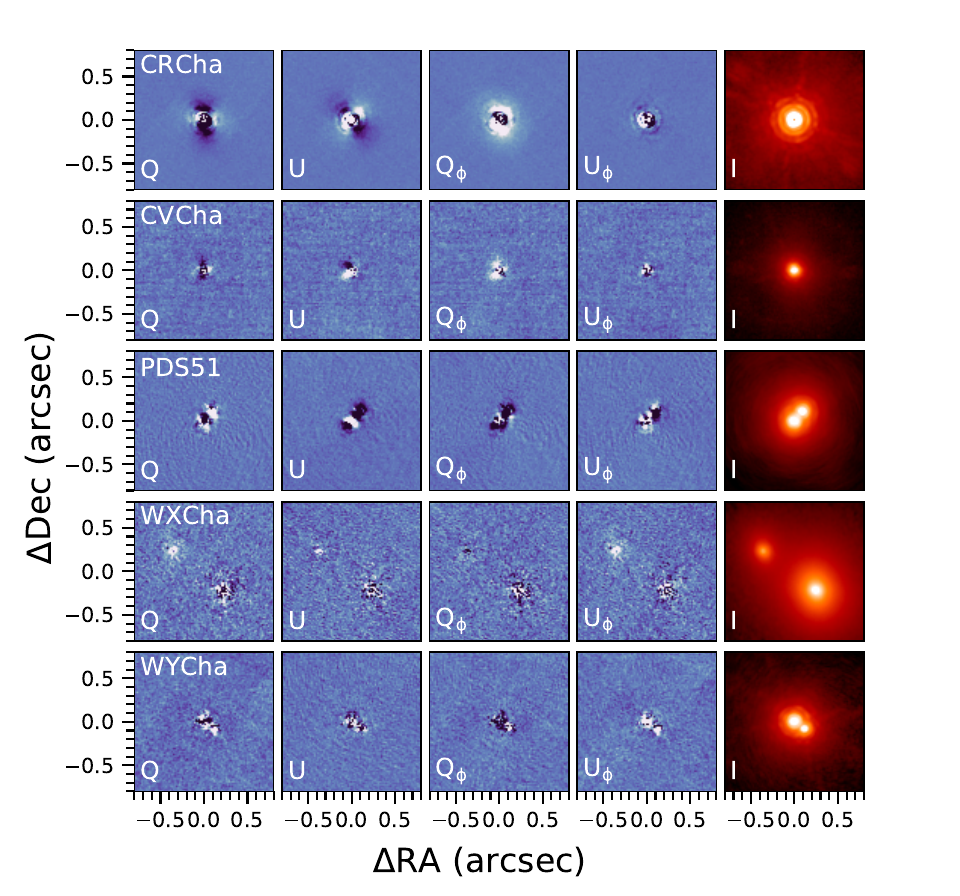}
\caption[]{All noncoronagraphic images of our target sample. The color scale is linear in all images with the same cuts for all images.
The color scale is symmetric around 0. The final column in each row in the red-orange hues shows the total intensity images in which the stellar light dominates. For WX\,Cha and WY\,Cha close stellar binary companions are visible in the data.}
\label{fig: nocoro_sample}
\end{figure*}

\begin{figure}
\centering
\includegraphics[width=0.49\textwidth]{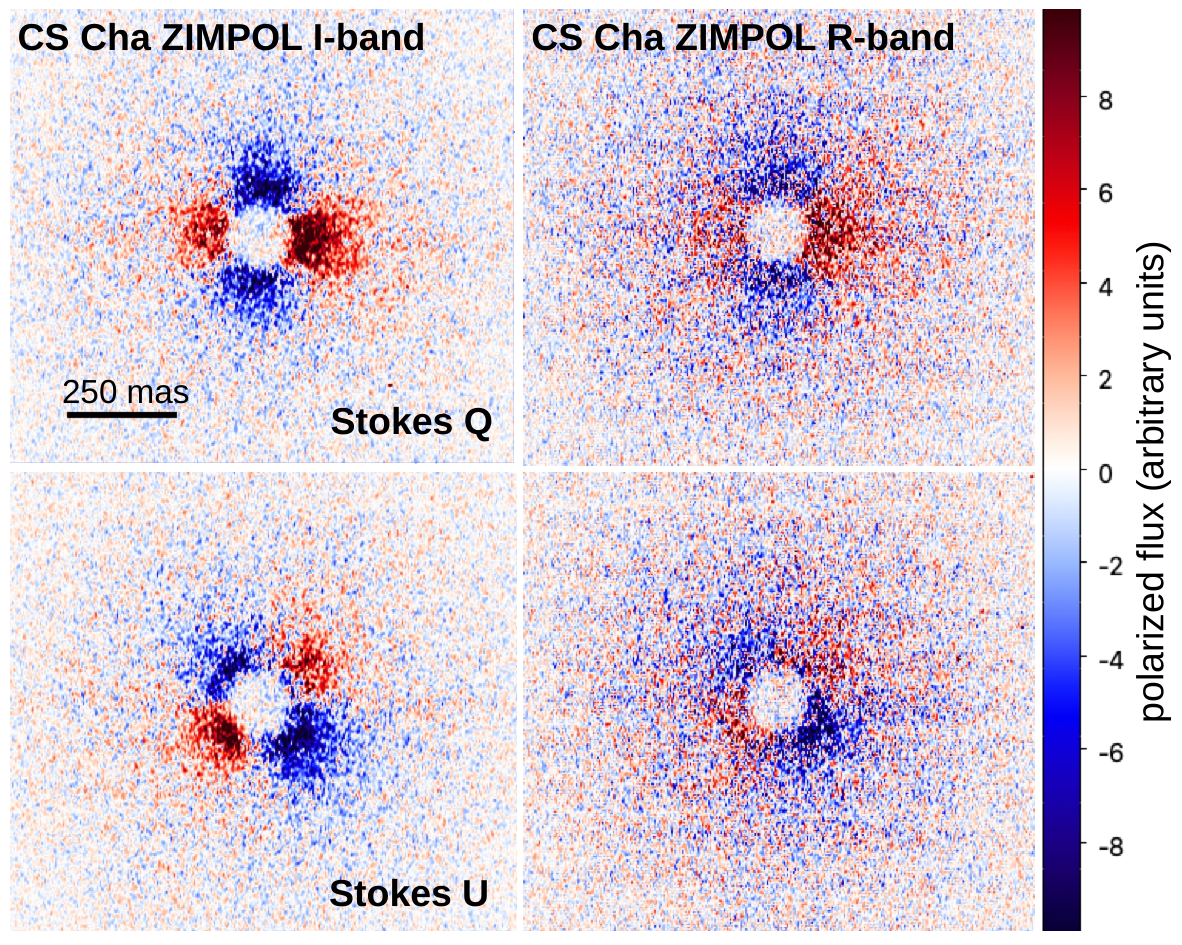}
\caption[]{Optical (ZIMPOL) Stokes Q and U images for our observations of the CS\,Cha system. The I-band data are on the left and the R-band data on the right. All images use the same (linear) color scale.}
\label{fig: cscha-zimpol-stokes}
\end{figure}

\begin{figure*}
\centering
\includegraphics[width=0.999\textwidth]{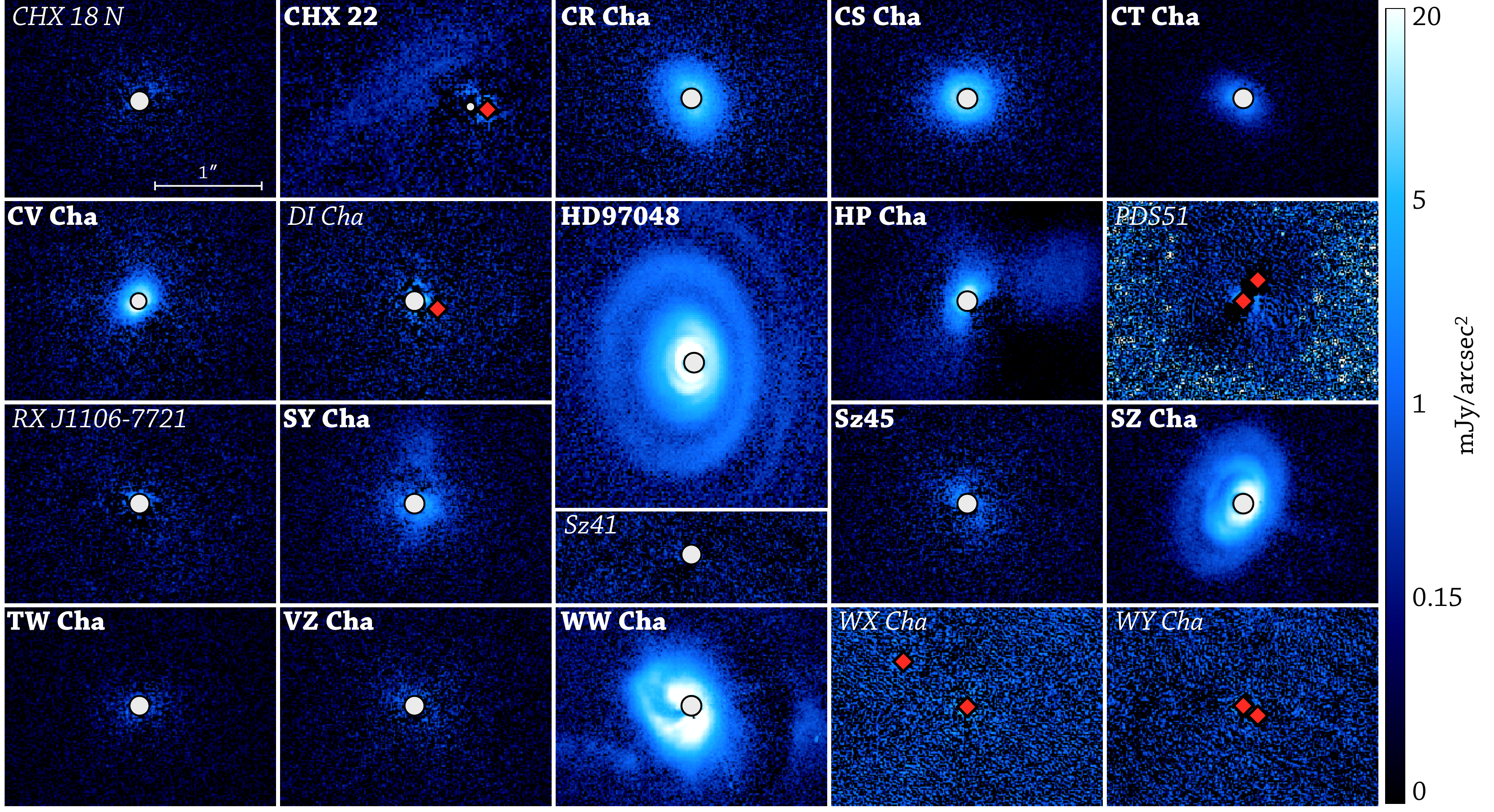}
\caption[]{Imagery of the sample. The Q$_\phi$ image of all targets is shown at the same logarithmic, flux scale, and physical scale (as indicated in the first panel). Detections are indicated by a name in bold, nondetections in italics. The gray circles mark the size of the coronagraphic mask, while the red diamonds give the position of any unmasked star (both primary in noncoronagraphic observations and companions).}
\label{fig: scaled_brightness_gallery}
\end{figure*}

In this section we show the Stokes Q and U images that are directly measured in our observations. For all coronagraphic observations, we summarize them in figure~\ref{fig: coro_sample} and for noncoronagraphic observations in figure~\ref{fig: nocoro_sample}.
Due to the azimuthal orientation of the angle of polarization of single scattered light, we expect a typical "butterfly" pattern in the Q and U images in which we detect resolved disk signal. This is indeed the case for all our coronagraphic observations, as well as the noncoronagraphic observations of CR\,Cha and CV\,Cha. In addition to Stokes Q and U we also show the derived Q$_\phi$ and U$_\phi$ images. As predicted from single scattering models the U$_\phi$ images contain little signal.
For the coronagraphic observations we show the Q$_\phi$ images with different scaling to bridge the dynamic range between bright inner and fainter outer structures. In the 5th column of figure~\ref{fig: coro_sample} we compensate for the drop in stellar illumination by scaling the signal with the squared distance from the central star. We take into account the inclination of the disk to prevent the introduction of nonphysical asymmetries between major and minor disk axes. However, we caution that we did not take into account the disk scale height for the correction, as the surface height profile is uncertain for most objects. Alternatively in the 6th panel of figure~\ref{fig: coro_sample} we show the same images with a logarithmic scaling, which allows us to display a larger dynamic range without a priory knowledge of the disk geometry. 
In figure~\ref{fig: nocoro_sample} we additionally show the total intensity images of the noncoronagraphic system. The binary stellar companions to WX\,Cha and WY\,Cha are clearly visible.\\
In figure~\ref{fig: cscha-zimpol-stokes} we show the Stokes Q and U images of the CS\,Cha system taken with SPHERE/ZIMPOL in the I and R-band.\\
Finally in figure~\ref{fig: scaled_brightness_gallery} we show all Q$_\phi$ images of the entire sample (detections and nondetections) scaled to the same surface brightness.

\begin{figure*}
\centering
\includegraphics[width=17.0cm]{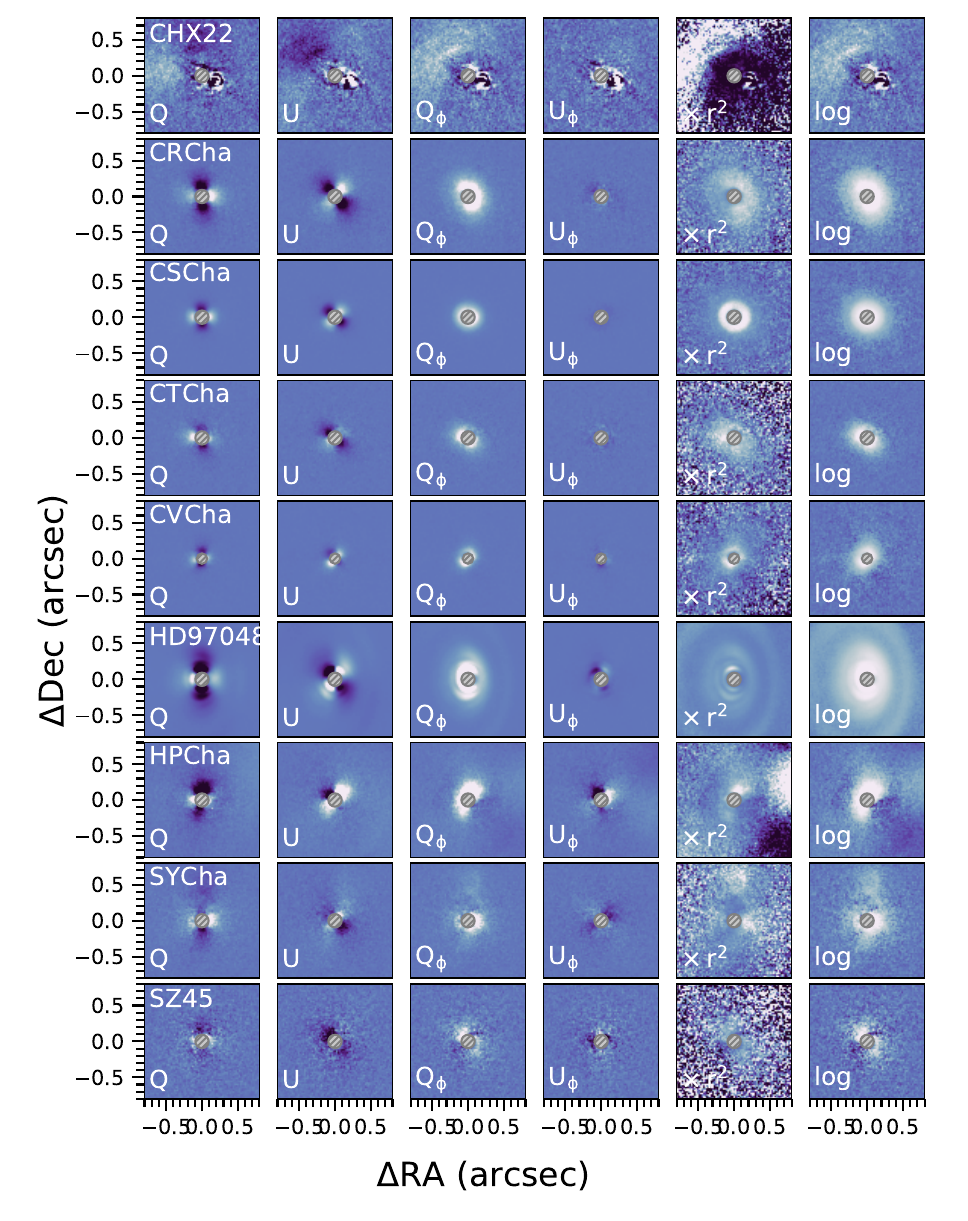}
\caption[]{All coronagraphic images of our target sample. The color scale is linear in all images with the same cuts in each row, but different cuts for different sources.
The color scale is symmetric around 0. The gray hashed circles show the coronagraph size and position. The Stokes Q and U images are in the first two columns and the derived Q$_\phi$ and U$_\phi$ images in columns 3 and 4. In    column 5 is shown the Q$_\phi$ image corrected for the separation-dependent drop-off in illumination (taking into account the system inclination). In the last column,   the Q$_\phi$ image is shown on a log scale to highlight the fainter parts of the disks.}
\label{fig: coro_sample}
\end{figure*}

\begin{figure*}[h!]
\ContinuedFloat
\centering
\includegraphics[width=17.0cm]{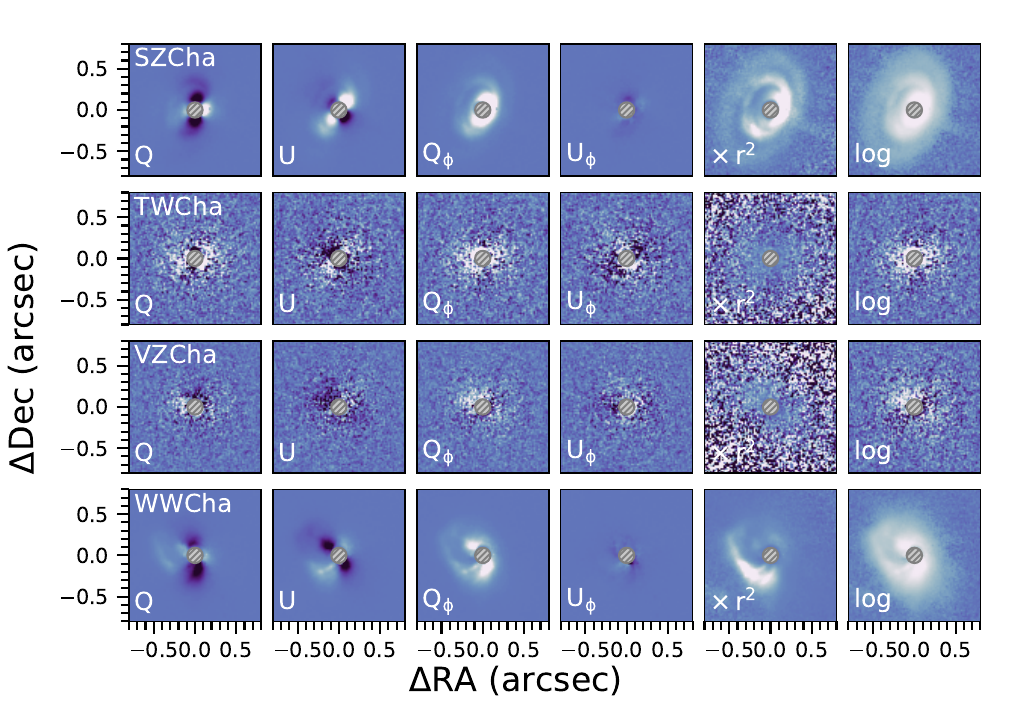}
\caption[]{ \textbf{(continued)} All coronagraphic images of our target sample. The color scale is linear in all images with the same cuts in each row, but different cuts for different sources.
The color scale is symmetric around 0. The gray hashed circles show the coronagraph size and position. The Stokes Q and U images are in the first two columns and the derived Q$_\phi$ and U$_\phi$ images in columns 3 and 4. In    column 5 is shown the Q$_\phi$ image corrected for the separation-dependent drop-off in illumination (taking into account the system inclination). In the last column,   the Q$_\phi$ image is shown on a log scale to highlight the fainter parts of the disks.}
\label{fig: coro_sample}
\end{figure*}

\section{Geometric fitting approach and results}
\label{app: geometric fitting}

In this section we briefly demonstrate the validity of our fitting approach as described in section~\ref{sec: ellipse-fitting} and illustrate the extracted data points and best-fitting ellipse solutions from the initial fitting step.

\begin{figure*}
\centering
\includegraphics[width=0.999\textwidth]{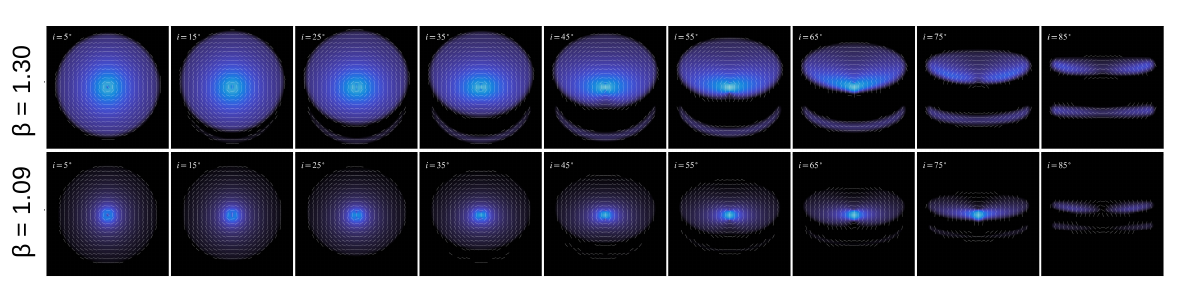}
\caption[]{Radiative transfer polarized intensity model images with different flaring exponents $\beta$ seen at different inclinations. The full gallery of model images are shown up to an inclination of 85$^\circ$. However,  only   models up to 65$^\circ$ are included in our test fits, as for higher inclinations our fitting approach breaks down. }
\label{app: fig: models}
\end{figure*}

\begin{figure*}
\centering
\subfloat[Fits for $\beta = 1.09$]{
\includegraphics[width=0.999\textwidth]{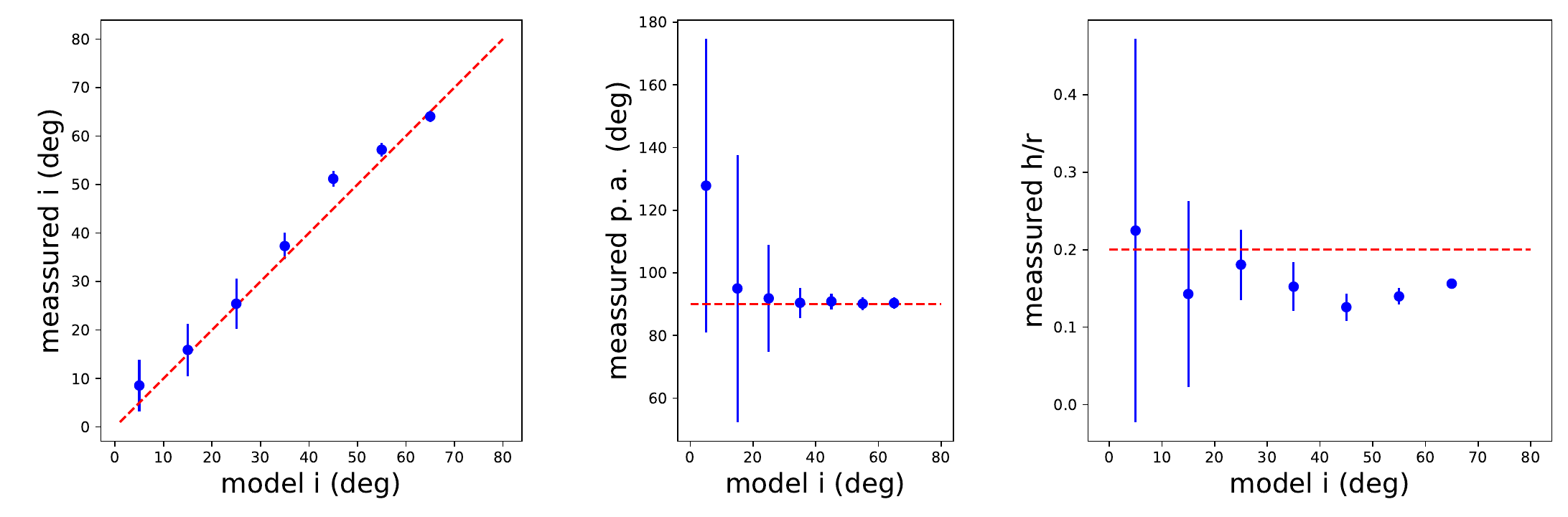}
\label{beta_1.09_results}
}

\subfloat[Fits for $\beta = 1.30$]{
\includegraphics[width=0.999\textwidth]{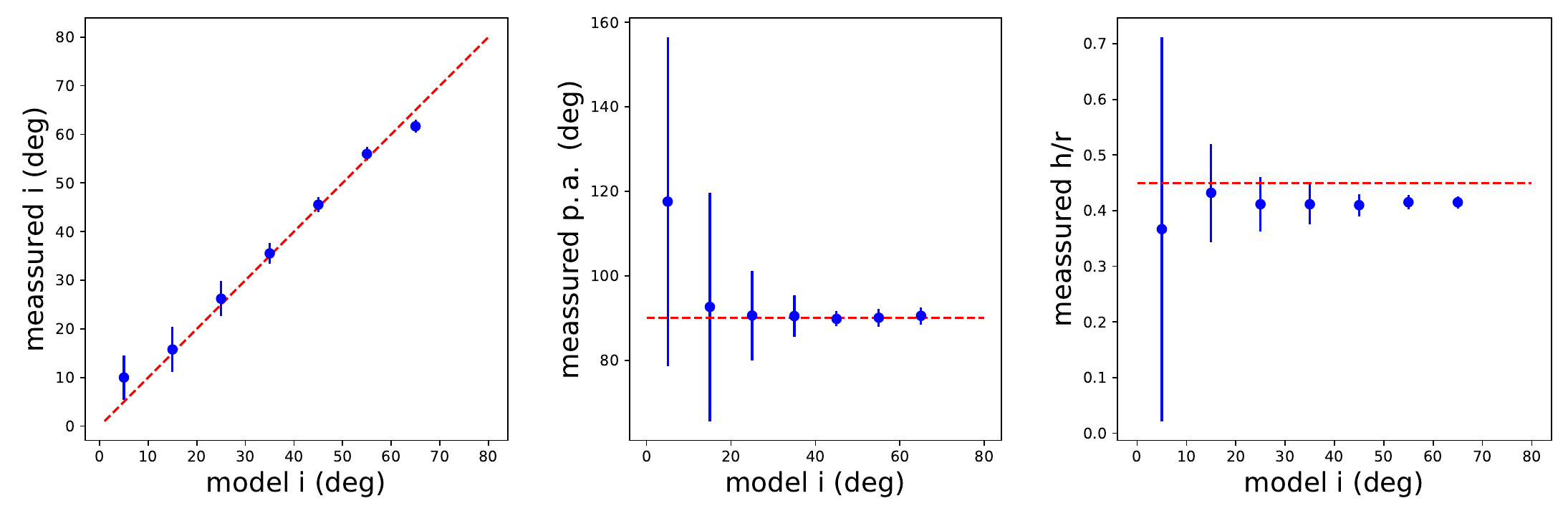}
\label{beta_1.30_results}
}

\caption[]{Ellipse fit to the outer disk edge for our radiative transfer model images. Fitting results are shown as a function of model inclination. The expected results are indicated by the red dashed lines. }
\label{fig: ellipse-fit-model-results}
\end{figure*}

While it was demonstrated in the literature that equidistant rings can be used to constrain the disk geometry in scattered light images, this was not  done for small, essentially feature-less (or in any case "ring-less") disks. For this purpose we created a grid of radiative transfer models with two different flaring exponents (1.09 and 1.3), simulating relatively vertically thin and strongly flared disks. The disks are smooth and featureless. Within the grid, we vary the inclination of the disk to demonstrate how the recovery of disk parameters is dependent on the viewing geometry. We show the model images in Figure~\ref{app: fig: models}.
Before testing our fitting procedure we convolved each model image with a Gaussian to simulate the resolution of VLT/SPHERE in the H-band. We then added noise to the images by co-adding the model image with a typical reduced image of a disk nondetection. The scaling of the disk forward scattering side and the background noise level was such that the resulting S/N were comparable to those of the observed disks in our sample, in particular the case of CR\,Cha, which is faint but still well detected.  
We then ran the two-step fitting procedure, as described in section~\ref{sec: ellipse-fitting} on all model images. We show the extracted inclination, position angle, and disk aspect ratio as a function of the model disk inclination in Figure~\ref{fig: ellipse-fit-model-results}. Extracted inclination and position angle are from the first LSMC step. The extracted aspect ratio is measured with the second aperture photometry step, with inclination and position angle of the model fixed. 
We find that for inclinations larger than $\sim10^\circ$ the measured inclination and position angle are well consistent with the expected results. For small inclinations below $\sim30^\circ$ the uncertainty on the position angle in particular is large and appears somewhat inflated for values between 10$^\circ$ and 30$^\circ$. The measured aspect ratio shows little variation for the case of the strongly flaring disk ($\beta$=1.3) as a function of model inclination. We do however observe a small systematic offset of $\sim0.05$ (i.e., the retrieved aspect ratio is slightly too low). This is likely an effect of the polarized phase function, which leads to a weak signal on the far side of the disk, which can lead to systematically smaller offset values found by the aperture method; we note that the same would be expected for the pure edge tracing method for the same reason.
For the less flaring disks ($\beta$=1.09) we find small variability within the retrieved aspect ratios as a function of model inclination. This is to be expected as the offset values are smaller for vertically thinner disks and thus inherently harder to measure. Within the uncertainties, all values are however consistent with each other. We find the same systematic offset for this case as for the strong flaring case (i.e., aspect ratios are roughly too low by $\sim0.05$).

\begin{figure*}
\centering
\subfloat[CR\,Cha]{
\includegraphics[scale=0.4]{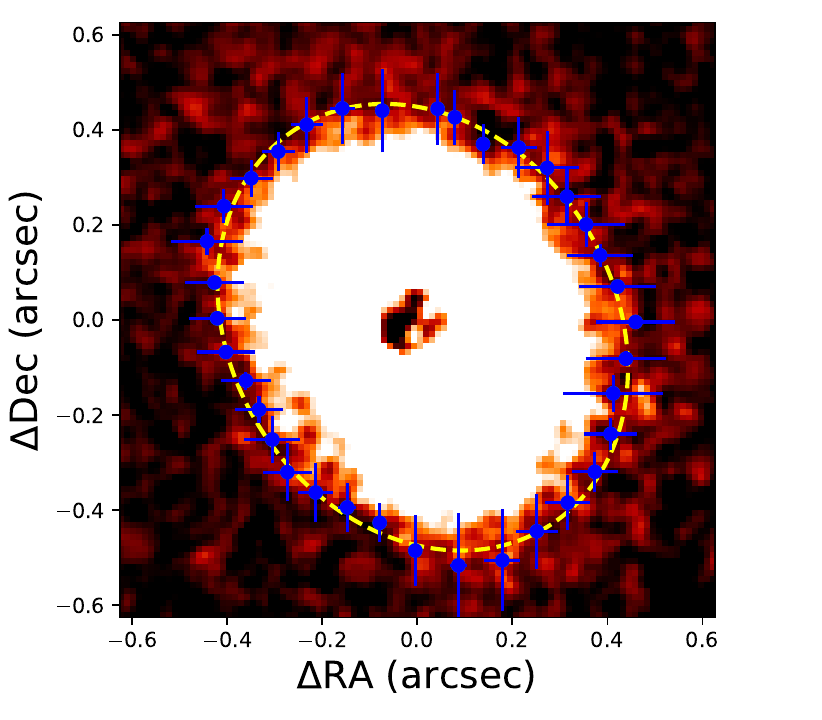}
\label{CRCha-ellipse}
}
\subfloat[CS Cha]{
\includegraphics[scale=0.4]{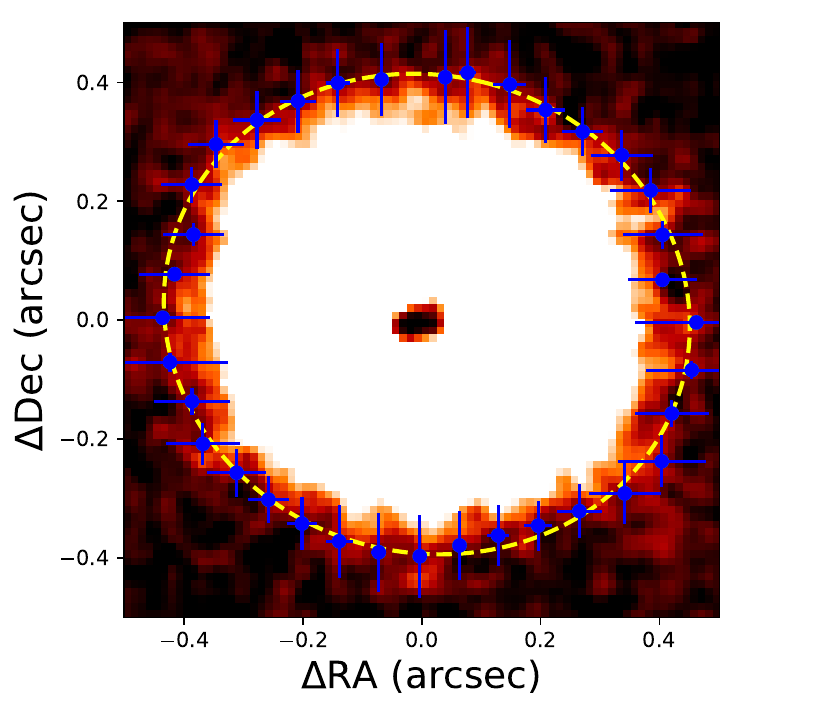}
\label{CSCha-ellipse}
}
\subfloat[CT Cha]{
\includegraphics[scale=0.4]{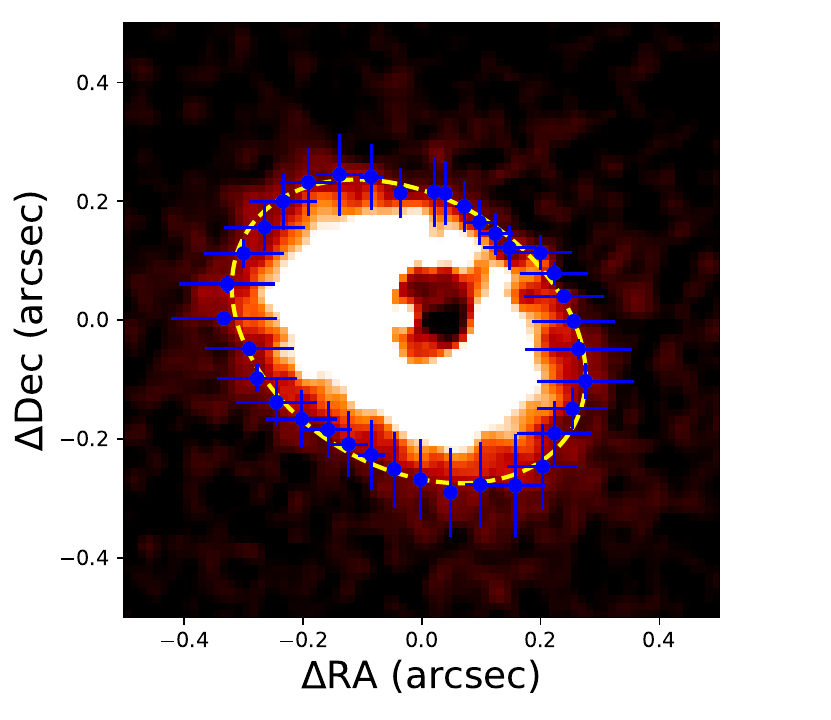}
\label{CTCha-ellipse}
}

\subfloat[CV Cha]{
\includegraphics[scale=0.4]{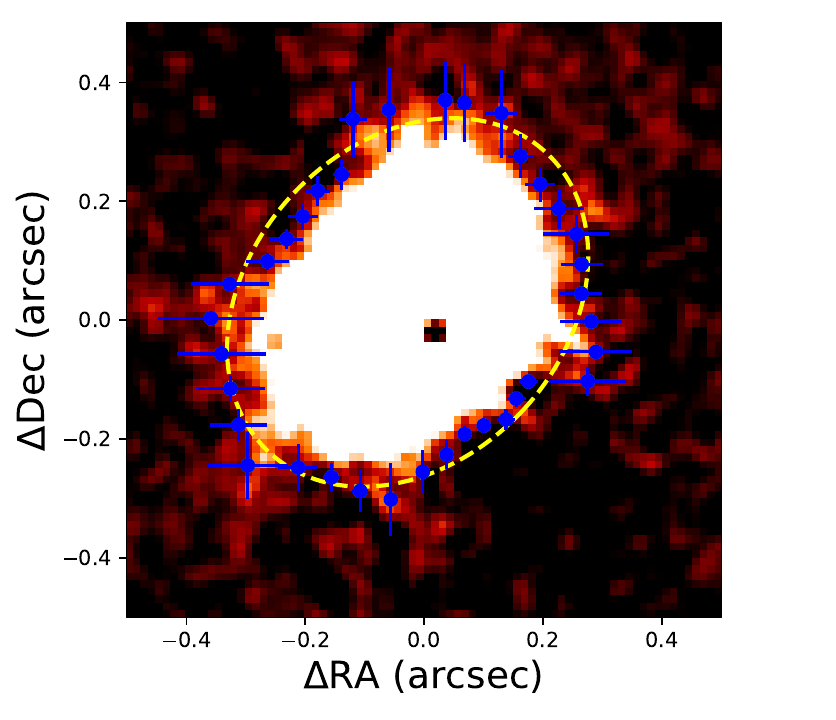}
\label{CVCha-ellipse}
}
\subfloat[HP Cha]{
\includegraphics[scale=0.4]{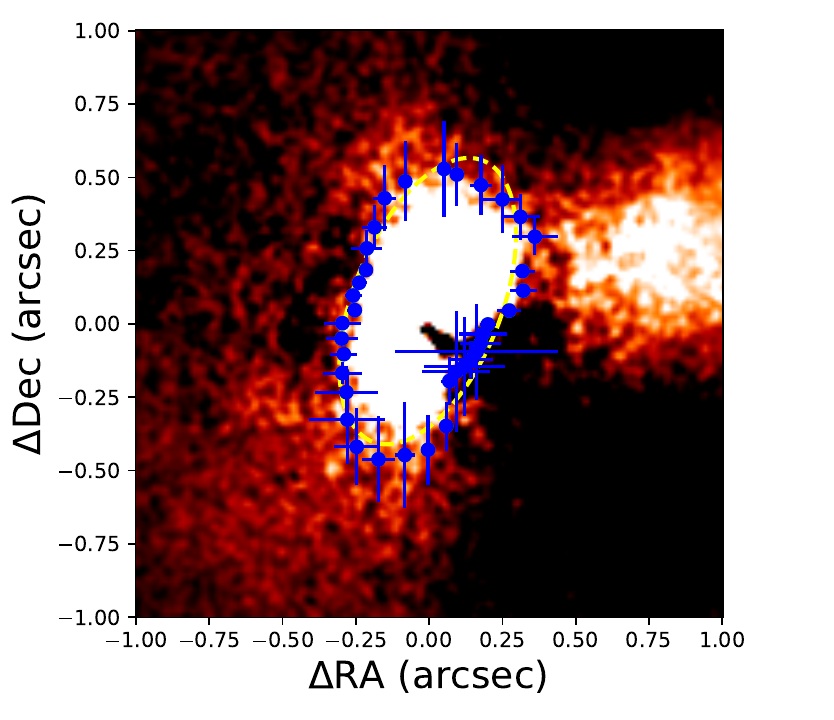}
\label{HPCha-ellipse}
}
\subfloat[SZ 45]{
\includegraphics[scale=0.4]{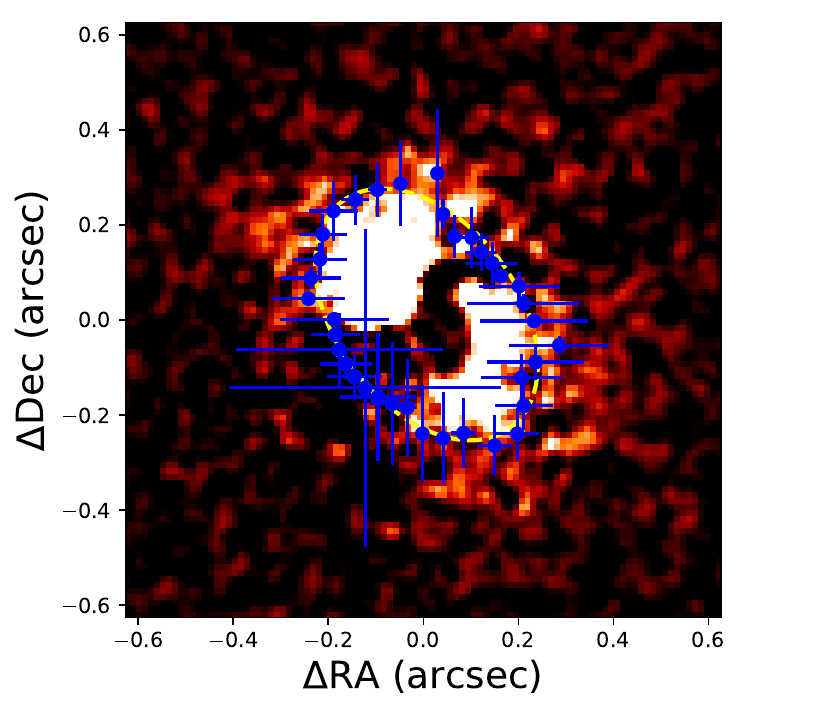}
\label{SZ45-ellipse}
}

\subfloat[TW Cha]{
\includegraphics[scale=0.4]{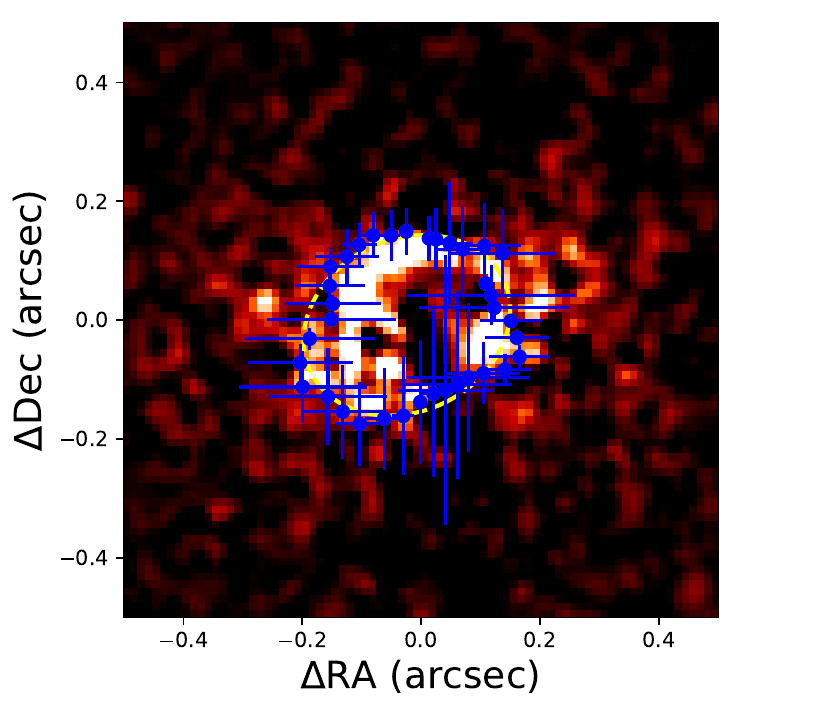}
\label{TWCha-ellipse}
}
\subfloat[VZ Cha]{
\includegraphics[scale=0.4]{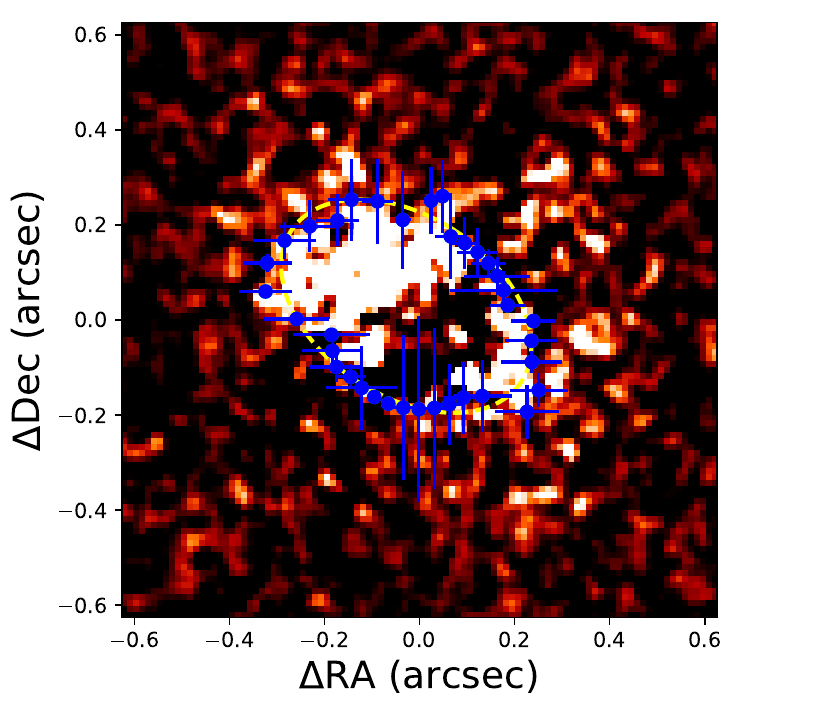}
\label{VZCha-ellipse}
}
\subfloat[WW Cha]{
\includegraphics[scale=0.4]{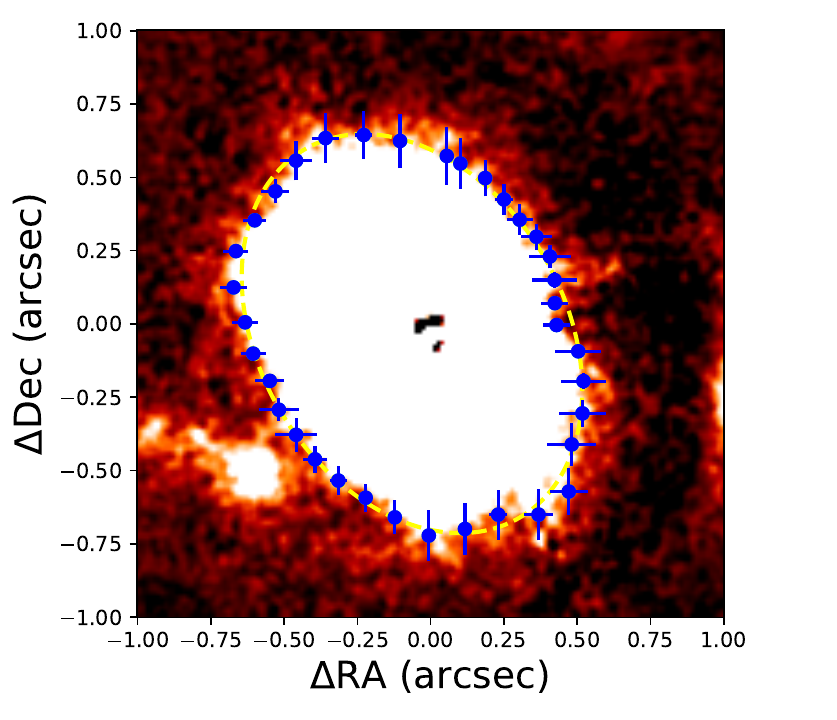}
\label{WWCha-ellipse}
}

\caption[]{Ellipse fit to the outer disk edge in 9 of the 20 systems in this study. Disk images are shown on a saturated color map to highlight the edge (i.e., the region where the  disk signal drops below 3$\sigma$ above the sky background). The yellow dashed lines show the final fitted ellipse. }
\label{fig: ellipse-fit}
\end{figure*}

\begin{figure*}
\centering
\subfloat[HD\,97048 inner ring]{
\includegraphics[scale=0.4]{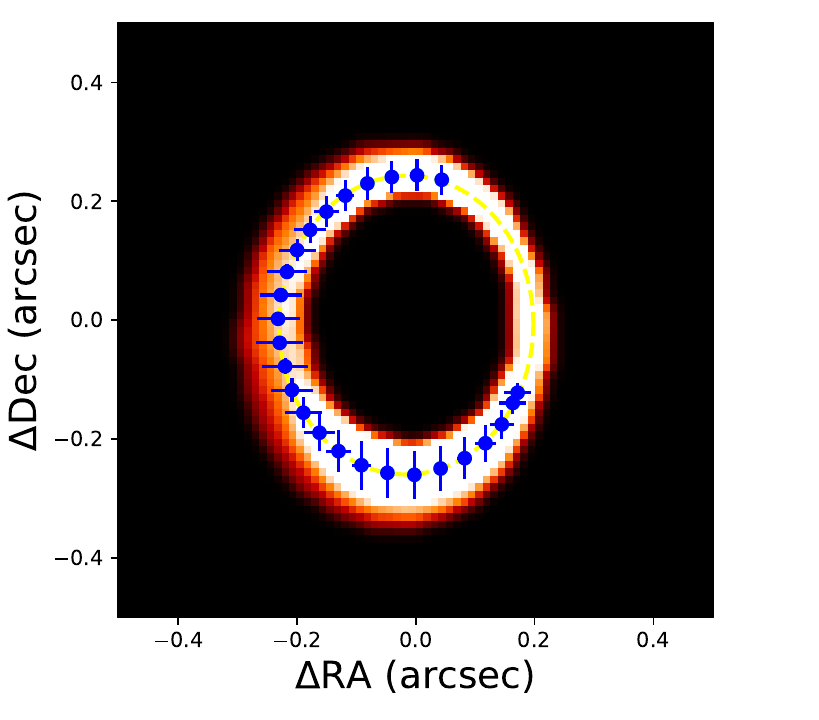}
\label{HD97048-ellipse-inner-ring}
}
\subfloat[HD\,97048 middle ring]{
\includegraphics[scale=0.4]{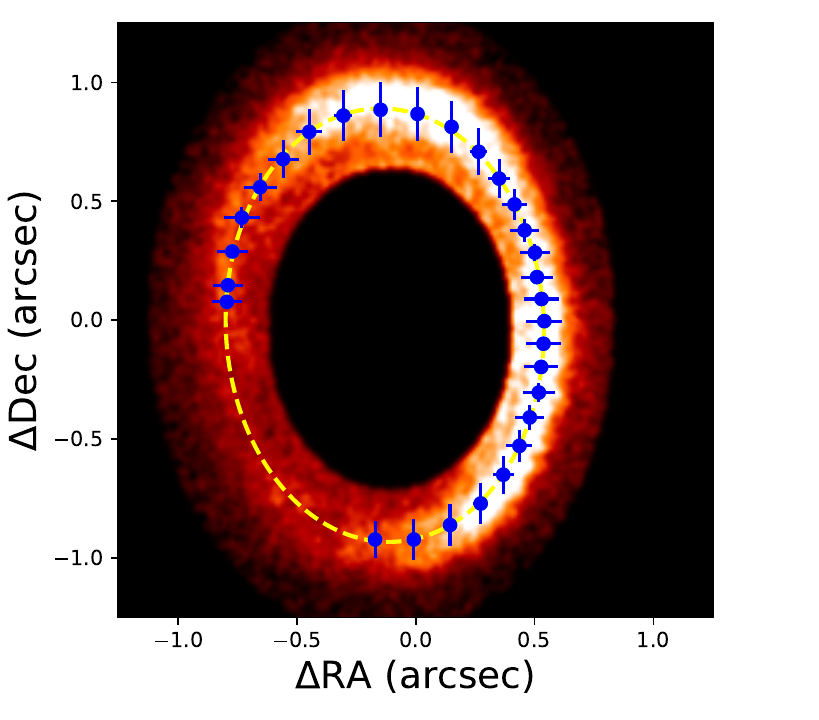}
\label{HD97048-ellipse-middle-ring}
}
\subfloat[HD\,97048 outer ring]{
\includegraphics[scale=0.4]{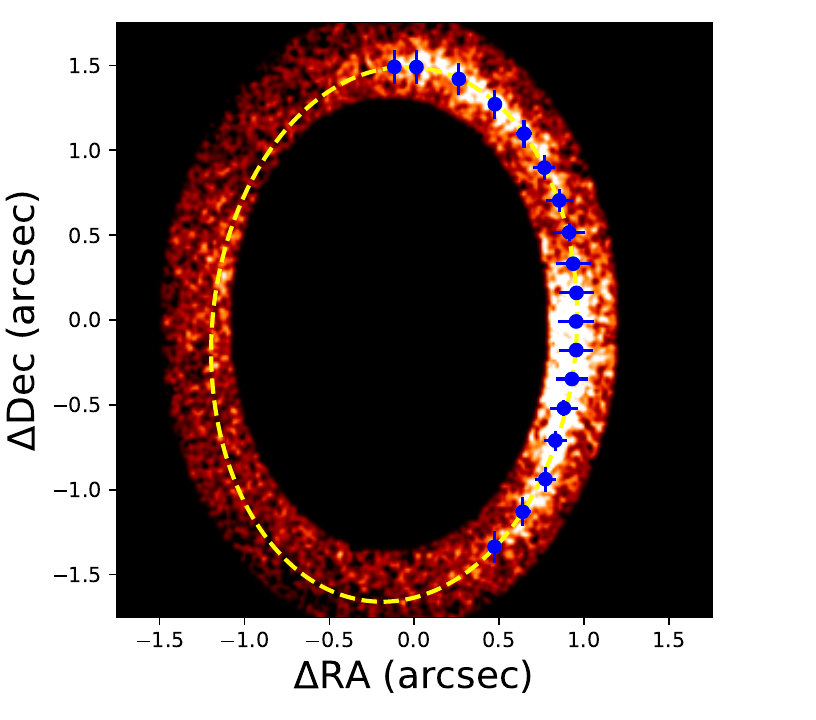}
\label{HD97048-ellipse-outer-ring}
}

\subfloat[SY\,Cha]{
\includegraphics[scale=0.4]{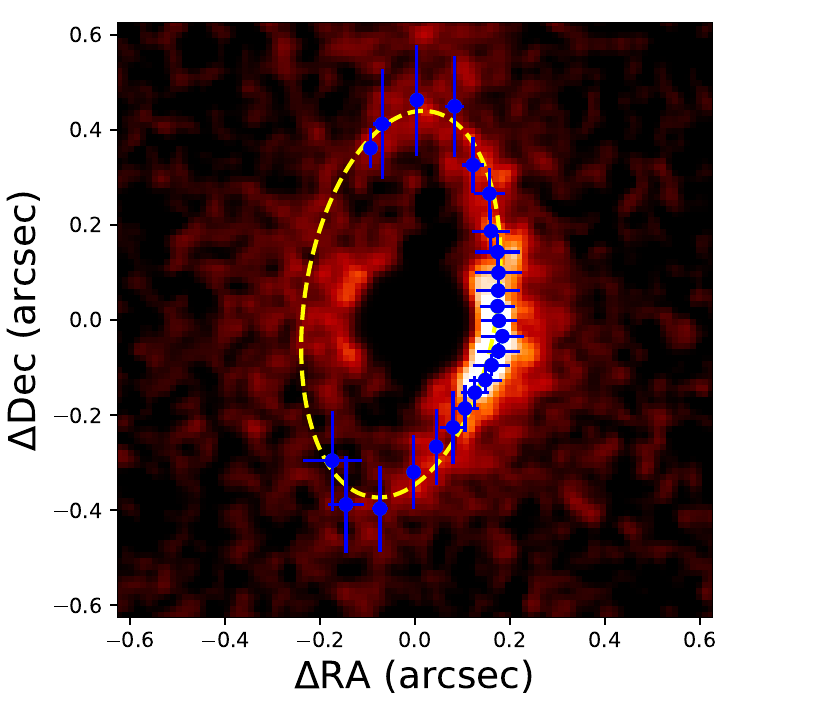}
\label{SYCha-ellipse}
}
\subfloat[SZ\,Cha inner ring]{
\includegraphics[scale=0.4]{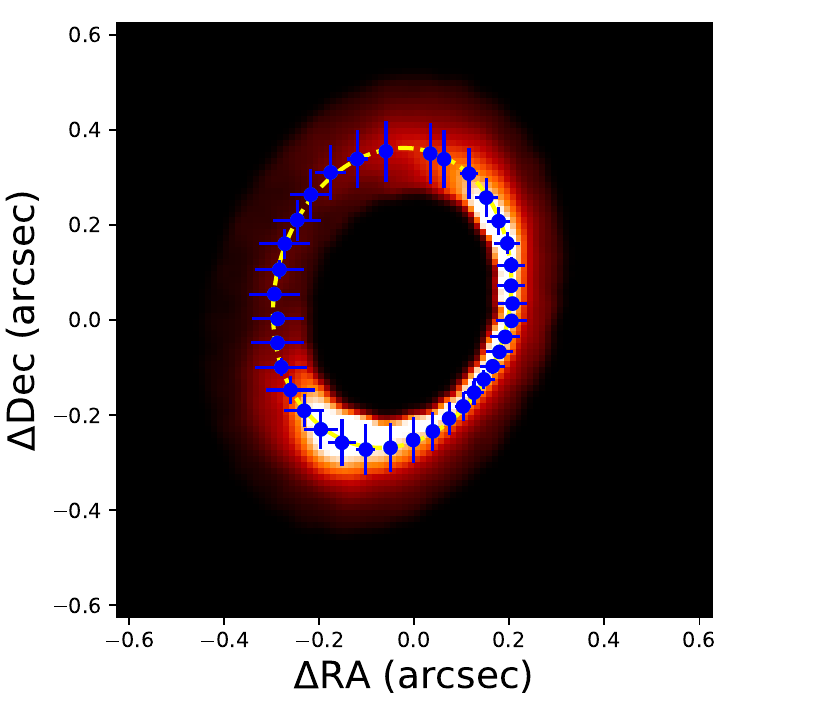}
\label{SZCha-ellipse-inner-ring}
}
\subfloat[SZ\,Cha outer ring]{
\includegraphics[scale=0.4]{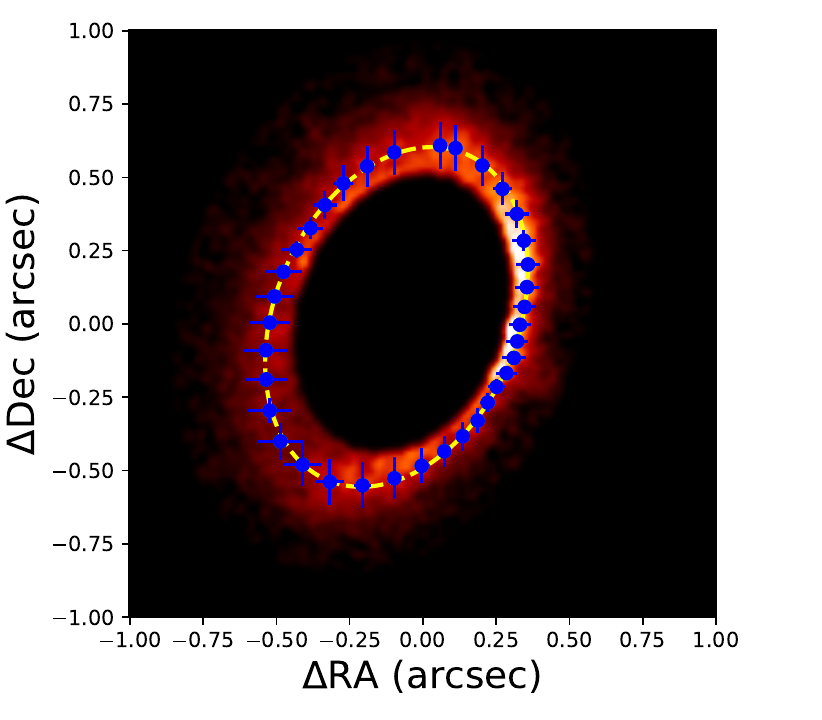}
\label{SZCha-ellipse-outer-ring}
}

\caption[]{Ellipse fit to the detected ring substructure in HD\,97048, SY\,Cha, and SZ\,Cha. Each ring was fitted individually, while other features were masked. The displayed fitted ellipses (yellow dashed lines) represent the nominal best fit to the data, not taking into account any prior knowledge of the disk inclination and position angle. }
\label{fig: ellipse-ring-fit}
\end{figure*}

In Figures~\ref{fig: ellipse-fit} and \ref{fig: ellipse-ring-fit} we show the extracted data points and best fitting LSMC ellipses for our data. We note that we intentionally display the data in a saturated color map to highlight the disk edge.

\section{Spectral energy distributions for sample systems}

In this appendix we show the spectral energy distributions for all sources included in our survey. The displayed data includes broad-band photometry (black data points), Spitzer/IRS spectra (red lines) and, if available, Herschel/SPIRE spectra (green lines). The data collection is taken from \cite{Ribas2017} (with references therein) for all sources but SY\,Cha, for which we did a similar data collection in this work.

\begin{figure*}
\centering
\includegraphics[width=\textwidth]{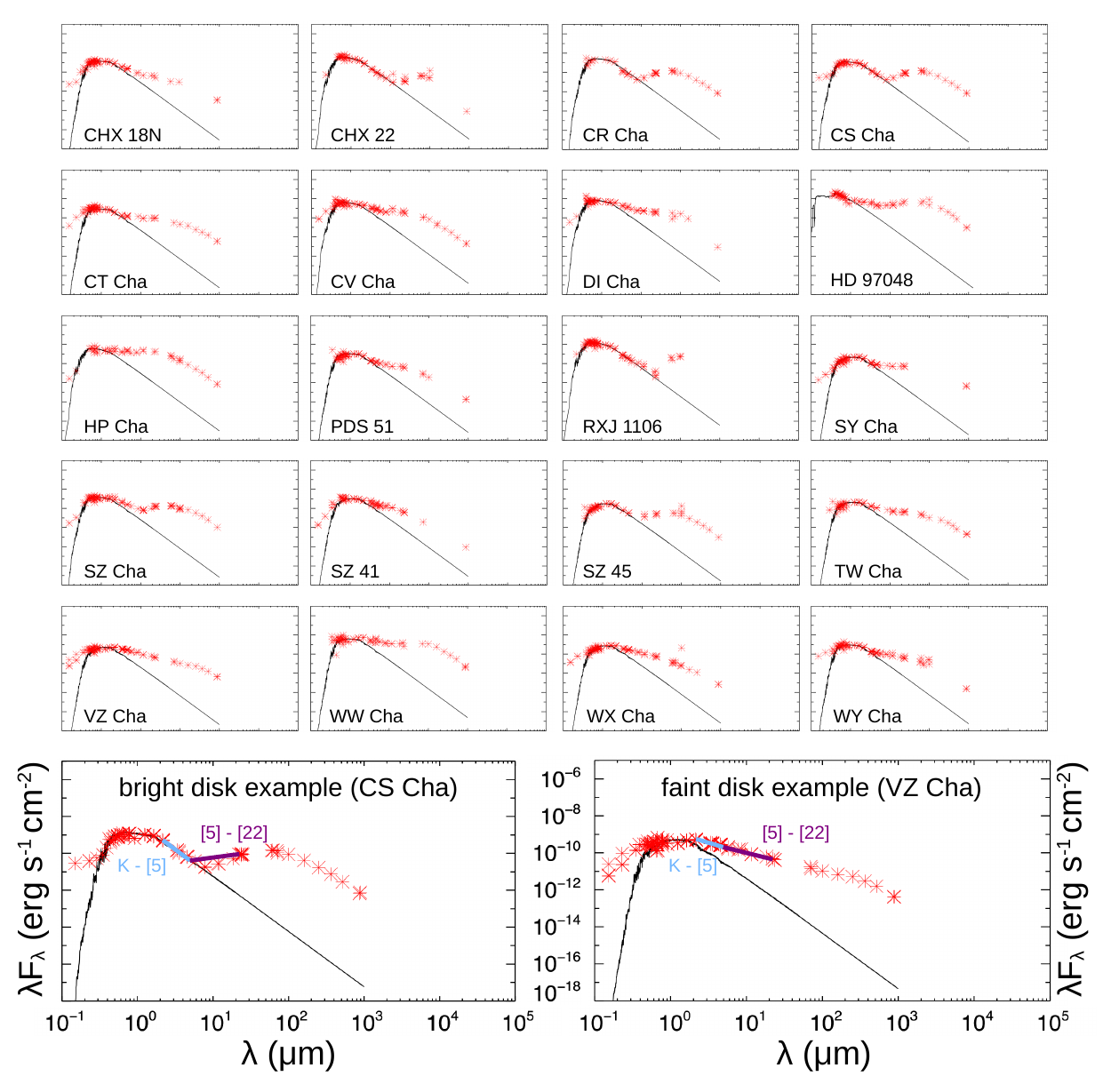}
\caption[]{Spectral energy distributions for all our target systems. Photometric data points are marked as red crosses (uncertainties are smaller than the symbol size).
The black lines mark the best fitting stellar models. 
In the last row  an example for a disk bright in scattered light and a disk faint in scattered light are highlighted. The blue and purple lines give the slopes used in figure~\ref{fig: slope-contrast} }
\label{fig: sed}
\end{figure*}

\section{CS\,Cha radiative transfer models}

We show the simulated scattered light images produced from the radiative transfer models utilized in the CS\,Cha color analysis discussed in section~\ref{cscha: disk-color}. 

\begin{figure*}
\centering
\includegraphics[width=0.999\textwidth]{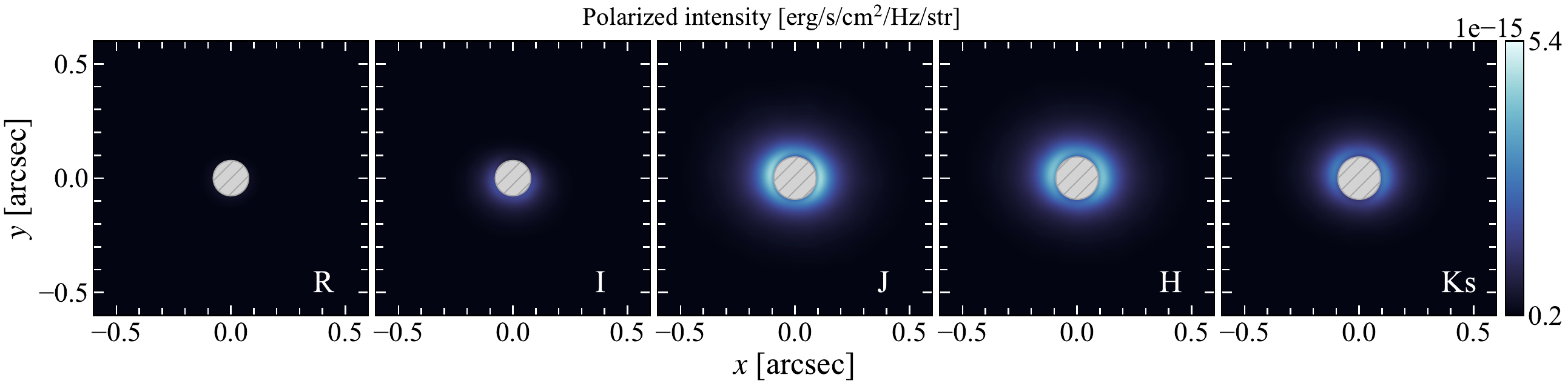}
\caption[]{Radiative transfer models of CS\,Cha created with RADMC3d using compact dust grains with a maximum size of 0.32$\mu m$. The system inclination is set to 22$^\circ$ with a flaring exponent of the disk surface of 1.09. The models closely match the lack of forward scattering also absent in the data, as well as the sharp drop in brightness between the J-band and the I-band.}
\label{fig: CSCha-RT}
\end{figure*}

\section{CS\,Cha\,Aab orbit fit}

\begin{figure*}
\centering
\includegraphics[width=20.0cm]{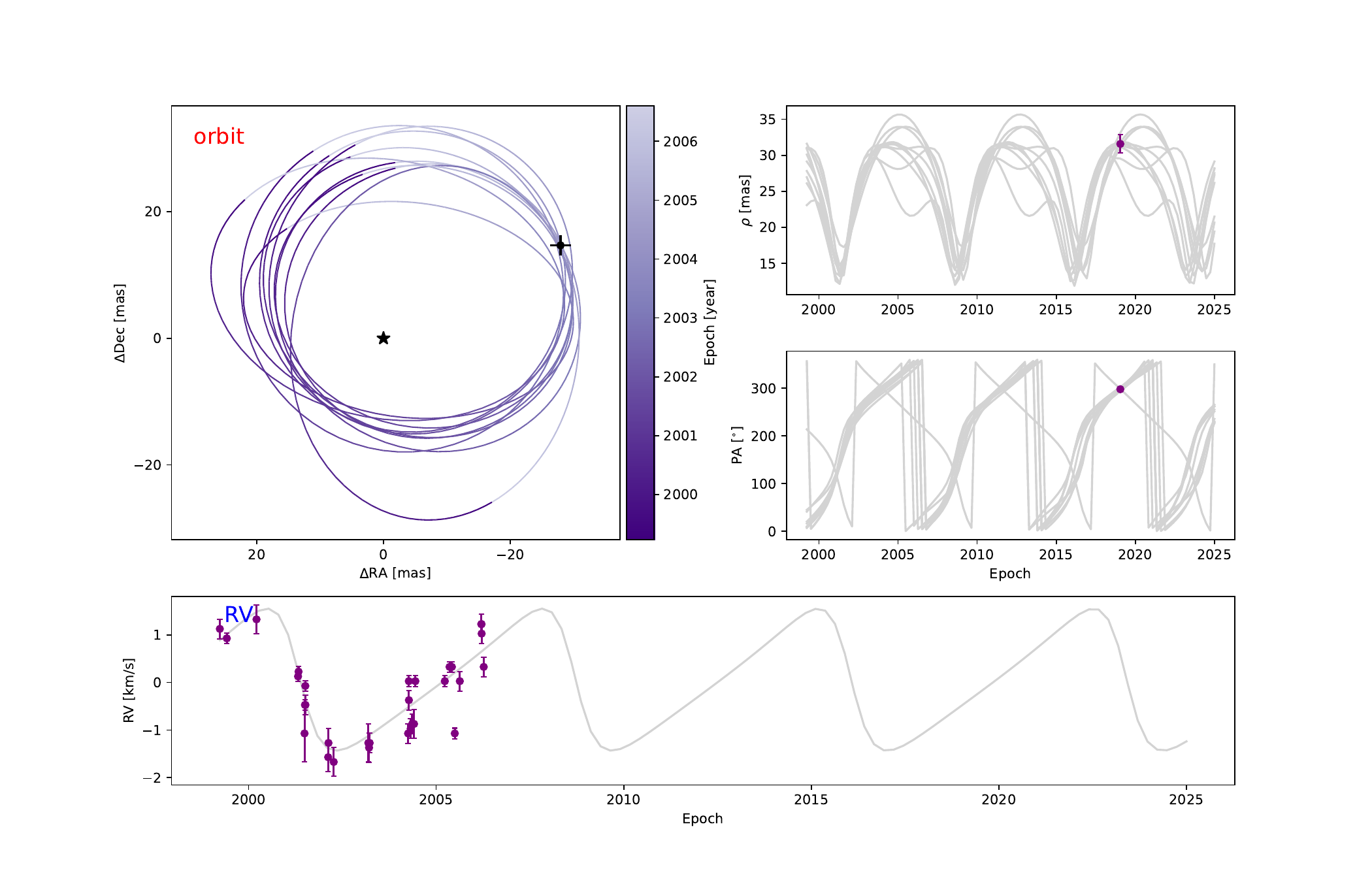}
\caption[]{Results of \emph{orbitize!} MCMC fit of the CS\,Cha stellar binary. The astrometry point (top panels) was retrieved in this study. The radial velocity data points are from \cite{Guenther2007}. The radial velocity data are plotted after subtraction of the radial velocity offset of 14.6\,km s$^{-1}$, as retrieved from our orbit fit.}
\label{fig: CSCha-orbit-appendix}
\end{figure*}

\begin{figure}
\centering
\includegraphics[width=9.0cm]{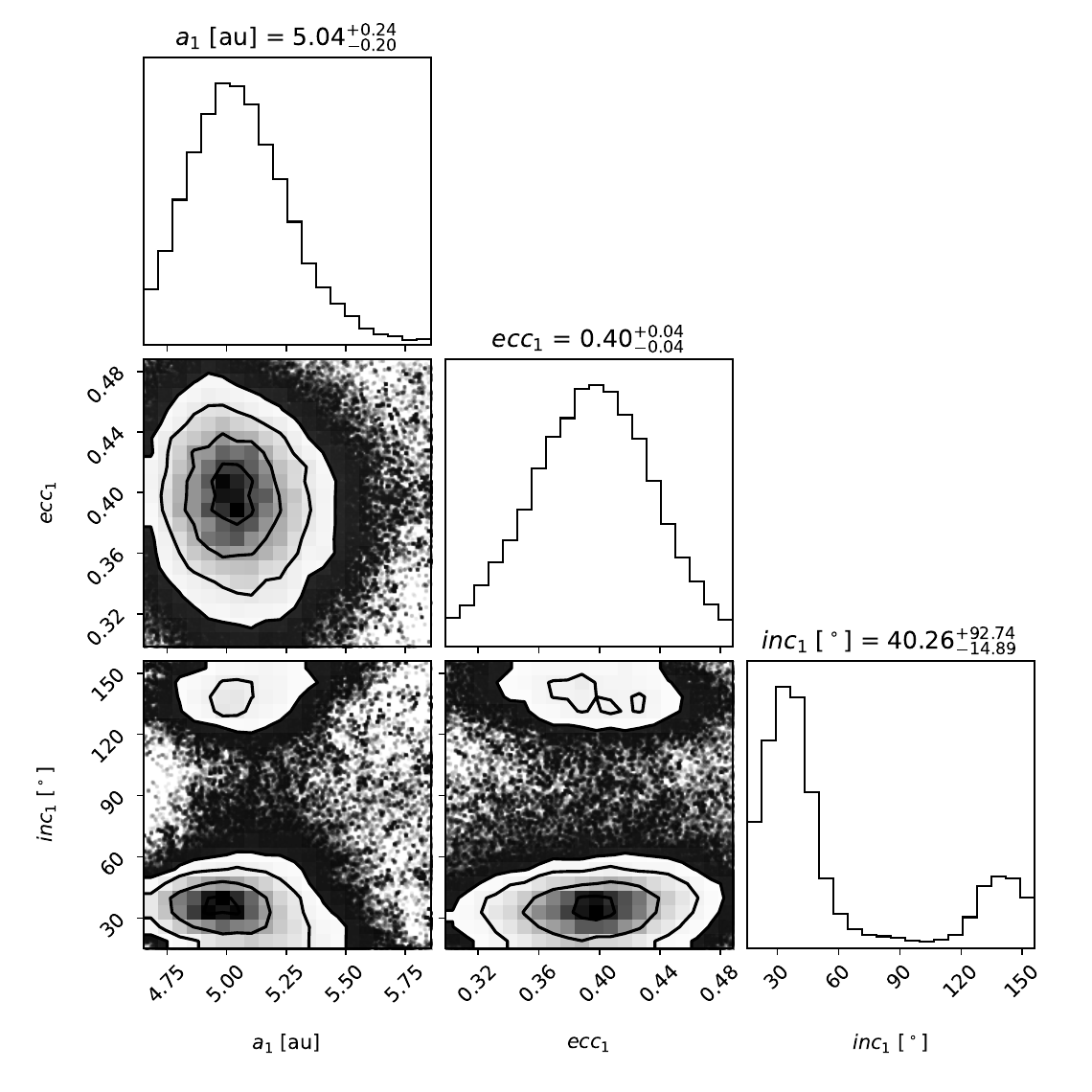}
\caption[]{Results of \emph{orbitize!} MCMC fit of the CS\,Cha stellar binary. Shown are the inclinations, eccentricities, and semi-major axes of the resulting orbit distributions.}
\label{fig: CSCha-orbit}
\end{figure}

We show the results of the orbit fits of the inner stellar binary in the CS\,Cha system, discussed in section~\ref{sec: CSCha-orbit}. In figure~\ref{fig: CSCha-orbit-appendix} we show a random selection of 10 possible orbits drawn from the posterior distribution of orbit elements. The top panels show the apparent orbits in R.A. - Dec space as well as separation - position angle space. The bottom panel shows the radial velocity of the primary star.  
In figure~\ref{fig: CSCha-orbit} we show the resulting posterior distributions of the orbit fit for the semi-major axis, inclination, and eccentricity. Displayed are the 95\,\% intervals for each parameter. 

\end{appendix}

\end{document}